\documentclass[journal,a4paper,10pt]{IEEEtran}
\usepackage[T1]{fontenc}
\usepackage[latin9]{inputenc}
\usepackage{epsfig,psfrag}
\usepackage{amsmath,amsfonts,amssymb,amsxtra,bm}
\usepackage{graphicx}
\usepackage{color}
\usepackage{cite}
\usepackage{tabularx, enumerate}
\usepackage{multirow}

\newcommand{\bd}[1]{\mathbf{#1}}
\newcolumntype{L}[1]{>{\raggedright\arraybackslash}p{#1}}
\newcolumntype{C}[1]{>{\centering\arraybackslash}p{#1}}
\newcolumntype{R}[1]{>{\raggedleft\arraybackslash}p{#1}}

\newcommand{\be}{\begin{equation}}
\newcommand{\ee}{\end{equation}}
\newcommand{\ist}{\hspace*{.3mm}}
\newcommand{\rmv}{\hspace*{-.3mm}}

\newcommand{\rrmv}{\hspace*{-1mm}}
\newcommand{\nn}{\nonumber}
\renewcommand{\aa}{\lambda}
\newcommand{\bb}{\nu}
\newcommand{\cl}[1]{\mathcal{#1}}

\newcommand{\internode}[1]{#1}

\allowdisplaybreaks

\newcommand{\remove}[1]{}

\usepackage{tikz}
\usetikzlibrary{arrows,backgrounds,calc,positioning,shapes,shadows}

\begin{document}
\title{Cooperative Simultaneous Localization and Synchronization in Mobile Agent Networks\vspace*{1mm}}

\author{Bernhard Etzlinger, Florian Meyer, Franz Hlawatsch, Andreas Springer, and Henk Wymeersch
\thanks{B.\ Etzlinger and A.\ Springer are with the Institute for Communications Engineering and RF-Systems, Johannes Kepler University, Linz, Austria, 
(email:\{bernhard.etzlinger,{\ist}andreas.springer\}@jku.at).
F.\ Meyer is with the NATO STO Centre for Maritime Research and Experimentation, La Spezia, Italy (e-mail: florian.meyer@cmre.nato.int).
F.\ Hlawatsch is with the Institute of Telecommunications, TU Wien, Vienna, Austria (email: franz.hlawatsch@tuwien.ac.at).
H.\ Wymeersch is with the Department of Signals and Systems, Chalmers University of Technology, Gothenburg, Sweden (email: henk.wymeersch@ieee.org).
This work was supported in part by the Linz Center of Mechatronics (LCM) in the framework of the Austrian COMET-K2 programme, 
by the NATO Supreme Allied Command Transformation under project SAC000601,
by the FWF under Grants S10603-N13 and P27370-N30, by the Newcom\# Network of Excellence in Wireless Communications of the European Commission, by the National Sustainability Program of the European Commission under Grant LO1401, by the European Research Council under Grant 
258418 (COOPNET), and by the EU HIGHTS project (High Precision Positioning for Cooperative ITS Applications) MG-3.5a-2014-636537.
Parts of this work were previously presented at the 47th Asilomar Conference on Signals, Systems and Computers, Pacific Grove, CA, Nov.\ 2013.}}

\maketitle

\begin{abstract}
Cooperative localization in agent networks based on interagent time-of-flight measurements is closely related to synchronization.
To leverage this relation, we propose a 
Bayesian factor graph framework for \emph{cooperative simultaneous localization and synchronization} (CoSLAS).
This framework is suited to 
mobile agents and time-varying local clock parameters.
Building on the CoSLAS factor graph, we develop a distributed (decentralized) belief propagation 
algorithm for CoSLAS in the practically important case of an affine clock model and asymmetric time stamping.
Our algorithm allows for
real-time operation and is suitable for
a time-varying network 
connectivity.
To achieve high accuracy at reduced complexity and communication cost, the algorithm combines particle implementations with parametric message representations
and takes advantage of a conditional independence property.
Simulation results demonstrate the good performance of the proposed algorithm in a challenging scenario with time-varying network connectivity.

\end{abstract}

\begin{IEEEkeywords}
Agent network,
network synchronization, 
cooperative localization, 
belief propagation, 
message passing, 
factor graph, 
CoSLAS.
\vspace{-.5mm}
\end{IEEEkeywords}


\section{Introduction}
\label{sec:introduction}

\vspace{.5mm}

\subsection{Background and State of the Art}

Location information in agent
networks  
enables a multitude
of location-aware applications \cite{corke10,zhao07,ko10,taranto14}.
In many systems,
the location information is obtained
from
interagent time measurements: 
each interagent distance is related to the time-of-flight of a signal and can thus be estimated from time-of-arrival 
measurements, and the
agent locations can then be estimated in a distributed (decentralized) manner via
cooperative localization techniques
\cite{patwari}. This scheme
presupposes a common time base at all the agents
and, thus, accurate synchronization throughout the network.
Accordingly, several methods for \emph{simultaneous localization and synchronization} (SLAS) have been developed recently.
These methods can be classified into six groups as follows.
Estimation of static clock
and 
location parameters is considered 
(i) for a single agent in \cite{zhu, chepuri, wang11, wu2}, 
(ii) for multiple agents with centralized computation in \cite{zachariah14,Vaghefi15,yerdir_SPAWC14}, and 
(iii) for multiple agents with distributed computation in \cite{yerdir_SPAWC14,benoit,meyer13asilomar,etzlinger13asilomar,yuanICCC14}. 
For a single agent,
(iv) estimation of dynamic clock
parameters and static location parameters is considered in \cite{Ahmad13}, and 
(v) estimation of static clock
parameters and dynamic location parameters is considered in \cite{li15journal}.
(vi) Distributed estimation of dynamic clock and location parameters of multiple agents is considered in \cite{yuanTVT16}.

Hereafter, we 
consider only \emph{distributed} SLAS methods for \emph{multiple} agents, i.e., methods from groups (iii) and (vi).
In these methods, the
local clocks differ either only in a clock offset \cite{yuanICCC14, yuanTVT16, yerdir_SPAWC14} 
or in both a clock offset and a clock skew 
\cite{benoit, meyer13asilomar, etzlinger13asilomar}. 
Considering also clock skews is important for accurate localization
when multiple time measurements are combined for each communication link \cite{etzlinger14sam}. 

To account for the nonlinear measurement model of the SLAS problem,
the distributed methods mentioned above 
use distributed least-squares (LS) or maximum likelihood 
estimation methods \cite{yerdir_SPAWC14,benoit} or Bayesian message passing methods  \cite{meyer13asilomar,etzlinger13asilomar,yuanICCC14,yuanTVT16}. 
Typically, message passing methods require significantly fewer iterations
than distributed LS methods \cite{yerdir_SPAWC14,benoit,meyer13asilomar}.
Despite this advantage, to the best of our knowledge, only \cite{yuanTVT16} previously proposed the message passing approach for SLAS
in mobile, dynamic agent networks. However, the method in \cite{yuanTVT16} is limited in practical scenarios 
in that no clock skews are considered, spatial references (anchors) must also serve as temporal references, and  
a linearization of the likelihood function is used that requires a dense deployment of anchors in the network.

Bayesian message passing methods 
are a powerful approach to 
cooperative estimation in agent networks and  
have been widely used for cooperative localization and cooperative synchronization individually \cite{ihler,wymeersch,caceres,meyer14sigma,etzlinger}.
To deal with nonlinearities in the message passing schemes, \cite{meyer13asilomar} and \cite{etzlinger13asilomar} use particle representations of messages 
whereas \cite{yuanICCC14} and \cite{yuanTVT16} use Gaussian messages based on the linearization of a specific 
term in the likelihood function.
The particle-based methods outperform the linearized Gaussian method if only few agents with a spatial reference are available;
this comes at the cost of higher communication requirements.
In cooperative localization, the communication requirements of message passing can be reduced by using a parametric message approximation \cite{caceres}
or a sigma point implementation \cite{meyer14sigma}.
In cooperative synchronization, Gaussian messages can be used because the measurement equations are approximately linear \cite{etzlinger}.

\vspace{-1mm}

\subsection{Contributions and Paper Organization}

Here,
we present a unified belief propagation (BP) message passing framework and algorithm
for distributed \emph{cooperative SLAS} (CoSLAS) in mobile agent networks with time-varying local clocks. BP 
methods 
provide accurate and computationally efficient solutions in many applications \cite{loeliger2007factor,wainwright,ihler,wymeersch,Wymeersch07_iterativeReceiver,etzlinger,meyer2016distributed}. 
In the proposed BP framework,
a low dimension of the involved state variables
is achieved by exploiting the conditional independence of 
time measurements and 
location-related parameters given the interagent distances, which leads to a detailed factorization of the joint posterior 
probability density function (pdf). In this factorization, the dimension of the 
state variables does not depend on the number of agents in the network, thus yielding excellent scalability.

The proposed BP 
algorithm 
enables each agent to determine its own clock 
and location parameters in a distributed, cooperative, and sequential manner. 
The
algorithm is a hybrid---both
particle-based and 
parametric---implementation of BP
that relies
on a specific, practically relevant model for
the clocks, state evolutions, and measurements. 
This model supports parametric representations of all messages, which strongly
reduces computation and communication requirements compared
to
purely particle-based methods
\cite{etzlinger13asilomar}.
The 
algorithm extends state-of-the-art methods in that it is suited to
time-varying clock and location parameters, time-varying network connectivity, and networks where
the sets of spatial and temporal reference agents may be different or even disjoint.

This paper is organized as follows. 
The agent network, clock model, and state evolution model
are described in Section \ref{sec:systemModel}. The
measurement model and 
corresponding likelihood function
are developed
in Section \ref{sec:likelihood}.
In Section \ref{sec:estimation}, we present a ``low-dimensional'' factorization of the joint posterior pdf 
and the corresponding factor graph, and we review the BP 
scheme for approximate marginalization.
The parametric message representations used by our algorithm are described in Section \ref{sec:messagerep}.
Section \ref{sec:extended} develops the proposed CoSLAS algorithm.
Finally, Section \ref{sec:simulations} presents simulation results.

This paper advances beyond the results 
reported in our conference publication \cite{etzlinger13asilomar} in that
(i) it extends the CoSLAS factor graph framework and BP message passing algorithm of
\cite{etzlinger13asilomar} to a time-dependent senario and a sequential (time-recursive) operation;
(ii) it presents a 
BP algorithm for mobile agents with time-varying local clocks; 
(iii) it proposes parametric representations for all messages.
  


\section{Network and States}
\label{sec:systemModel}

\vspace{1mm}

\subsection{Agent Network, Clock Model, and States} \label{ssec:net_clock_model}
\label{sec:systemModel-1}


We consider a connected
time-varying network of $I$ mobile, asynchronous agents $i \rmv\in\rmv \mathcal{I} \rmv\triangleq\rmv \{1,\ldots,I\}$. 
The reference time, $t$, is slotted into intervals $[ n T, (n \rmv+\! 1)\ist T)$, $n \rmv\in\rmv \{0,1,\ldots\}$.
The agents know the interval duration $T$ but, due to their imprecise clocks,
are not able to autonomously determine the beginning of a new time interval. 
At time step $n$, i.e., during the $n$th time interval, 
two agents $i,j \!\in\! \mathcal{I}$, $i \!\not=\! j$ are able to communicate if $(i,j) \!\in\rmv \mathcal{C}^{\internode{(n)}} \!\rmv\subseteq\rmv \mathcal{I} \!\times\! \mathcal{I}$ 
(and, by symmetry, $(j,i) \!\in\rmv \mathcal{C}^{\internode{(n)}}$).
The \emph{neighborhood} $\mathcal{T}^{\internode{(n)}}_i \!\subseteq \mathcal{I} \rmv\setminus\! \{i\}$ of agent $i \!\in\! \mathcal{I}$ consists of all agents
$j \!\in\! \mathcal{I}\rmv\setminus\! \{i\}$ that communicate with agent $i$ at time step $n$, i.e.,
$\mathcal{T}^{\internode{(n)}}_i\! \triangleq \big\{ j \!\in\! \mathcal{I} \rmv\setminus\! \{i\} \ist\big|\ist (i,j) \!\in\! \mathcal{C}^{\internode{(n)}} \big\}$. 
Note that $\mathcal{C}^{\internode{(n)}}$ and $\mathcal{T}^{\internode{(n)}}$
are assumed 
constant within the $n$th time interval. 
Some of the agents $i$ are spatial and/or temporal references, which have perfect knowledge of their own location and/or clock, respectively, at all times.
In particular, a temporal reference agent is able to determine the beginning of a new time interval.

Each agent $i \!\in\! \mathcal{I}$ has an internal/local clock $c_i$, whose dependence on the reference time $t$ is modeled as
  \begin{equation}
    c_i \big(t;\bm{\vartheta}^{\internode{(n)}}_{i}\big) \ist=\ist \alpha_i^{(n)} t + \beta_i^{(n)} . 
    \label{eq:aff_clk}
\vspace{-.5mm}
  \end{equation}
Here, $\alpha_i^{(n)} \!>\rmv 0$ and $\beta_i^{(n)} \!\in\rmv \mathbb{R}$ are the \emph{clock skew} and \emph{clock\linebreak 
phase}, respectively, which 
define the \emph{clock state}
${\bm{\vartheta}}_{i}^{(n)}  \!\triangleq \big[\bb_{i}^{(n)} \; \aa_{i}^{(n)}\big]^\text{T}\rmv$ with 
$\bb_{i}^{(n)} \!\triangleq \beta_{i}^{(n)}/\alpha_{i}^{(n)}\!$ and $\aa_i^{(n)} \!\triangleq 1/\alpha_{i}^{(n)}\rmv$. (This parameter transformation
leads to
an approximately Gaussian 
likelihood function, cf.\ Section \ref{ssec:LLF}.)
Each agent $i$ has a \emph{location-related state} $\bd{x}^{(n)}_{i}\rmv\triangleq \big[ \bd{p}^{(n)\ist\text{T}}_{i} \; \dot{\bd{p}}^{(n)\ist\text{T}}_{i} \big]^\text{T}\rmv$, 
where $\bd{p}^{\internode{(n)}}_i \!\triangleq \big[ x^{(n)}_{1,i}\; x^{(n)}_{2,i}\big]^\text{T}\rmv$ is the location vector and
$\dot{\bd{p}}^{(n)}_{i} \rmv\triangleq \big[\dot{x}^{(n)}_{1,i} \; \dot{x}^{(n)}_{2,i} \big]^\text{T}$ is the velocity vector (relative to $t$).
The \emph{state} of agent $i$ at time step $n$ is thus given by 
$\bm{\theta}^{\internode{(n)}}_{i} \triangleq \big[ \bm{\vartheta}^{\internode{(n)}\ist\text{T}}_{i} \,\ist\ist \bd{x}^{\internode{(n)}\ist\text{T}}_{i}\big]^\text{T}\!$. 
We note that $\bd{p}_i^{(n)} \!=\rmv \bd{P} \bd{x}_i^{(n)}$ with $\bd{P} = \big[\bd{I}_2 \,\ist \bd{0}_2 \big]$, where $\bd{I}_2$ is the $2 \!\times\! 2$ identity matrix and
$\bd{0}_2$ is the $2 \!\times\! 2$ zero matrix.

\vspace{-1mm}

\subsection{State-Evolution Model and Prior Distribution} 
\label{ssec:evolut}

For the temporal evolution of the clock state $\bm{\vartheta}_i^{(n)}\!$, 
we use a standard random walk model as in \cite{Ahmad13}, i.e.,
\be
  \bm{\vartheta}_{i}^{(n)} =\, \bm{\vartheta}_{i}^{(n-1)} \rmv+ \bd{u}_{1,i}^{(n)} \ist, \quad\;\; n \rmv=\rmv 1,2,\dots \,, 
\label{eq:clockEvol}
\vspace{-.5mm}
\ee
where $\bd{u}_{1,i}^{(n)} \rmv\sim\rmv \mathcal{N}\big(\bd{u}_{1,i}^{(n)};\bd{0}, \bm{\Sigma}_{u_{1,i}} \big)$ with $\bm{\Sigma}_{u_{1,i}} = \mathrm{diag}\big\{ \sigma^2_{1,i},
\linebreak
\sigma^2_{2,i}\big\}$ 
is Gaussian process noise that is independent across $n$ and $i$.
The state-evolution pdf corresponding to \eqref{eq:clockEvol} is
\[
f\big(\bm{\vartheta}_i^{(n)} \big| \bm{\vartheta}_i^{(n-1)} \big) \propto\ist \exp\rmv\bigg({-\frac{1}{2}} \ist \big\|\bm{\vartheta}_i^{(n)} \rmv\! - \bm{\vartheta}_i^{(n-1)}\big\|^2_{\bm{\Sigma}_{u_{1,i}}^{-1}} \bigg) \ist ,
\]
where $\| \mathbf{v} \|^2_{\bd{A}} \!\triangleq\rmv \mathbf{v}^\text{T} \rmv \bd{A} \mathbf{v}$.
The temporal evolution of the location-related state $\bd{x}^{(n)}_{i}$
is modeled as \cite{sathyan13} 
  \begin{equation}
  \bd{x}_{i}^{(n)} =\, \bd{G}_1\bd{x}^{(n-1)}_{i} \rmv+ \bd{u}_{2,i}^{(n)} \ist, \quad\;\; n \rmv=\rmv 1,2,\dots \,,
  \label{eq:posEvol}
\vspace{-.5mm}
  \end{equation}
  where  $\bd{u}_{2,i}^{(n)} {\sim}\ist \mathcal{N}\big(\bd{u}_{2,i}^{(n)};\bd{0}, \bm{\Sigma}_{u_{2,i}}\big)$ with $\bm{\Sigma}_{u_{2,i}} {=}\ist \sigma_{u_2,i}^2 \bd{G}_2$;
  here, $\bd{G}_1$ and $\bd{G}_2$ are as in \cite{sathyan13}.
The state-evolution pdf corresponding to \eqref{eq:posEvol} is
\[
f\big(\mathbf{x}_i^{(n)} \big| \mathbf{x}_i^{(n-1)} \big) \propto\ist \exp\rmv\bigg({-\frac{1}{2}} \ist \big\|\mathbf{x}_i^{(n)} \! -\rmv \mathbf{G}_1 \mathbf{x}_i^{(n-1)}\big\|^2_{\bm{\Sigma}_{u_{2,i}}^{-1}} \bigg) \ist .
\vspace{-1mm}
\]
Furthermore, $\bd{u}_{1,i}^{(n)}$ 
and $\bd{u}_{2,i}^{(n)}$ 
are assumed independent and also independent across $i$ and $n$.
The initial states ${\bm{\vartheta}}^{\internode{(0)}}_{i}\rmv$ and $\bd{x}^{\internode{(0)}}_{i}\rmv$ are modeled
as independent, independent across $i$, and Gaussian with independent entries,
i.e., 
\begin{align}
  {\bm{\vartheta}}^{\internode{(0)}}_{i} & \rmv\sim f\big({\bm{\vartheta}}_i^{(0)}\big) 
  \ist=\ist \mathcal{N}\big({\bm{\vartheta}}^{\internode{(0)}}_{i};\bm{\mu}_{{f}_i\to {\vartheta}_i}^{(0)}, \bm{\Sigma}_{{f}_i\to {\vartheta}_i}^{(0)} \big) \ist , \label{eq:prior_theta}
  \\[.8mm]
  \bd{x}^{\internode{(0)}}_{i} & \rmv\sim f\big(\bd{x}_i^{(0)}\big) 
  \ist=\ist \mathcal{N}\big(\bd{x}^{\internode{(0)}}_{i};\bm{\mu}_{l_i\to x_i}^{(0)}, \bm{\Sigma}_{l_i\to x_i}^{(0)} \big) \ist , \label{eq:prior_x} \\[-6mm]
\nonumber
\end{align}
with $\bm{\Sigma}_{{f}_i\to {\vartheta}_i}^{(0)} \!= \mathrm{diag}\big\{
\sigma^2_{\nu_i}, \sigma^2_{\lambda_i}\big\}$ and 
$\bm{\Sigma}_{l_i\to x_i}^{(0)} \!= \mathrm{diag}\big\{
\sigma^2_{x_i}, \sigma^2_{x_i},$\linebreak 
$\sigma^2_{\dot{x}_i},\sigma^2_{\dot{x}_i}\big\}$.
It follows that the joint prior pdf 
\pagebreak 
of all the states up to time $n$ factors as
\begin{align}
&\hspace{-2mm} f\big({\bm{\theta}}^{\internode{(0:n)}}\big) =\ist \prod_{i \in \mathcal{I}}   f\big({\bm{\vartheta}}_i^{(0)}\big) f\big(\bd{x}_i^{(0)} \big) \nonumber\\[-1mm]
&\hspace{17mm} \times \prod_{n'=1}^n \! f\big({\bm{\vartheta}}_i^{(n')}\big| {\bm{\vartheta}}_i^{(n'-1)}\big) \, f\big({\bd{x}}_i^{(n')}\big| {\bd{x}}_i^{(n'-1)}\big)  \, .
\label{eq:prior}\\[-6.5mm]
\nn
\end{align}
Here, ${\bm{\theta}}^{\internode{(0:n)}}$ collects all 
$\bm{\theta}_{i}^{\internode{(n')}}$ for $i \rmv\rmv\in\rmv\rmv \mathcal{I}$ and $n' \rmv\rmv\in\rmv\rmv \{0,\ldots,n\}$.

\section{Measurements
and Likelihood Function}
\label{sec:likelihood}

\vspace{1mm}


\subsection{Time-Stamping Measurement Model} 
\label{ssec:asymTimeStamp}


Each time 
interval $[nT,(n+1)T)$ contains a ``measurement phase'' in which the agents acquire 
measurements.
Each measurement phase consists of an \emph{initialization} in which the temporal reference agents inform
the other agents about the beginning of the measurement phase, and a \emph{packet exchange} during which the agents obtain time measurements using
the asymmetric time-stamped communication scheme proposed in \cite{chepuri13}. 
The 
measurement phase
is short compared to the time interval duration $T$,
so that the clock parameters are
approximately constant during the measurement phase. 

\textit{1. Initialization:} 
The agents are not able to 
determine autonomously 
the start of a new time interval and, in turn, of a packet exchange.
This information is provided by the temporal reference agents 
via
the following 
protocol:
(i) After time $T$ has passed since the beginning of the last measurement phase, each temporal reference agent initializes a new time interval
by broadcasting a ``start packet exchange'' message to its neighbors.
(ii) When an agent receives a ``start packet exchange'' message from one of its neighbors, it starts the packet exchange with that neighbor and itself
broadcasts a ``start packet exchange'' message to its neighbors.

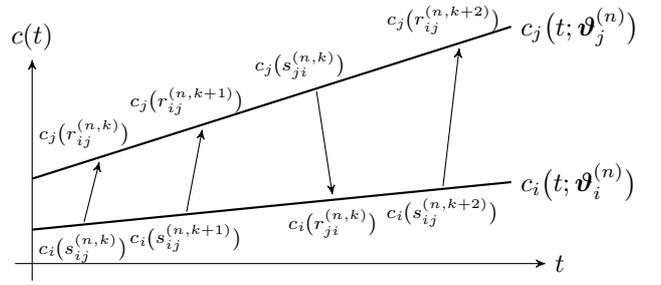
\begin{figure}[t!]
  \centering
	    \begin{tikzpicture}[scale=4.5, >=stealth',
		  axis/.style={->},
		  counter line/.style={thick},
		  packet/.style={->, shorten <=1pt, shorten >=2pt},
		  measure/.style={<->, thin ,draw=black!30, fill=black!30},
		  ]
	

		\draw[axis] (-0.05,0)  -- (1.5,0) node(xline)[right] {$t$};
		\draw[axis] (0,-0.05) -- (0,0.6) node(yline)[above] {$c(t)$};
	  
		\def\alphaM{0.1}; \def\betaM{0.1};
		\def\alphaS{0.32}; \def\betaS{0.25};
		\draw[domain=0:1.4,samples=2,counter line] plot (\x, {\alphaS*\x+\betaS});
		  \node [right] (S) at (1.4,{\alphaS*1.4+\betaS}) {$c_j\big(t;\bm{\vartheta}_j^{(n)}\big)$};
		\draw[domain=0:1.4,samples=2,counter line] plot (\x, {\alphaM*\x+\betaM});
		  \node [right] (M) at (1.4,{\alphaM*1.4+\betaM}) {$c_i\big(t;\bm{\vartheta}_i^{(n)}\big)$};
		
		\def\Del{0.05};
		\foreach \x/\pnum in {0.15/{k}, 0.45/{k+1}, 1.2/{k+2}}{
			\draw[packet] (\x ,\alphaM*\x+\betaM) node (TXm)[below]{\scriptsize$c_i\rmv\big(s_{ij}^{\internode{(n,\pnum)}}\big)$} -- (\x+\Del ,\alphaS*\x+\alphaS*\Del+\betaS) node (RXs)[above]{\scriptsize \hspace{-0.5cm} $c_j\rmv\big(r_{ij}^{\internode{(n,\pnum)}}\big)$};
		}
		\foreach \x/\pnum in {0.83/{k}}{
			\draw[packet] (\x,\alphaS*\x+\betaS) node (TXs)[above]{\scriptsize \hspace{-0.5cm} $c_j\rmv\big(s_{ji}^{\internode{(n,\pnum)}}\big)$} -- (\x+\Del  ,\alphaM*\x+\alphaM*\Del+\betaM) node (RXm)[below]{\scriptsize$c_i\rmv\big(r_{ji}^{\internode{(n,\pnum)}}\big)$};
		}

	    \end{tikzpicture}
	  \caption{\label{fig:2_way_msg} Local clock functions $c_i\big(t,\bm{\vartheta}_i^{(n)}\big)$ and $c_j\big(t,\bm{\vartheta}_j^{(n)}\big)$, 
	  packet transmissions, and 
	  local time measurements (time stamps) for agents $i$ and $j$.}
\vspace{-2mm}

  \end{figure}

\textit{2. Packet exchange:}
Consider a
communicating agent pair $(i,j) \!\in\rmv \mathcal{C}^{(n)}$ with distance $\big\|\bd{p}_i^{(n)} \!\rmv-\rmv \bd{p}_j^{(n)}\big\|$.
Agent $i$ transmits $K_{ij}\!\ge\! 1$ packets to agent $j$, and agent $j$ transmits $K_{ji}\!\ge\! 1$ packets to agent $i$. 
The communication is termed asymmetric if $K_{ij} \neq K_{ji}$ \cite{chepuri}.
At time $n\geq 1$, the $k$th ``$i \rmv\to\! j$'' packet 
(where $k \in \{1,\ldots, K_{ij}\}$) departs from
agent $i$ at time $s_{ij}^{\internode{(n,k)}}$ and arrives at agent $j$ at measured time 
  \begin{equation}
  r_{ij}^{\internode{(n,k)}} \!= s_{ij}^{\internode{(n,k)}} \rmv+ \delta_{ij}^{\internode{(n,k)}}\!, \;\text{with}\;\, 
  \delta_{ij}^{\internode{(n,k)}} \!\triangleq\ist 
  \frac{\|\mathbf{p}_i^{(n)}\rrmv-\rmv\mathbf{p}_j^{(n)}\|}{c} 
  + v_{ij}^{\internode{(n,k)}}\!.
  \label{eq:mess_rel}
  \end{equation}
Here, $\delta_{ij}^{\internode{(n,k)}}$ is the delay expressed in true time, 
$c$ is the speed of light, and 
$v_{ij}^{\internode{(n,k)}} \!\!\sim\! 
\mathcal{N}\big(v_{ij}^{\internode{(n,k)}};0,\sigma^2_v\big)$ is Gaussian measurement noise that is 
independent and identically distributed (iid) across $i$, $j$, $k$, and $n$.
The transmit times $s_{ij}^{\internode{(n,k)}}$ and receive times $r_{ij}^{\internode{(n,k)}}$ are recorded at agent $i$ and $j$, respectively 
in local time according to \eqref{eq:aff_clk}. This results in the \emph{time stamps} 
\begin{align}
c_{i}\big(s_{ij}^{\internode{(n,k)}}\big) &\ist=\ist \alpha^{\internode{(n)}}_i \rmv s_{ij}^{\internode{(n,k)}} + \beta^{\internode{(n)}}_i , \label{eq:timestamp_i} \\[.5mm]
c_{j}\big(r_{ij}^{\internode{(n,k)}}\big) &\ist=\ist \alpha^{\internode{(n)}}_j r_{ij}^{\internode{(n,k)}} + \beta^{\internode{(n)}}_j . \label{eq:timestamp_j}\\[-6.3mm]
\nonumber
\end{align}
Plugging \eqref{eq:mess_rel} into \eqref{eq:timestamp_j} and inserting in the resulting expression the expression of $s_{ij}^{\internode{(n,k)}}$ obtained from \eqref{eq:timestamp_i},
we find
\be
  c_{j}\big(r_{ij}^{\internode{(n,k)}}\big) \,=\, 
      \psi^{\internode{(n,k)}}_{i\to j}\big(\bm{\theta}^{\internode{(n)}}_{i}\!,\bm{\theta}^{\internode{(n)}}_{j}\big) 
        \ist+\ist v_{ij}^{\internode{(n,k)}}\alpha^{\internode{(n)}}_{j} \rmv,
      \label{eq:basicmeasurement} 
\pagebreak 
\ee
with
  \begin{align}
  & 
  \hspace{-3mm}\psi^{\internode{(n,k)}}_{i\to j}\big(\bm{\theta}^{\internode{(n)}}_{i}\!,\bm{\theta}^{\internode{(n)}}_{j}\big) 
  \nonumber \\[1mm]
  &\hspace{-1.5mm}
  \triangleq \Bigg( \rmv\frac{c_{i}(s_{ij}^{\internode{(n,k)}}) \rmv-\rmv \beta^{\internode{(n)}}_{i}}{\alpha^{\internode{(n)}}_{i}} 
    \ist+\ist \frac{\|\mathbf{p}_i^{(n)}\rrmv-\rmv\mathbf{p}_j^{(n)}\|}{c} \Bigg) \ist \alpha^{\internode{(n)}}_{j} \rmv+\ist \beta^{\internode{(n)}}_{j} \rmv. \!\!
  \label{eq:basicmeasurement_1}
  \end{align}
Similarly, the transmission of the $k$th packet from agent $j$ to agent $i$ 
(where $k \in \{1,\ldots,K_{ji}\}$) yields the time stamps $c_{j}\big(s_{ji}^{\internode{(n,k)}}\big)$ and $c_{i}\big(r_{ji}^{\internode{(n,k)}}\big)$;
expressions of these time stamps are obtained by exchanging $i$ and $j$ in \eqref{eq:timestamp_i}--\eqref{eq:basicmeasurement_1}.
The clock functions $c_i \big(t;\bm{\vartheta}^{\internode{(n)}}_{i}\big)$ and $c_j \big(t;\bm{\vartheta}^{\internode{(n)}}_{j}\big)$ 
and time stamps are visualized in Fig.~\ref{fig:2_way_msg}.
A communication protocol ensures that these time stamps are available at both agents $i$ and $j$.

The aggregated measurement of agents $i$ and $j$
comprises
all\linebreak 
``received'' time stamps, i.e.,
$\bd{y}^{\internode{(n)}}_{ij} \!\triangleq\rmv \big[\bd{y}^{\internode{(n)}\ist\text{T}}_{i \to j} \,\ist\ist \bd{y}^{\internode{(n)\ist\text{T}}}_{j \to i} \big]^\text{T}$
with $\bd{y}^{\internode{(n)}}_{i\to j}$\linebreak 
$\triangleq \big[c_{j}\big(r_{ij}^{\internode{(n,1)}}\big)\ist\cdots\ist c_{j}\big(r_{ij}^{\internode{(n,K_{ij})}}\big) \big]^\text{T}\rmv$ and
$\bd{y}^{\internode{(n)}}_{j\to i} \triangleq \big[c_{i}\big(r_{ji}^{\internode{(n,1)}}\big) \ist\cdots$\linebreak 
$c_{i}\big(r_{ji}^{\internode{(n,K_{ji})}}\big) \big]^\text{T}\!$.
We also define the (recorded, not measured) ``transmitted'' time stamp vectors
$\tilde{\bd{y}}^{(n)}_{i\to j} \triangleq \big[c_{i}\big(s_{ij}^{\internode{(n,1)}}\big) \ist\cdots$\linebreak 
$c_{i}\big(s_{ij}^{\internode{(n,K_{ij})}}\big) \big]^\text{T}\rmv$ and 
$\tilde{\bd{y}}_{j\to i} \triangleq \big[c_{j}\big(s_{ji}^{\internode{(n,1)}}\big) \ist\cdots\ist c_{j}\big(s_{ji}^{\internode{(n,K_{ji})}}\big) \big]^\text{T}\!$.

\subsection{Likelihood Function}
\label{ssec:LLF}

We first consider the ``single-packet'' likelihood function of the $k$th $i \rmv \to \rmv j$ packet at time $n$, 
$f\big( c_{j}\big(r_{ij}^{\internode{(n,k)}}\big) \big| \bm{\theta}^{\internode{(n)}}_{i}\rrmv,\bm{\theta}^{\internode{(n)}}_{j}\big)$.
From \eqref{eq:basicmeasurement} with $v_{ij}^{\internode{(n,k)}} \!\sim\mathcal{N}\big(v_{ij}^{\internode{(n,k)}};0,\sigma^2_v\big)$, we obtain
\begin{align*}
  &\hspace{-.1mm}f\big( c_{j}\big(r_{ij}^{\internode{(n,k)}}\big) \big| \bm{\theta}^{\internode{(n)}}_{i}\rrmv,\bm{\theta}^{\internode{(n)}}_{j}\big)  \nonumber \\[-1mm]
  & \hspace{-.1mm}
    = \frac{1}{\sqrt{2\pi} \ist \alpha^{\internode{(n)}}_{j} \sigma_v} 
    \exp \!\Bigg( \rmv {-\ist \frac{\Big(  c_{j}\big(r_{ij}^{\internode{(n,k)}}\big) - \psi^{\internode{(n,k)}}_{i\to j}\big(\bm{\theta}^{\internode{(n)}}_{i}\!,\bm{\theta}^{\internode{(n)}}_{j}\big) \Big)^2}{2 \ist\alpha^{\internode{(n)}\ist 2}_{j} \sigma_v^2}} \Bigg) \ist . 
\end{align*}
The single-packet likelihood function for the $k$th $j \rmv \to \rmv i$ packet,
$f\big( c_{i}\big(r_{ji}^{\internode{(n,k)}}\big) \big| \bm{\theta}^{\internode{(n)}}_{i}\rrmv,\bm{\theta}^{\internode{(n)}}_{j} \big)$, 
is obtained by exchanging $i$ and $j$.
Because 
$v_{ij}^{\internode{(n,k)}}$ was assumed iid across $i$, $j$, and $k$,
the measurements 
between any agents $i$ and $j$ with $(i,j)\in\mathcal{C}^{(n)}$ (cf.\ \eqref{eq:basicmeasurement}) are conditionally independent given the respective 
agent states $\bm{\theta}_i^{(n)}$ and $\bm{\theta}_j^{(n)}\!$, and thus we have
\begin{align}
  & \hspace{-1.7mm}f\big(\bd{y}^{\internode{(n)}}_{ij} \big|\bm{\theta}^{\internode{(n)}}_{i}\!,\bm{\theta}^{\internode{(n)}}_{j} \big) \nonumber \\[1mm]
  & \hspace{-2mm} =  f\big(\bd{y}^{\internode{(n)}}_{i\rightarrow j} \big|\bm{\theta}^{\internode{(n)}}_{i}\!,\bm{\theta}^{\internode{(n)}}_{j} \big) 
    \,f\big(\bd{y}^{\internode{(n)}}_{j\rightarrow i} \big|\bm{\theta}^{\internode{(n)}}_{i}\!,\bm{\theta}^{\internode{(n)}}_{j} \big)\nonumber \\[.5mm]
  & \hspace{-2mm} = \prod_{k=1}^{K_{ij}} f\big( c_{j}\big(r_{ij}^{\internode{(n,k)}}\big) \big|\bm{\theta}^{\internode{(n)}}_{i}\!,\bm{\theta}^{\internode{(n)}}_{j} \big) 
  \rmv \prod_{k'=1}^{K_{ji}} f\big(c_{i}\big(r_{ji}^{\internode{(n,k')}}\big) \big| \bm{\theta}^{\internode{(n)}}_{i}\!,\bm{\theta}^{\internode{(n)}}_{j}\rmv \big)\nonumber \\[0mm]
  & \hspace{-2mm}= \, G^{\internode{(n)}}_{ij} \rmv 
    \exp\!\Bigg(\rrmv-\rmv\frac{\| \bd{y}^{\internode{(n)}}_{i\to j} \rmv-\bm{\psi}^{\internode{(n)}}_{i\to j}\|^{2} \rmv}{2\ist \alpha^{\internode{(n)}2}_{j}\sigma^2_v} 
    - \rmv \frac{\| \bd{y}^{\internode{(n)}}_{j\to i} \rmv-\bm{\psi}^{\internode{(n)}}_{j\to i} \|^{2} \rmv}{2\ist \alpha^{\internode{(n)}2}_{i}\sigma^2_v} \Bigg) , \rrmv
    \label{eq:LH_pair}
\end{align}
where 
$G^{\internode{(n)}}_{ij} \! \triangleq \rmv \big(\sqrt{2\pi}\, \alpha^{\internode{(n)}}_{j} \sigma_v\big)^{\rmv-K_{ij}}
\big(\sqrt{2\pi}\, \alpha^{\internode{(n)}}_{i}\sigma_v\big)^{\rmv-K_{ji}}\rmv$,
$\bm{\psi}^{\internode{(n)}}_{i\to j} \triangleq \big[\psi^{\internode{(n,1)}}_{i\to j}\big(\bm{\theta}^{\internode{(n)}}_{i}\!,\bm{\theta}^{\internode{(n)}}_{j}  \big) 
  \ist\cdots\ist \psi^{\internode{(n,K_{ij})}}_{i\to j}\big( \bm{\theta}^{\internode{(n)}}_{i}\!,\bm{\theta}^{\internode{(n)}}_{j}\big) \big]^\text{T}\!$,
and 
$\bm{\psi}^{\internode{(n)}}_{j\to i} \triangleq $ $\big[\psi^{\internode{(n,1) }}_{j\to i}\big(\bm{\theta}^{\internode{(n)}}_{i}\!,\bm{\theta}^{\internode{(n)}}_{j} \big) 
\ist\cdots\ist \psi^{\internode{(n,K_{ji})}}_{j\to i}\big(\bm{\theta}^{\internode{(n)}}_{i}\!,\bm{\theta}^{\internode{(n)}}_{j} \big) \big]^\text{T}\!$.
As analyzed in \cite{etzlinger}, if the difference of successive 
packet 
transmit times is much larger than the noise standard deviation, i.e., 
$s_{ij}^{(n,k)} \!-
s_{ij}^{(n,k-1)}\!\rmv\gg \rmv \sigma_v$ for $k \!\in\! \{ 2,\ldots,K_{ij} \}$, then the following accu- rate approx\-imation of the likelihood function \eqref{eq:LH_pair} 
is obtained by approximating 
$\alpha_{j}^{\rmv(n)}\rmv\sigma_v$ and $\alpha_{i}^{\rmv(n)}\rmv\sigma_v$ (involved in $G^{\internode{(n)}}_{ij}$) by $\sigma_v$: 
  \begin{align}
  & f\big(\bd{y}^{\internode{(n)}}_{ij} \big|\bm{\theta}^{\internode{(n)}}_{i}\!,\bm{\theta}^{\internode{(n)}}_{j} \big) \nonumber \\[.7mm]
  & \;\;\, \approx \tilde{f}\big(\bd{y}_{ij}^{(n)} \big|\bm{\theta}_{i}^{(n)}\!, \bm{\theta}_{j}^{(n)} \big)  \label{eq:apprx_LH_clock_0} \\[1mm]
  & \;\;\, \propto \exp\rmv\Bigg( \!\rmv-\rmv \frac{\big\|\bd{A}_{ij}^{(n)}{\bm{\vartheta}}_{i}^{(n)} \rmv+\ist \bd{B}_{ij}^{(n)} {\bm{\vartheta}}_{\rmv j}^{(n)} 
  \rmv+\ist \bd{a}_d \ist \|\mathbf{p}_i^{(n)} \!\rmv-\! \mathbf{p}_j^{(n)}\rmv\| \big\|^{2}}{2\ist \sigma^2_v}\Bigg) \ist , 
  \label{eq:apprx_LH_clock}\\[-7mm]
  \nonumber
  \end{align}
where 
the symbol $\propto$ indicates equality up to a constant normalization factor (i.e., not depending on $\alpha_{i}^{(n)}$ or $\alpha_{j}^{(n)}$), 
\vspace{.5mm}
and 
$\bd{A}_{ij}^{(n)} \!\triangleq\rmv {\small \begin{bmatrix} 
	\bd{1}_{K_{ij}} \!\!\!\! & \!\! -\tilde{\bd{y}}_{i\to j}^{(n)} \\[.7mm]
	-\bd{1}_{K_{ji}} \!\!\!\! & \!\! \bd{y}_{j\to i}^{(n)} 
	\end{bmatrix}}$,
$\bd{B}_{ij}^{(n)} \!\triangleq\rmv {\small \begin{bmatrix} 
	-\bd{1}_{K_{ij} }\!\!\!\! & \!\! \bd{y}_{i\to j}^{(n)}\\[.7mm]
	\bd{1}_{K_{ji}}\!\!\!\! & \!\! -\tilde{\bd{y}}_{j\to i}^{(n)} 
	\end{bmatrix}}$,
\vspace{.5mm}
and $\bd{a}_{d} \rmv\triangleq\rmv -\frac{1}{c}\bd{1}_{K_{ij} + K_{ji}}$
with $\bd{1}_{K}$ denoting the all-ones vector of dimension $K$. 
In
\eqref{eq:apprx_LH_clock}, $f\big(\bd{y}^{\internode{(n)}}_{ij} \big|\bm{\theta}^{\internode{(n)}}_{i}\!,\bm{\theta}^{\internode{(n)}}_{j} \big)$ is approximated by a Gaussian function in the agent distance $\|\mathbf{p}_i^{(n)} \!-\rmv \mathbf{p}_i^{(n)}\|$ and the 
clock states $\bm{\vartheta}_{i}^{(n)}\rmv$ and $\bm{\vartheta}_{j}^{(n)}\!$. As in \cite{etzlinger}, this approximation will allow us to develop a BP 
message passing scheme where the clock messages are represented by Gaussian parameters.

Finally, because
$v_{ij}^{(n,k)}$ 
was 
assumed 
independent across 
$n$, we obtain the approximate joint likelihood function
\be
\hspace{.1mm} \tilde{f}\big( \bd{y}^{\internode{(1:n)}} \big| {\bm{\theta}}^{\internode{(1:n)}}\rmv \big) 
  = \prod_{n'=1}^n \!\! \prod_{\begin{array}{c} \rule{1mm}{0mm}\\[-5.2mm]{\scriptstyle (i,\ist j) \in \mathcal{C}^{\internode{(n')}}}\\[-1.5mm]{\scriptstyle i > j} \end{array}} 
  \hspace*{-3mm}
   \tilde{f}\big(\bd{y}^{\internode{(n')}}_{ij} \big| {\bm{\theta}}_{i}^{\internode{(n')}}\!,{\bm{\theta}}_{j}^{\internode{(n')}}\big)
\ist , \!\!
\label{eq:likelihood}
\vspace{-1mm}
\ee
where $\bd{y}^{\internode{(1:n)}}$ collects all 
$\bd{y}^{\internode{(n')}}_{ij}\!$, $(i,j)\rmv\rmv\in \mathcal{C}^{\internode{(n')}}$, $i \!>\! j$  
and ${\bm{\theta}}^{\internode{(1:n)}}$ collects all 
$\bm{\theta}_{i}^{\internode{(n')}}\!$, $i \rmv\rmv\in\rmv\rmv \mathcal{I}$, 
both for $n' \rmv\rmv\in\rmv\rmv \{1,\ldots,n\}$.


\section{Sequential
State
Estimation Using BP}
\label{sec:estimation}

\vspace{.5mm}

At each time step $n$, each agent $i \!\in\! \mathcal{I}$ estimates its current clock state $\bm{\vartheta}_{i}^{(n)}$ and location-related state $\bd{x}^{\internode{(n)}}_{i}$ 
from all past and present measurements, $\bd{y}^{\internode{(1:n)}}\rmv$.
This is based
on the minimum mean-square error (MMSE) estimates \cite{kay} 
  \begin{align}
  \hat{{\bm{\vartheta}}}^{\internode{(n)}}_{i,\text{MMSE}} &\ist\triangleq 
  \int \rmv {\bm{\vartheta}}^{\internode{(n)}}_{i}  f\big({\bm{\vartheta}}^{\internode{(n)}}_{i} \big|\bd{y}^{\internode{(1:n)}}\big) \, 
    \mathrm{d}{\bm{\vartheta}}^{\internode{(n)}}_{i} \ist, 
  \label{eq:mmse_theta}
  \\
  \hat{\bd{x}}^{\internode{(n)}}_{i,\text{MMSE}} &\ist\triangleq 
  \int \rmv \bd{x}^{\internode{(n)}}_{i}  f\big(\bd{x}^{\internode{(n)}}_{i} \big|\bd{y}^{\internode{(1:n)}} \big) \, \mathrm{d}\bd{x}^{\internode{(n)}}_{i} \ist. 
  \label{eq:mmse_x} \\[-6.5mm]
  \nonumber
  \end{align}
Here, the \textit{marginal posterior pdfs}
$f\big({\bm{\vartheta}}^{\internode{(n)}}_{i} \big|\bd{y}^{\internode{(1:n)}}\big)$ and 
$f\big(\bd{x}^{\internode{(n)}}_{i} \big|\bd{y}^{\internode{(1:n)}}\big)$ can
be obtained from the \textit{joint posterior pdf} 
$f\big(\bm{\theta}^{\internode{(0:n)}}\big|\bd{y}^{\internode{(1:n)}}\big) \propto f\big(\bd{y}^{\internode{(1:n)}}\big|\bm{\theta}^{\internode{(1:n)}}\big)\,f\big(\bm{\theta}^{\internode{(0:n)}}\big) $
by marginalizations. 
Because
these marginalizations are typically computationally infeasible, we resort to approximate MMSE estimation by means of iterative BP
\cite{loeliger2007factor, wainwright,Wymeersch07_iterativeReceiver}.
BP 
provides approximations of the marginal posterior pdfs, 
$b\big({\bm{\vartheta}}^{\internode{(n)}}_{i}\big) \approx f\big({\bm{\vartheta}}^{\internode{(n)}}_{i} \big|\bd{y}^{\internode{(1:n)}}\big)$ 
and
$b\big(\bd{x}^{\internode{(n)}}_{i}\big) \approx f\big(\bd{x}^{\internode{(n)}}_{i} \big|\bd{y}^{\internode{(1:n)}}\big)$,
so-called \emph{beliefs}, which can be calculated in a sequential (time-recursive), distributed manner. The means of these beliefs then provide approximations of the
MMSE estimates $\hat{{\bm{\vartheta}}}^{\internode{(n)}}_{i,\text{MMSE}}$ and $\hat{\bd{x}}^{\internode{(n)}}_{i,\text{MMSE}}$.


\vspace{-1.5mm}

\subsection{Joint Posterior pdf and Factor Graph}
\label{ssec:factorgraph}

BP 
is based on a factor graph (FG), which represents the factorization structure of the 
joint posterior pdf \cite{loeliger2007factor, wainwright,Wymeersch07_iterativeReceiver}.  
In our case, using the approximation \eqref{eq:apprx_LH_clock_0} and  the factorizations in \eqref{eq:prior} and \eqref{eq:likelihood}, the joint posterior pdf 
\vspace{.5mm}
is 
\begin{align}
  & f\big({\bm{\theta}}^{\internode{(0:n)}} \big|\bd{y}^{\internode{(1:n)}}\big)\nn\\[.5mm]
  & \;\propto f\big({\bm{\theta}}^{\internode{(0:n)}}\big) \ist \tilde{f}\big(\bd{y}^{\internode{(1:n)}} \big|{\bm{\theta}}^{\internode{(1:n)}}\big) \nn \\
  & \;= \prod_{i \in \mathcal{I}} \rmv 
     f\big({\bm{\vartheta}}^{\internode{(0)}}_{i}\big) f\big(\bd{x}^{\internode{(0)}}_{i}\big)
    \rmv\rmv \prod_{n' = 1}^n \rmv\rmv 
    f\big({\bm{\vartheta}}^{\internode{(n')}}_{i} \big| {\bm{\vartheta}}^{\internode{(n'\!-1)}}_{i}\big) \ist
    f\big(\bd{x}^{\internode{(n')}}_{i} \big| \bd{x}^{\internode{(n' \!-1)}}_{i}\big) \nn\\[.5mm]
  &\hspace{7mm}\times \hspace{-3mm} \prod_{\begin{array}{c} \rule{1mm}{0mm}\\[-5.2mm]{\scriptstyle (i,j) \in \mathcal{C}^{\internode{(n')}}}\\[-1.5mm]{\scriptstyle i > j} \end{array}} \hspace{-3mm} \tilde{f}\big(\bd{y}^{\internode{(n')}}_{i j} \big| {\bm{\theta}}_{i}^{\internode{(n')}}\!,{\bm{\theta}}_{j}^{\internode{(n')}}\big) \, .
   \label{eq:fac}  \\[-7mm] \nn
\end{align}
%

In a direct application of BP,
the maximum dimension of the messages would be the dimension of $\bm{\theta}_i^{(n)}\!$, i.e., six.
To obtain lower-dimensional messages,
we apply the ``opening nodes'' principle
\cite[Sec.~5.2.2]{Wymeersch07_iterativeReceiver}, i.e., 
we augment \eqref{eq:fac} by additional variables that depend deterministically on certain variables in \eqref{eq:fac}. 
More specifically, we introduce location variable replicas $\tilde{\bd{p}}_i^{\internode{(n)}} \!\rmv \triangleq \rmv \mathbf{P} \mathbf{x}_i^{(n)}\!$ 
(note that formally  $\tilde{\bd{p}}_i^{\internode{(n)}} \! = \rmv \bd{p}_i^{\internode{(n)}}$)
and interagent distances involving these location replicas, 
$d^{\internode{(n)}}_{ij} \!\triangleq\rmv \big\|\tilde{\bd{p}}_i^{(n)} \!-\rmv \tilde{\bd{p}}_j^{(n)} \big\|$. 
In this way, the  joint posterior pdf $f\big({\bm{\theta}}^{\internode{(0:n)}} \big|\bd{y}^{\internode{(1:n)}}\big)$ in \eqref{eq:fac} is extended 
\vspace{.5mm}
to
\begin{align}
&\hspace{-3mm}f\big({\bm{\theta}}^{\internode{(0:n)}}\rmv, \tilde{\bd{p}}^{\internode{(1:n)}}\rmv,\bd{d}^{\internode{(1:n)}} \big|\bd{y}^{\internode{(1:n)}}\big) \nn\\[0mm]
&\hspace{-3.5mm}\propto \rmv f\big({\bm{\theta}}^{\internode{(0:n)}}, \tilde{\bd{p}}^{\internode{(1:n)}}\rmv,\bd{d}^{\internode{(1:n)}} \big) \ist 
  f\big(\bd{y}^{\internode{(1:n)}} \big| {\bm{\theta}}^{\internode{(1:n)}}\rmv, \tilde{\bd{p}}^{\internode{(1:n)}}\rmv,\bd{d}^{\internode{(1:n)}} \big) \ist, \!\!\!
\label{eq:post-pdf-ext} \\[-5.5mm]
\nonumber
\end{align}
where $\tilde{\bd{p}}^{\internode{(1:n)}}\rmv$ consists of all $\tilde{\bd{p}}_i^{\internode{(n')}}\rmv$ for $i \rmv\in\rmv \mathcal{I}$, 
and $\bd{d}^{\internode{(1:n)}}\rmv$ consists of all $d^{\internode{(n')}}_{ij}\rmv$ for $(i,j) \rmv\in\rmv \mathcal{C}^{(n')}$ ($i \!>\! j$),  
both for
$n' \rmv\rmv\in\rmv\rmv \{1,\ldots,n\}$.
%
The new likelihood function (cf.\ \eqref{eq:likelihood}) 
\vspace{.5mm}
is
\begin{align}
   & f\big(\bd{y}^{\internode{(1:n)}} \big| {\bm{\theta}}^{\internode{(1:n)}}\rmv, \tilde{\bd{p}}^{\internode{(1:n)}}\rmv,\bd{d}^{\internode{(1:n)}} \big) \nn \\
   & \qquad = \prod_{n'=1}^n \!\! \prod_{\begin{array}{c} \rule{1mm}{0mm}\\[-5.2mm]{\scriptstyle (i,\ist j) \in \mathcal{C}^{\internode{(n')}}}\\[-1.5mm]{\scriptstyle i > j} \end{array}}
    \hspace{-3mm} \tilde{f}\big(\bd{y}^{\internode{(n')}}_{i j} \big| {\bm{\vartheta}}_{i}^{\internode{(n')}}\!,{\bm{\vartheta}}_{j}^{\internode{(n')}}\!, d^{\internode{(n')}}_{i j}\big) \ist , \label{eq:extended_likelihood} \\[-6mm]
    \nn
\end{align}
where $\!\tilde{f}\big(\bd{y}^{\internode{(n)}}_{ij} \big|\bm{\vartheta}^{\internode{(n)}}_{i}\!\!,\bm{\vartheta}^{\internode{(n)}}_{j}\!\!,d^{\internode{(n)}}_{ij} \big)\!$ is 
given by \eqref{eq:apprx_LH_clock} with $\big\|\bd{p}_i^{(n)}\!- \bd{p}_j^{(n)}\big\|$ replaced by $d^{\internode{(n)}}_{ij}$.
Here, we exploited the fact that the measurements $\bd{y}^{\internode{(1:n)}}$ are conditionally independent of the location-related states given the interagent distances, i.e., 
$f\big(\bd{y}^{\internode{(1:n)}} \big| {\bm{\theta}}^{\internode{(1:n)}}\rmv, \tilde{\bd{p}}^{\internode{(1:n)}}\rmv,\bd{d}^{\internode{(1:n)}} \big) 
= f\big(\bd{y}^{\internode{(1:n)}} \big| {\bm{\vartheta}}^{\internode{(1:n)}}\rmv, {\bd{d}}^{\internode{(1:n)}}\big)$.
Furthermore, using the deterministic relations mentioned above, the extended prior pdf (cf.\ \eqref{eq:prior}) is obtained 
\vspace{.5mm}
as
\begin{align}
  & \hspace{-.1mm}f\big({\bm{\theta}}^{\internode{(0:n)}}\rmv, \tilde{\bd{p}}^{\internode{(1:n)}}\rmv,\bd{d}^{\internode{(1:n)}} \big) \nn \\
  & \, = \prod_{i \in \mathcal{I}} \rmv 
    f\big({\bm{\vartheta}}^{\internode{(0)}}_{i}\big) f\big(\bd{x}^{\internode{(0)}}_{i}\big)
    \rmv\rmv \prod_{n' = 1}^n \rmv\rmv 
    f\big({\bm{\vartheta}}^{\internode{(n')}}_{i} \big| {\bm{\vartheta}}^{\internode{(n'\!-1)}}_{i}\big) \ist
    f\big(\bd{x}^{\internode{(n')}}_{i} \big| \bd{x}^{\internode{(n' \!-1)}}_{i}\big)
    \nn\\[.5mm]    
  & \hspace{8mm}\times
  f\big( \tilde{\bd{p}}^{\internode{(n')}}_{i} \big| \bd{x}^{\internode{(n')}}_{i} \big)
  \hspace{-3mm} \prod_{\begin{array}{c} \rule{1mm}{0mm}\\[-5.2mm]{\scriptstyle (i,j) \in \mathcal{C}^{\internode{(n')}}}\\[-1.5mm]{\scriptstyle i > j} \end{array}} \hspace{-3mm} 
  f\big( d^{\internode{(n')}}_{i j} \big|\tilde{\bd{p}}^{\internode{(n')}}_{i}\!, \tilde{\bd{p}}^{\internode{(n')}}_{j} \big) \, , \label{eq:extended_prior}
  \\[-7.5mm]
\nonumber
\end{align}
where
$f\big(d^{\internode{(n)}}_{ij} \big|\tilde{\bd{p}}^{\internode{(n)}}_i\!, \tilde{\bd{p}}^{\internode{(n)}}_j\big) 
= \delta\big(d^{\internode{(n)}}_{ij} \!- \big\|\tilde{\bd{p}}_i^{(n)} \!\rmv-\rmv \tilde{\bd{p}}_j^{(n)} \big\|\big)$ and 
$f\big(\tilde{\bd{p}}^{\internode{(n)}}_i \big| \bd{x}^{\internode{(n)}}_i \big) = \delta\big(\tilde{\bd{p}}_i^{(n)} \!- \bd{P}\bd{x}_i^{(n)}\big)$ 
express the deterministic relations $d^{\internode{(n)}}_{ij} \!= \big\|\tilde{\bd{p}}_i^{(n)} \!\rmv-\rmv \tilde{\bd{p}}_j^{(n)} \big\|$ and
$\tilde{\bd{p}}_i^{\internode{(n)}} \!\rmv = \rmv \mathbf{P} \mathbf{x}_i^{(n)}\!$, respectively. 
%
Inserting \eqref{eq:extended_likelihood} and \eqref{eq:extended_prior} into \eqref{eq:post-pdf-ext}, we obtain for the extended joint posterior 
\vspace{.5mm}
pdf
\begin{align}
  & f\big({\bm{\theta}}^{\internode{(0:n)}}\rmv, \tilde{\bd{p}}^{\internode{(1:n)}}\rmv, \bd{d}^{\internode{(1:n)}} \big|\bd{y}^{\internode{(1:n)}}\big) \nonumber\\[0mm]
   & \;\propto \ist \prod_{i \in \mathcal{I}} \rmv 
    f\big({\bm{\vartheta}}^{\internode{(0)}}_{i}\big) f\big(\bd{x}^{\internode{(0)}}_{i}\big)
    \rmv\rmv \prod_{n' = 1}^n \rmv\rmv 
    f\big({\bm{\vartheta}}^{\internode{(n')}}_{i} \big| {\bm{\vartheta}}^{\internode{(n'\!-1)}}_{i}\big) \ist
    f\big(\bd{x}^{\internode{(n')}}_{i} \big| \bd{x}^{\internode{(n' \!-1)}}_{i}\big)
    \nn\\[.5mm]
  &\hspace{8mm}\times
  f\big( \tilde{\bd{p}}^{\internode{(n')}}_{i} \big| \bd{x}^{\internode{(n')}}_{i} \big)
  \hspace{-3mm} \prod_{\begin{array}{c} \rule{1mm}{0mm}\\[-5.2mm]{\scriptstyle (i,j) \in \mathcal{C}^{\internode{(n')}}}\\[-1.5mm]{\scriptstyle i > j} \end{array}} \hspace{-3mm} \tilde{f}\big(\bd{y}^{\internode{(n')}}_{i j} \big| {\bm{\vartheta}}_{i}^{\internode{(n')}}\!,{\bm{\vartheta}}_{j}^{\internode{(n')}}\!, d^{\internode{(n')}}_{i j}\big)
  \nn \\[0mm]   
  & \hspace*{8mm} \times 
  f\big( d^{\internode{(n')}}_{i j} \big|\tilde{\bd{p}}^{\internode{(n')}}_{i}\!, \tilde{\bd{p}}^{\internode{(n')}}_{j} \big)
  \ist.  \label{eq:facEx} \\[-5mm]
  \nonumber
  \end{align}
This extended joint posterior pdf is related to the original
joint posterior pdf $f\big(\bm{\theta}^{\internode{(0:n)}} \big|\bd{y}^{\internode{(1:n)}} \big)$ (cf.\ \eqref{eq:fac}) via the 
marginalization
 $f(\bm{\theta}^{\internode{(0:n)}}|\bd{y}^{\internode{(1:n)}}) = \int \!\int f\big({\bm{\theta}}^{\internode{(0:n)}}\rmv, \tilde{\bd{p}}^{\internode{(1:n)}}\rmv, \bd{d}^{\internode{(1:n)}} \big|\bd{y}^{\internode{(1:n)}}\big)$\linebreak 
$\times\ist \mathrm{d}\tilde{\bd{p}}^{\internode{(1:n)}} \mathrm{d}\bd{d}^{\internode{(1:n)}}\rmv$.
In the factorization \eqref{eq:facEx}, all factors involve only state variables with a maximum dimension of four.

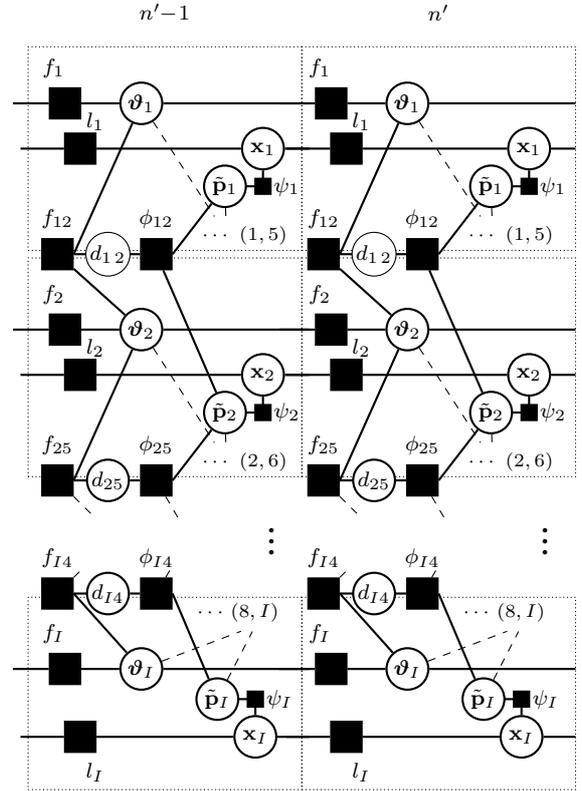
\begin{figure}[t!]
\vspace{-.5mm}
	  \centering
	  \begin{tikzpicture}[scale=1,
	    solid line/.style={thick},
	    variable node/.style={circle, draw, font=\footnotesize, minimum size=5.5mm, inner sep=0.2mm, fill=white!100},
	    factor node/.style={draw, font=\footnotesize, minimum size=4mm, inner sep=0.2mm, fill=black!100},
	    tinyfactor node/.style={draw, font=\footnotesize, minimum size=2mm, inner sep=0.2mm, fill=black!100}
	    ]
	    
	   \foreach \n/\textn in {0/{\rule{6mm}{0mm}n'\!\rmv-\!1}, 1/{\rule{8mm}{0mm}n'}}{
	      \node () at (3.5*\n+0.5,1.7) {\footnotesize$\textn$};
	      \foreach \ioffset/\i/\j/\k/\l in {0/1/2/5/0,1/2/5/6/1}{	
		\ifnum\i>1
		  \draw[densely dotted] (3.6*\n-1,-\ioffset*3+1.45) rectangle (3.6*\n+2.6,-\ioffset*3-1.45);
		\else
		  \draw[densely dotted] (3.6*\n-1,-\ioffset*3+1.25) rectangle (3.6*\n+2.6,-\ioffset*3-1.45);
		\fi
		\draw[solid line] 
			(3.5*\n-0.3,-\ioffset*3+0.5) node[factor node] (t01) [left, label=above:\hspace{-3mm}\footnotesize${f}_{\i}$] {} 
			-- 
			(3.5*\n+0.2,-\ioffset*3+0.5) node[right,variable node] (v00) {${\bm{\vartheta}}_{\i}$};
		\draw[solid line] (3.5*\n-1.2,-\ioffset*3+0.5) -- (t01);
		\draw[solid line] (v00) -- (3.5*\n+2.7,-\ioffset*3+0.5);
		\draw[solid line] 
			(3.5*\n-0.1,-\ioffset*3-0.1) node[factor node] (l01) [left, label={[label distance=-0.1cm]90:\hspace{4mm}\footnotesize${l}_{\i}$}] {} 
			-- 
			(3.5*\n+1.8,-\ioffset*3-0.1) node[right,variable node] (x01) {$\bd{x}_{\i}$};
		\draw[solid line] (x01) -- ($(x01) + (0cm,-0.5cm)$) node[tinyfactor node, label={[label distance=-0.1cm]0:\footnotesize$\psi_{\i}$}] (a01) {} -- ($(x01) + (-0.5cm,-0.5cm)$) node[variable node] (v01) {$\tilde{\bd{p}}_{\i}$};
		\draw[solid line] (3.5*\n-1.1,-\ioffset*3-0.1) -- (l01);
		\draw[solid line] (x01) -- (3.5*\n+2.7,-\ioffset*3-0.1);
		\ifnum\i>1
		  \draw[solid line] (v00) -- (t12);
		  \draw[solid line] (v01) -- (l12);
		  \node[variable node, fill=white!100] (v12r) at (v12) {$d_{\l\, \i}$}; 
		\fi
		\draw[solid line] (v00) -- (3.5*\n-0.4,-\ioffset*3-1.5) node[factor node] (t12) [left, label=above:\hspace{-0mm}\footnotesize${f}_{\i\j}$] {} ;
		\draw[solid line] (v01) -- (3.5*\n+0.9,-\ioffset*3-1.5) node[factor node] (l12) [left, label=above:\hspace{-0mm}\footnotesize$\phi_{\i\j}$] {} ;
		\draw[solid line] (t12) -- (3.5*\n+0.05,-\ioffset*3-1.5) node[variable node, fill=white!100] (v12) {$d_{\i\j}$};
		\draw[solid line] (v12) -- (l12);
		\draw[dashed] (v00) -- (3.5*\n+1.45,-\ioffset*3-1) node (l1k) [below] {\hspace{0.8cm}\scriptsize$\cdots \hspace{0.1cm} (\i,\k)$};
		\draw[dashed] (v01) -- ($(l1k) + (0.15,0.25)$);
		\ifnum\i=2
		  \draw[dashed] (t12) -- ++(-45:0.7);
		  \draw[dashed] (l12) -- ++(-60:0.6);
		\fi
	      }
	   }

	   
	   \foreach \n/\textn in {0/{n-1}, 1/{n}}{
	      \node at (3.6*\n+2.2,-2.5*3+2.3) {$\bm{\vdots}$};
	      \foreach \ioffset/\i/\j/\k in {2.5/I/4/8}{	
		\draw[densely dotted] (3.6*\n-1,-\ioffset*3+1.45) rectangle (3.6*\n+2.6,-\ioffset*3-1.1);
		\draw[solid line] 
			(3.5*\n-0.3,-\ioffset*3+0.5) node[factor node] (t01) [left, label=above:\hspace{-3mm}\footnotesize${f}_{\i}$] {} 
			-- 
			(3.5*\n+0.2,-\ioffset*3+0.5) node[right,variable node] (v00) {${\bm{\vartheta}}_{\i}$};
		\draw[solid line] (3.5*\n-1.2,-\ioffset*3+0.5) -- (t01);
		\draw[solid line] (v00) -- (3.5*\n+2.7,-\ioffset*3+0.5);
		\draw[solid line] 
			(3.5*\n-0.1,-\ioffset*3-0.4) node[factor node] (l01) [left, label=below:\hspace{4mm}\footnotesize${l}_{\i}$] {} 
			-- 
			(3.5*\n+1.7,-\ioffset*3-0.4) node[right,variable node] (x01) {$\bd{x}_{\i}$};
		\draw[solid line] (3.5*\n-1.1,-\ioffset*3-0.4) -- (l01);
		\draw[solid line] (x01) -- ($(x01) + (0cm,0.5cm)$) node[tinyfactor node, label={[label distance=-0.1cm]0:\footnotesize$\psi_{\i}$}] (a01) {} -- ($(x01) + (-0.5cm,0.5cm)$) node[variable node] (v01) {$\tilde{\bd{p}}_{\i}$};
		\draw[solid line] (x01) -- (3.5*\n+2.7,-\ioffset*3-0.4);
		\draw[solid line] (v00) to (3.5*\n-0.4,-\ioffset*3+1.5) node[factor node] (t12) [left, label=above:\hspace{-0mm}\footnotesize${f}_{\i\j}$] {} ;
		\draw[solid line] (v01) -- (3.5*\n+0.9,-\ioffset*3+1.5) node[factor node] (l12) [left, label=above:\hspace{-0mm}\footnotesize$\phi_{\i\j}$] {} ;
		\draw[solid line] (t12) -- (3.5*\n+0.05,-\ioffset*3+1.5) node[variable node, fill=white!100] (v12) {$d_{\i\j}$};
		\draw[solid line] (v12) -- (l12);
		\draw[dashed] (v00) -- (3.5*\n+1.8,-\ioffset*3+1) node (l1k) [above] {\hspace{-0.2cm} \scriptsize$\cdots\hspace{0.05cm} (\k,\i)$};
		\draw[dashed] (v01) -- ($(l1k) + (0.15,-0.25)$);
		  \draw[dashed] (t12) -- ++(45:0.5);
		  \draw[dashed] (l12) -- ++(60:0.4);
	      }
	    }

	  \end{tikzpicture}
  
	  \caption{CoSLAS factor graph for 
	  a network with agents $i\rmv\in\rmv\{1,2,\ldots,I\}$, 
	  where $(1,2)$, $(1,5)$, $(2,5)$, $(2,6)$, $(4,I)$, and $(8,I)$  belong to both $\mathcal{C}^{(n'\!-1)}$ and $\mathcal{C}^{(n')}\!$.
	  Only the time steps $n'\!\rmv-\!1$ and $n'$ are shown. Time indices are omitted for simplicity (e.g., $\bd{x}_i$ is short for 
	  $\bd{x}_i^{\internode{(n'\!-1)}}$ or
	  $\bd{x}_i^{\internode{(n')}}$). 
	  Each dotted box corresponds to an agent $i\rmv\in\rmv\mathcal{I}$ at 
	  time step $n'\!-1$ or $n'$; calculations within the box are performed locally by that agent. 
	  Connections between dotted boxes at the same time imply communication between agents.}
	  \label{fig:FG_extend_factorization}

	  \vspace*{-1mm}
  \end{figure}

The FG representing the factorization \eqref{eq:facEx} is shown in Fig.~\ref{fig:FG_extend_factorization}.
Each factor function 
in \eqref{eq:facEx} is represented by a square
factor node, and each 
variable 
by a circular variable node. A variable node is connected to a factor node by an edge if the corresponding variable is an argument of
the corresponding factor function. 
In Fig.~\ref{fig:FG_extend_factorization} and hereafter, we use the following short notations:
	  $f_i \triangleq f\big({\bm{\vartheta}}_i^{(n')} \big| {\bm{\vartheta}}_i^{(n'\!-1)}\big)$, 
	  $l_i \triangleq f\big(\mathbf{x}_i^{(n')} \big| \mathbf{x}_i^{(n'\!-1)}\big)$,	   
	  ${f}_{ij} \triangleq \tilde{f}\big(\bd{y}_{ij}^{(n')} \big| {\bm{\vartheta}}_{i}^{(n')}\!,{\bm{\vartheta}}_{j}^{(n')}\!, d_{ij}^{(n')}\big)$, 
	  $\phi_{ij} \triangleq 
	  f\big( d^{\internode{(n')}}_{i j} \big|\tilde{\bd{p}}^{\internode{(n')}}_{i}\!, \tilde{\bd{p}}^{\internode{(n')}}_{j} \big)$, and
	  \vspace{-1mm}
	  $\psi_i \triangleq f\big( \tilde{\bd{p}}^{\internode{(n')}}_{i} \big| \bd{x}^{\internode{(n')}}_{i} \big)$.

\subsection{BP Message Passing} 
\label{ssec:messagepassing}

The proposed sequential CoSLAS algorithm applies
BP 
\cite{loeliger2007factor,wainwright} to the FG in Fig.~\ref{fig:FG_extend_factorization}. 
Before presenting our 
algorithm in Section \ref{sec:extended}, we 
review 
the BP message update rules for a generic factor function $f$ and a generic variable $\bd{z}$.
Let $\mathcal{Z}_f$ denote the set of arguments of 
$f$, and assume $\bd{z} \rmv\in\rmv \mathcal{Z}_f$, i.e., $f \rmv=\rmv f(\bd{z},\ldots)$. 
Furthermore, let $\mathcal{F}_z$ denote the set of all functions $f'$ of which $\bd{z}$ is an argument, i.e., $\bd{z} \rmv\in\rmv \mathcal{Z}_{f'}$
if and only if $f'\!\in\rmv \mathcal{F}_z$.
In message passing iteration $q\in\{1,\ldots,Q\}$, the message from factor node $f$ to variable node $\bd{z}$---denoted by $\zeta^{\internode{(q)}}_{f}(\bd{z})$
---and the message from variable node $\bd{z}$ to factor node $f$---denoted by $\eta^{\internode{(q)}}_{f}(\bd{z})$---are calculated recursively as
  \begin{align}
    \zeta^{\internode{(q)}}_{f}(\bd{z}) & \ist=\rmv \int \rmv f(\bd{z},\ldots) \Bigg( \prod_{\bd{z}' \in \mathcal{Z}_f \rmv\setminus \{\bd{z}\}} \!\! \eta^{\internode{(q-1)}}_{f}(\bd{z}') \rmv\Bigg)
    \ist \mathrm{d}\!\sim\!\bd{z} \, , \label{eq:BP_zeta}\\[1mm]
    \eta^{\internode{(q)}}_{f}(\bd{z}) & \ist= \! \prod_{ f' \in \mathcal{F}_z \rmv\setminus \{f\} } \!\! \zeta^{\internode{(q)}}_{f'}(\bd{z})\, ,  \label{eq:BP_eta}\\[-6mm]\nn
  \end{align}
where ${\sim\!\bd{z}}$ denotes all $\bd{z}' \rmv\rmv\in\rmv\rmv \mathcal{Z}_f$ except $\bd{z}$. 
After the final iteration $q \rmv=\rmv Q$, the belief for variable $\bd{z}$ is obtained (up to a normalization) as
  \be
    b(\bd{z}) \ist\propto \rmv \prod_{ f \in \mathcal{F}_z } \! \zeta^{\internode{(Q)}}_{f}(\bd{z}) \, .  
  \label{eq:BP_be}
  \ee
For a function $f(\bd{z},\bd{z}')$ with only two 
arguments $\bd{z},\bd{z}'\rmv$,
\eqref{eq:BP_zeta} simplifies to 
$\zeta^{\internode{(q)}}_{f}(\bd{z}) =\rmv \int \rmv f(\bd{z},\bd{z}') \ist \eta^{\internode{(q-1)}}_{f}(\bd{z}') 
    \ist \mathrm{d}\bd{z}'\rmv$.
\pagebreak 
 If $\eta^{\internode{(q-1)}}_{f}(\bd{z}')$ is a weighted $S$-component mixture distribution,
    then $\zeta^{\internode{(q)}}_{f}(\bd{z})$ is again a weighted $S$-component mixture distribution, with the same weights.
  
If the FG is a tree,
then the BP algorithm is noniterative ($Q \rmv=\! 1$), there is a well-defined order of calculating the messages
(message schedule),
and the 
beliefs are exactly equal to the respective marginal posterior pdfs \cite{loeliger2007factor, wainwright}. 
However,
if the FG has loops, as in the case of the FG in Fig.~\ref{fig:FG_extend_factorization}, the beliefs are only approximations of the marginal posterior 
pdfs \cite{loeliger2007factor,wainwright}. 
Moreover, 
BP operates iteratively, and convergence 
is not guaranteed for general non-Gaussian joint posterior pdfs. Finally, there exist many possible message schedules, which may lead to different beliefs. 
Nevertheless, loopy BP 
provides accurate approximations of the marginal posterior pdfs in many applications \cite{loeliger2007factor,wainwright,ihler,wymeersch,Wymeersch07_iterativeReceiver,etzlinger,meyer2016distributed}. 

The sequential BP algorithm proposed in Section \ref{sec:extended} follows a specific schedule that was observed to converge for the scenarios studied
in Section \ref{sec:simulations}. 
The scheduling 
of the BP operations 
\eqref{eq:BP_zeta}--\eqref{eq:BP_be} 
is chosen such that
messages are not passed backward
in time \cite{wymeersch} and 
uninformative messages are censored \cite{das12} (i.e., not used in message calculations).
Since the messages are not passed backward
in time, our
algorithm can cope with a changing network connectivity and its complexity does not increase with time $n$; moreover, the beliefs 
are directly
equal to the messages passed to the 
\pagebreak 
next time.
The algorithm consists of the following main steps:

\begin{enumerate}


 \item \textit{Prediction:} 
 Each agent $i$ locally 
 converts the previous belief of its  clock state, $b\big(\bm{\vartheta}_i^{(n-1)}\big)$, and 
 of its location-related state, $b\big(\bd{x}_i^{(n-1)} \big)$, 
 into messages $\zeta_{f_i}\big(\bm{\vartheta}_i^{(n)}\big)$ and $\zeta_{l_i}\big(\bd{x}_i^{(n)}\big)$ for the current time interval $n$.
 This corresponds to messages passed from the $n' \!\rmv-\! 1$ section to the $n'$ section along the horizontal edges of the FG in Fig.~\ref{fig:FG_extend_factorization}.
 
 \item \textit{Iterative message passing:} Each agent $i$ exchanges messages related to its 
 states $\bm{\vartheta}_i^{(n)}$ and $\tilde{\bd{p}}_i^{(n)}$ with neighboring agents, 
 and uses the received messages to update its own 
 messages
 according to \eqref{eq:BP_zeta} and \eqref{eq:BP_eta}. Only messages that are informative according to some 
 criterion (see Section \ref{ssec:Msg_itMsgPass}) are used for further calculations.
 In Fig.~\ref{fig:FG_extend_factorization}, these messages are passed along the vertical edges connecting different agents.
 This step requires communication (packet exchanges) with neighboring agents; it is repeated during a predefined number of iterations $Q$.

 \item \textit{Belief calculation and estimation:} Each agent calculates its beliefs
 by multiplying according to \eqref{eq:BP_be} the appropriate messages calculated in Steps 1 and 2. 
 It then uses these beliefs 
 for state estimation according to \eqref{eq:mmse_theta} and \eqref{eq:mmse_x}, and as messages for the next prediction
 (Step 1).

\end{enumerate}

These steps will be worked out
in Section \ref{sec:extended} after the introduction of parametric message representations.

\section{Parametric Message Representations}
\label{sec:messagerep}



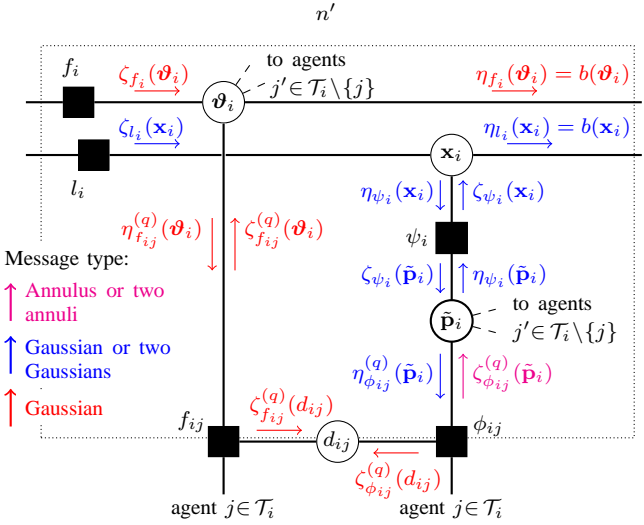
\begin{figure}[t!]
	  \centering
	  \begin{tikzpicture}[scale=1,
	    solid line/.style={thick},
	    variable node/.style={circle, draw, font=\footnotesize, minimum size=5.5mm, inner sep=0.2mm, fill=white!100},
	    factor node/.style={draw, font=\footnotesize, minimum size=4mm, fill=black!100, align=center},
	    tinyfactor node/.style={draw, font=\footnotesize, minimum size=2mm, inner sep=0.2mm, fill=black!100}
	    ]
	    
	   \foreach \n/\textn in {0/{n'}}{
	      \node () at (4.5*\n+2.75,1.7) {\footnotesize$\textn$};
	      \foreach \ioffset/\i/\j/\texti/\textj in {0/1/2/i/j}{	
		\ifnum\i<2
		  \draw[densely dotted] (4.5*\n-1,-\ioffset*5+1.25) rectangle (4.5*\n+6.8,-\ioffset*5-3.95);
		\else
		  \draw[densely dotted] (4.5*\n-1,-\ioffset*5+2.45) rectangle (4.5*\n+6.8,-\ioffset*5-1.25);
		\fi
		\node[factor node] (t01) at (4.5*\n-0.3,-\ioffset*5+0.5)  [left, label=above:\hspace{-2mm}\footnotesize${f}_{\texti}$] {};
		\node[variable node] (v00) at (4.5*\n+1.4,-\ioffset*5+0.5) {$\bm{\vartheta}_{\texti}$};
		\draw[solid line] (t01) -- (v00);
		\draw[solid line] (4.5*\n-1.2,-\ioffset*5+0.5) -- (t01);
		\draw[solid line] (v00) -- (4.5*\n+6.9,-\ioffset*5+0.5);
		  \draw[->, shorten <=1pt, shorten >=2pt,color=red] ($(t01) + (0.7,0.15)$) node[right,xshift=-3mm,yshift=2mm] (m) {{\footnotesize{$\zeta_{{f}_{\texti}}({\bm{\vartheta}_{\texti}})$}}} -- ($(t01) + (1.4,0.15)$);
		  \draw[->, shorten <=1pt, shorten >=2pt,color=red] ($(t01) + (5.4,0.15)$) node[right,xshift=-3mm,yshift=2mm] (m) {{\footnotesize{$\eta_{{f}_{\texti}}({\bm{\vartheta}_{\texti}})=b({\bm{\vartheta}_{\texti}})$}}} -- ($(t01) + (6.1,0.15)$);
		  
		\node[factor node] (l01) at (4.5*\n-0.1,-\ioffset*5-0.2) [left, label=below:\hspace{-4mm}\footnotesize${l}_{\texti}$] {};
	\node[variable node] (v01) at (4.5*\n+4.4,-\ioffset*5-0.2) {$\bd{x}_{\texti}$};
		\draw[solid line] (l01) -- (v01);
		\draw[solid line] (4.5*\n-1.2,-\ioffset*5-0.2) -- (l01);
		\draw[solid line] (v01) -- (4.5*\n+6.9,-\ioffset*5-0.2);
		  \draw[->, shorten <=1pt, shorten >=2pt,color=blue] ($(l01) + (5.4,0.15)$) node[right,xshift=-4mm,yshift=2mm] (m) {{\footnotesize{$\eta_{l_{\texti}}({\bd{x}_{\texti}}) = b({\bd{x}_{\texti}})$}}} -- ($(l01) + (6.1,0.15)$);
		  \draw[->, shorten <=1pt, shorten >=2pt,color=blue] ($(l01) + (0.5,0.15)$) node[right,xshift=-3mm,yshift=2mm] (m) {{\footnotesize{$\zeta_{l_{\texti}}({\bd{x}_{\texti}})$}}} -- ($(l01) + (1.2,0.15)$);
		\ifnum\i>1
		  \draw[solid line] (v00) -- (t12);
		  \draw[ultra thick,color=white!100] (v01) -- (l12);	
		  \draw[solid line] (v01) -- ($(v01) + (0,0.7)$) node[tinyfactor node,label={[label distance=-0.05cm]180:\footnotesize$A$}](A1) {} -- ($(v01) + (0,1.2)$) node[variable node] (x01) {$\tilde{\bd{p}}_{\texti}$} -- (l12); 
		  \node[variable node, fill=white!100] (v12r) at (v12) {$d_{\textj, \texti}$}; 
		    \draw[->, shorten <=1pt, shorten >=2pt,color=red] ($(v00) + (-0.15,0.8)$) node[left,yshift=6mm] (m) {{\footnotesize{$\eta^{\internode{(q)}}_{{f}_{\textj \texti}}(\bm{\vartheta}_\texti)$}}} -- ($(t12) + (-0.15,-0.7)$);
		    \draw[<-, shorten <=1pt, shorten >=2pt,color=red] ($(v00) + (0.15,0.8)$) node[right,yshift=2mm] (m) {{\footnotesize{$\zeta^{\internode{(q)}}_{{f}_{\textj \texti}}(\bm{\vartheta}_\texti)$}}} -- ($(t12) + (0.15,-0.7)$);
		    \draw[->, shorten <=1pt, shorten >=2pt,color=magenta] ($(x01) + (-0.15,0.3)$) node[left,yshift=2mm,xshift=0.5mm] (m) {{\footnotesize{$\eta^{\internode{(q)}}_{\phi_{\textj \texti}}(\tilde{\bd{p}}_\texti)$}}} -- ($(l12) + (-0.15,-0.7)$);
		    \draw[<-, shorten <=1pt, shorten >=2pt,color=magenta] ($(x01) + (0.15,0.3)$) node[right,yshift=2mm] (m) {{\footnotesize{$\zeta^{\internode{(q)}}_{\phi_{\textj \texti}}(\tilde{\bd{p}}_\texti)$}}} -- ($(l12) + (0.15,-0.7)$);
		    \draw[dashed] (v00) -- ++(-45:0.6);
		    \draw[dashed] (x01) -- ++(-45:0.6);
		\else
		  \node[factor node] (t12) at (4.5*\n+1.4,-\ioffset*5-4)  [label=left:\raisebox{6mm}{\footnotesize${f}_{\texti \textj}$\hspace{-1mm}}] {} ;
		  \draw[ultra thick,color=white!100] (v00) -- (t12); 
		  \draw[solid line] (v00) -- (t12);
		    \draw[->, shorten <=1pt, shorten >=2pt,color=red] ($(v00) + (-0.15,-1.5)$) node[left,yshift=-2mm] (m) {{\footnotesize{$\eta^{\internode{(q)}}_{{f}_{\texti \textj}}(\bm{\vartheta}_\texti)$}}} -- ($(v00) + (-0.15,-2.3)$);
		    \draw[<-, shorten <=1pt, shorten >=2pt,color=red] ($(v00) + (0.15,-1.5)$) node[right,yshift=-2mm] (m) {{\footnotesize{$\zeta^{\internode{(q)}}_{{f}_{\texti \textj}}(\bm{\vartheta}_\texti)$}}} -- ($(v00) + (0.15,-2.3)$);
		  \node[factor node] (l12) at (4.5*\n+4.4,-\ioffset*5-4) [label=right:\hspace{-0.5mm}\raisebox{5mm}{\footnotesize$\phi_{\texti \textj}$}] {} ;
		  \draw[solid line] (v01) -- ($(v01) + (0,-1.1)$) node[factor node, label={[label distance=-0.05cm]180:\footnotesize$\psi_\texti$}](A1) {} -- ($(v01) + (0,-2.2)$) node[variable node] (x01) {$\tilde{\bd{p}}_{\texti}$} -- (l12); 
		    \draw[->, shorten <=1pt, shorten >=2pt,color=blue] ($(v01) + (-0.15,-0.3)$) node[left,yshift=-2mm,xshift=0.5mm] (m) {{\footnotesize{$\eta_{\psi_\texti}({\bd{x}}_\texti)$}}} -- ($(v01) + (-0.15,-0.8)$);
		    \draw[<-, shorten <=1pt, shorten >=2pt,color=blue] ($(v01) + (0.15,-0.3)$) node[right,yshift=-2mm] (m) {{\footnotesize{$\zeta_{\psi_i}({\bd{x}}_\texti)$}}} -- ($(v01) + (0.15,-0.8)$);
		    \draw[->, shorten <=1pt, shorten >=2pt,color=blue] ($(A1) + (-0.15,-0.3)$) node[left,yshift=-2mm,xshift=0.5mm] (m) {{\footnotesize{$\zeta_{\psi_\texti}(\tilde{\bd{p}}_\texti)$}}} -- ($(A1) + (-0.15,-0.8)$);
		    \draw[<-, shorten <=1pt, shorten >=2pt,color=blue] ($(A1) + (0.15,-0.3)$) node[right,yshift=-2mm] (m) {{\footnotesize{$\eta_{\psi_\texti}(\tilde{\bd{p}}_\texti)$}}} -- ($(A1) + (0.15,-0.8)$);
		    
		    \draw[->, shorten <=1pt, shorten >=2pt,color=blue] ($(x01) + (-0.15,-0.4)$) node[left,yshift=-3mm,xshift=0.5mm] (m) {{\footnotesize{$\eta^{\internode{(q)}}_{\phi_{\texti \textj}}(\tilde{\bd{p}}_\texti)$}}} -- ($(l12) + (-0.15,0.5)$);
		    \draw[<-, shorten <=1pt, shorten >=2pt,color=magenta] ($(x01) + (0.15,-0.4)$) node[right,yshift=-3mm] (m) {{\footnotesize{$\zeta^{\internode{(q)}}_{\phi_{\texti \textj}}(\tilde{\bd{p}}_\texti)$}}} -- ($(l12) + (0.15,0.5)$);
		  \node[variable node, fill=white!100] (v12) at (4.5*\n+2.9,-\ioffset*5-4) {$d_{\texti \textj}$};
		    \draw[->, shorten <=1pt, shorten >=2pt,color=red] ($(t12) + (0.4,0.15)$) node[above,yshift=-0.5mm,xshift=5mm] (m) {\footnotesize{$\zeta^{\internode{(q)}}_{{f}_{\texti \textj}}(d_{\texti \textj})$}} -- ($(v12) + (-0.4,0.15)$);
		    \draw[->, shorten <=1pt, shorten >=2pt,color=red] ($(l12) + (-0.4,-0.15)$) node[below,yshift=0.0mm,xshift=-3mm] (m) {\footnotesize{$\zeta^{\internode{(q)}}_{\phi_{\texti \textj}}(d_{\texti \textj})$}} -- ($(v12) + (0.4,-0.15)$);
		  \draw[solid line] (t12) -- (v12);
		  \draw[solid line] (v12) -- (l12);
		  \draw[solid line] (t12) -- ++(-90:0.7) node[yshift=-0.2cm] (tmp) {\footnotesize{agent $j\!\rmv \in \rmv\mathcal{T}_i$}};
		  \draw[solid line] (l12) -- ++(-90:0.7) node[yshift=-0.2cm] (tmp) {\footnotesize{agent $j\!\rmv \in \rmv\mathcal{T}_i$}};
		    \draw[dashed] (v00) -- ++(55:0.6) node[right,xshift=0.1cm,text width=2.05cm,yshift=-0.15cm,align=left] {\footnotesize{to agents\\[-0.5mm] $\, j' \!\rmv \in \rmv \mathcal{T}_i \! \setminus \! \{j\}$} };
		    \draw[dashed] (v00) -- ++(25:0.6);
		    \draw[dashed] (x01) -- ++(15:0.6) node[right,xshift=0.1cm,yshift=-0.15cm,text width=2.05cm,align=left] {\footnotesize{to agents\\[-0.5mm] $j' \!\rmv \in \rmv \mathcal{T}_i \! \setminus \! \{j\}$} };
		    \draw[dashed] (x01) -- ++(-15:0.6);
		\fi
		
	      \draw[->,thick,color=red] ($(t12) + (-2.8,0.2)$) -- node[midway,right,fill=white,xshift=0.06cm,yshift=0.0mm] {\footnotesize{Gaussian}} ($(t12) + (-2.8,0.7)$) ;
	      \draw[->,thick,color=blue] ($(t12) + (-2.8,0.9)$) -- node[midway,right,fill=white,xshift=0.06cm,yshift=-0.5mm,text width=2.4cm,align=left] {\footnotesize{Gaussian or two\\[-1mm]Gaussians}} ($(t12) + (-2.8,1.4)$) ;
	      \draw[->,thick,color=magenta] ($(t12) + (-2.8,1.6)$) -- node[midway,right,fill=white,xshift=0.06cm,text width=2.05cm,align=left] {\footnotesize{Annulus or two\\[-1mm] annuli}} ($(t12) + (-2.8,2.1)$) ;
	      \node[fill=white,right,xshift=-0.2cm] at ($(t12) + (-2.8,2.4)$) {\footnotesize{Message type:}};
	      }
	   }

	  \end{tikzpicture}
  \vspace*{-5mm}

	  \caption{Detail of the FG in Fig.~\ref{fig:FG_extend_factorization}, corresponding to agent $i$ and its connection to agent $j \rmv\in\rmv \mathcal{T}_i$ at time step $n'\rmv$.
	    All depicted messages are calculated by agent $i$. The messages $\eta^{\internode{(q)}}_{f_{ij}}(d_{ij})$ and $\eta^{\internode{(q)}}_{\phi_{ij}}(d_{ij})$ (which are 
	    \vspace{-.7mm}
equal to $\zeta^{\internode{(q)}}_{\phi_{ij}}(d_{ij})$ and $\zeta^{\internode{(q)}}_{f_{ij}}(d_{ij})$, respectively) are 
	    omitted to avoid visual clutter. Messages represented by an annulus or two annuli are drawn in 
	    magenta, messages represented by a Gaussian or a two-component Gaussian mixture 
	    in 
	    blue, and messages represented by a single Gaussian in 
	    red.}
	  \label{fig:FG_extend_factorization2}
\vspace*{-1mm}
  \end{figure}

The messages calculated at agent $i \rmv\in\rmv \mathcal{I}$ are displayed in the FG detail shown in Fig.~\ref{fig:FG_extend_factorization2}. 
Hereafter, for simplicity, we drop the time index $n$ in the superscript.
For the messages involved in the \textit{prediction}
and \textit{belief calculation} steps, we use Gaussian or Gaussian mixture representations.
More specifically, the clock
messages $\zeta_{{f}_{i}}({{\bm{\vartheta}}_{i}})$ and 
$\eta_{{f}_{i}}({{\bm{\vartheta}}_{i}})$ are represented by a Gaussian, e.g.,
$\zeta_{{f}_{i}}({{\bm{\vartheta}}_{i}}) \triangleq  \mathcal{N}\big({\bm{\vartheta}}_{i}; \bm{\mu}_{{{f}_{i} \to {\vartheta}}_{i}  }, \bm{\Sigma}_{{{f}_{i} \to {\vartheta}}_{i} }\big)$,
and the location-related messages  
$\zeta_{l_{i}}({\bd{x}_{i}})$, 
$\eta_{l_{i}}({\bd{x}_{i}})$,
$\zeta_{\psi_i}({\bd{x}_{i}})$,
$\eta_{\psi_i}({\tilde{\bd{p}}_{i}})$,
$\zeta_{\psi_i}({\tilde{\bd{p}}_{i}})$
and $\eta_{\psi_i}({\bd{x}_{i}})$ are represented by a Gaussian or a two-component Gaussian mixture \cite{caceres,meyer_ICASSP13}, e.g., 
\[
 \zeta_{l_{i}}({\bd{x}_{i}}) \ist\triangleq \sum_{s=1}^{ S_{{x}_{i}} } \rmv w_{{x}_{i},s} \,\ist \mathcal{N}\big(\bd{x}_i;\bm{\mu}_{l_{i}\to {x}_{i},s},
   \bm{\Sigma}_{l_{i} \to {x}_{i},s}\big) \ist, 
\]
with 
$S_{{x}_{i}} \!\!\in\rmv \{1,2\}$ and normalized weights $w_{{x}_{i},s}$. 
The latter representation is motivated by the observation 
that the location messages
tend to be unimodal or bimodal \cite{wymeersch}. Because $l_i$ (short for $f\big(\bd{x}_i^{(n)}|\bd{x}_i^{(n-1)}\big)$) has only two arguments, 
$\zeta_{l_{i}}({\bd{x}_{i}})$ has the same $w$ and $S$ parameters as $\eta_{l_{i}}({\bd{x}_{i}})$ from the previous time interval (cf.\ \eqref{eq:BP_zeta}). 
For the same reason, at function $\psi_i$ (short for $f\big(\tilde{\bd{p}}_i|\bd{x}_i \big)$),
$\eta_{\psi_i}({\bd{x}_{i}})$ and $\zeta_{\psi_i}({\tilde{\bd{p}}_{i}})$ have the same $w$ and $S$ parameters, and similarly for
$\eta_{\psi_i}({\tilde{\bd{p}}_{i}})$ and $\zeta_{\psi_i}({\bd{x}_{i}})$.
Moreover, since messages are not
passed backward
in time,
we have $b(\bm{\vartheta}_i) \rmv=\rmv \eta_{f_i}(\bm{\vartheta}_i)$ (cf.\ \eqref{eq:BP_be} with only 
$\zeta_{{f}_{i}}({\bm{\vartheta}}_i) \ist \zeta^{\internode{(Q)}}_{{f}_{ij}}({\bm{\vartheta}}_i)$ on the right-hand side, which equals $\eta_{{f}_{i}}({\bm{\vartheta}}_i)$ due to \eqref{eq:BP_eta}) 
and similarly 
$b(\bd{x}_i) \rmv=\rmv \eta_{l_i}(\bd{x}_i)$, 
and $\eta_{\psi_i}({\bd{x}_{i}}) \rmv= \rmv\zeta_{l_{i}}({\bd{x}_{i}})$ (cf.\ \eqref{eq:BP_eta} with only $\zeta_{l_{i}}({\bd{x}_{i}})$ on the right hand side). 
The notation used for the parameters of these messages is indicated in Table \ref{tab:gausspar-1}.

\begin{table}[t]
\vspace*{3mm}
\centering
{
\begin{tabular}{|c||c|c|c|c|}
  \hline
\rule[-1.7mm]{0mm}{5mm}Message & $\bm{\mu}$ & $\bm{\Sigma}$ & $w$ & $\!S\!\in\rmv \{1,2\}\!$ \\[-.01mm]
  \hline \hline
  \rule[-2mm]{0mm}{5.3mm} $\zeta_{{f}_{i}}({{\bm{\vartheta}}_{i}})$ & $\bm{\mu}_{{f}_{i} \to {{\vartheta}}_{i}}$ 
    & $\bm{\Sigma}_{{f}_{i} \to {{\vartheta}}_{i}}$ & ---  & --- \\[.2mm]
  \hline
  \rule[-2mm]{0mm}{5.3mm} \hspace*{-2mm} $\!\zeta_{l_{i}}({\bd{x}_{i}}) = \eta_{\psi_i}({\bd{x}_{i}})\!$ & $\bm{\mu}_{l_{i}\to {x}_{i},s}$ 
    & $\bm{\Sigma}_{l_{i}\to {x}_{i},s}$ &  
    $w_{{x}_{i},s}$  & $S_{{x}_{i}}$ \\[.2mm] 
  \hline
  \rule[-2mm]{0mm}{5.3mm} $\zeta_{\psi_i}({\tilde{\bd{p}}_{i}})$ & $\bm{\mu}_{\psi_i \to \tilde{p}_{i},s}$ 
    & $\bm{\Sigma}_{\psi_i \to \tilde{p}_{i},s}$ & 
    $w_{{x}_{i},s}$  & $S_{{x}_{i}}$ \\[.2mm] 
  \hline
  \rule[-2mm]{0mm}{5.3mm} $\eta_{\psi_i}({\tilde{\bd{p}}_{i}})$ & $\bm{\mu}_{\tilde{p}_{i} \to \psi_i,s}$ 
    & $\bm{\Sigma}_{\tilde{p}_{i} \to \psi_i,s}$ & 
    $w_{\tilde{{p}}_i,s}$  & $S_{\tilde{p}_i}$ \\[.2mm] 
  \hline  
  \rule[-2mm]{0mm}{5.3mm} $\zeta_{\psi_i}({\bd{x}_{i}})$ & $\bm{\mu}_{\psi_i \to x_{i},s}$ 
    & $\bm{\Sigma}_{\psi_i \to x_{i},s}$ & 
    $w_{\tilde{{p}}_i,s}$  & $S_{\tilde{p}_i}$ \\[.2mm] 
  \hline
  \rule[-2mm]{0mm}{5.3mm} $\eta_{{f}_{i}}({{\bm{\vartheta}}_{i}}) = b(\bm{\vartheta}_i)$ & $\bm{\mu}_{{{\vartheta}}_{i} \to {f}_{i}}$ 
    & $\bm{\Sigma}_{{{\vartheta}}_{i} \to {f}_{i}}$ & ---  & --- \\[.2mm]
  \hline
  \rule[-2mm]{0mm}{5.3mm} $\eta_{l_{i}}({\bd{x}_{i}}) = b({\bd{x}_{i}})$ & $\bm{\mu}_{{x}_{i} \to l_{i},s}$ 
    & $\bm{\Sigma}_{{x}_{i} \to l_{i},s}$ & 
    $w_{{b}_{i},s}$  & $S_{{b}_{i}}$ \\[.2mm] 
  \hline
\end{tabular}
}
\vspace{1.5mm}
\renewcommand{\baselinestretch}{.9}\small\normalsize
\caption{Parameters of the Messages Involved in the Prediction and Belief Calculation Steps.}
\label{tab:gausspar-1}
\vspace*{-4mm}
\end{table}


Regarding the messages involved in the \textit{iterative message passing} step,
we use Gaussian representations for 
$\eta^{\internode{(q)}}_{{f}_{ij}}({\bm{\vartheta}}_i)$, 
$\zeta^{\internode{(q)}}_{{f}_{ij}}({\bm{\vartheta}}_i)$,
$\zeta^{\internode{(q)}}_{{f}_{ij}}(d_{ij})$,
and $\zeta^{\internode{(q)}}_{\phi_{ij}}(d_{ij})$, and Gaussian or two-compo\-nent Gaussian mixture representations
for $\eta^{\internode{(q)}}_{\phi_{ij}}(\tilde{\mathbf{p}}_i)$ 
(here, $j \!\in\! \mathcal{T}_i\ist$, and $q \!\in\! \{1,\ldots,Q\}$ is the iteration index). The corresponding parameters are listed in Table \ref{tab:gausspar-2}. 
For $\eta^{\internode{(q)}}_{f_{ij}}(d_{ij})$ and $\eta^{\internode{(q)}}_{\phi_{ij}}(d_{ij})$,
the same Gaussian models as for, respectively, $\zeta^{\internode{(q)}}_{\phi_{ij}}(d_{ij})$ and $\zeta^{\internode{(q)}}_{f_{ij}}(d_{ij})$ are used,
because $\eta^{\internode{(q)}}_{f_{ij}}(d_{ij}) \rmv=\rmv \zeta^{\internode{(q)}}_{\phi_{ij}}(d_{ij})$ and 
$\eta^{\internode{(q)}}_{\phi_{ij}}(d_{ij}) \rmv=\rmv \zeta^{\internode{(q)}}_{f_{ij}}(d_{ij})$ according to \eqref{eq:BP_eta}. 
Finally, $\zeta^{\internode{(q)}}_{\phi_{ij}}(\tilde{\mathbf{p}}_i)$ is represented by an annulus or a 
mixture of two annuli defined as (cf.\ \cite{meyer_ICASSP13})
\begin{align}
&\hspace{-2mm}\zeta^{\internode{(q)}}_{\phi_{ij}}(\tilde{\mathbf{p}}_i) \ist\triangleq\rmv \sum_{s=1}^{S_{j\to i}^{(q-1)}} \!\!\rmv w_{j\to i,s}^{(q-1)} 
\,\ist
\text{exp} \! \left( \rmv-\ist \frac{ \big( r^{\internode{(q)}}_{\phi_{ij}} \rmv- 
    \big\|\tilde{\mathbf{p}}_i \rmv-\rmv  \bm{\mu}^{\internode{(q)}}_{\phi_{ij},s} \big\| \big)^2 }{2 \ist \sigma_{\phi_{ij},s}^{2\ist (q)} } \right)  \!. \label{eq:gauss_zeta_px} \\[-7.5mm]
\nonumber
\end{align}
Here, 
$S_{j\to i}^{(q-1)}$ and $w_{j\to i,s}^{(q-1)}$ equal the $S$ and $w$ parameters of 
$\eta_{\phi_{ji}}^{(q-1)}(\tilde{\bd{p}}_j)$ (cf.\ Section \ref{sec:zeta-phi-p}),
$r^{\internode{(q)}}_{\phi_{ij}}$ is
the nominal radius of the annulus or annuli, and $\bm{\mu}^{\internode{(q)}}_{\phi_{ij},s}$ and $\sigma_{\phi_{ij},s}^{2\ist \internode{(q)}}$ 
\pagebreak 
are, respectively, 
the midpoint and squared nominal width of annulus (mixture component) $s$.
In each message passing iteration $q$, the parameters of these messages (see Table \ref{tab:gausspar-2}) are calculated at agent $i$ for all $j \!\in\! \mathcal{T}_i\ist$, 
and the parameters of $\eta^{\internode{(q)}}_{\phi_{ij}}(\tilde{\mathbf{p}}_i)$ and 
$\eta^{\internode{(q)}}_{f_{ij}}({\bm{\vartheta}}_i)$ are transmitted to neighbor agent $j$.

\begin{table}[t]
\vspace*{2.5mm}
\centering
\hspace*{-.2mm}{
\begin{tabular}{|c||c|c|c|c|}
  \hline
\rule[-1.7mm]{0mm}{5mm}Message & $\bm{\mu}$ & $\bm{\Sigma}$ & $w$ & \!\!$S\!\in\rmv \{1,2\}$\!\! \\[-.01mm]
  \hline \hline
  \rule[-2.8mm]{0mm}{6.5mm} $\eta^{\internode{(q)}}_{{f}_{ij}}({\bm{\vartheta}}_i)$ & $\bm{\mu}^{\internode{(q)}}_{{\vartheta}_i \to {f}_{ij}}$ 
    & \!$\bm{\Sigma}^{\internode{(q)}}_{{\vartheta}_i \to {f}_{ij}}$\! & ---  & --- \\[.2mm]
  \hline
  \rule[-2.8mm]{0mm}{6.5mm} $\zeta^{\internode{(q)}}_{{f}_{ij}}({\bm{\vartheta}}_i)$ & $\bm{\mu}^{\internode{(q)}}_{{f}_{ij}\to {\vartheta}_i}$ 
    & \!$\bm{\Sigma}^{\internode{(q)}}_{{f}_{ij}\to {\vartheta}_i}$\! & ---  & --- \\[.2mm]
  \hline
  \rule[-2.5mm]{0mm}{6.2mm} $\eta^{\internode{(q)}}_{\phi_{ij}}(\tilde{\mathbf{p}}_i)$ & \!\!\!$\bm{\mu}^{\internode{(q)}}_{\tilde{p}_i \to \phi_{ij},s}$\!\!\! 
    & \!\!\!$\bm{\Sigma}^{(q)}_{\tilde{p}_i \to \phi_{ij},s}$\!\!\! & \!\!\rmv$w_{i\to j,s}^{(q)}$\!\!\!  & $S_{i\to j}^{(q)}$ \\[.2mm]
  \hline
  \rule[-2.8mm]{0mm}{6.5mm} \!\!\!\!$\eta^{\internode{(q)}}_{{f}_{ij}}(d_{ij}) \rmv=\rmv \zeta^{\internode{(q)}}_{{f}_{ij}}(d_{ij})$\!\! & \!${\mu}^{\internode{(q)}}_{{f}_{ij} \to d_{ij}}$\! 
    & \!${\sigma}^{2 \ist \internode{(q)}}_{\!{f}_{ij} \to d_{ij}}$\! & ---  & --- \\[.2mm]
  \hline
  \rule[-2.5mm]{0mm}{6.2mm} \!\!\!\!$\eta^{\internode{(q)}}_{\phi_{ij}}(d_{ij}) \rmv=\rmv \zeta^{\internode{(q)}}_{\phi_{ij}}(d_{ij})$\!\! & \!${\mu}^{\internode{(q)}}_{\phi_{ij} \to d_{ij}}$\!
    & \!${\sigma}^{\internode{2 \ist (q)}}_{\phi_{ij} \to d_{ij}}$\! & ---  & --- \\[.2mm]
  \hline
  \rule[-2.5mm]{0mm}{6.2mm} \!\!\!\!$\zeta^{\internode{(q)}}_{\phi_{ij}}(\tilde{\mathbf{p}}_i)$\!\! & $\bm{\mu}^{\internode{(q)}}_{\phi_{ij},s}$
    & \!$\sigma_{\phi_{ij},s}^{2 \ist (q)}$\! & \!\!\rmv$w_{j\to i,s}^{(q-1)}$\!\!\!  & $S_{j\to i}^{(q-1)}$ \\[.2mm]
  \hline
\end{tabular}
}
\vspace{2mm}
\renewcommand{\baselinestretch}{.9}\small\normalsize
\caption{Parameters of the Messages Involved in the Iterative Message Passing Step.}
\label{tab:gausspar-2}
\vspace*{-4.4mm}
\end{table}


\vspace{.5mm}

\section{The Proposed CoSLAS Algorithm}
\label{sec:extended}

\vspace{.5mm}

Although the BP 
algorithm reviewed in Section \ref{ssec:messagepassing} is 
less complex than straightforward marginalization of 
$f\big(\bm{\theta}^{\internode{(0:n)}}\big|\bd{y}^{\internode{(1:n)}}\big)$, 
a direct implementation of the BP rules \eqref{eq:BP_zeta}--\eqref{eq:BP_be}
in the considered CoSLAS scenario is still computationally infeasible.
Therefore, we next develop an approximate version of the BP algorithm
that has moderate complexity and low communication requirements.
This approximate algorithm is a hybrid particle-based and parametric implementation of 
\eqref{eq:BP_zeta}--\eqref{eq:BP_be}: 
it combines a nonparametric (particle-based) BP implementation, which is typically used for the 
nonlinear cooperative localization problem 
\cite{wymeersch}, with parametric
representations for messages and beliefs 
(see Section \ref{sec:messagerep}), 
which are suited to
the approximately linear-Gaussian synchronization problem \cite{etzlinger}. 
This combination is enabled by the extended factorization \eqref{eq:facEx} involving $\tilde{\bd{p}}_i$ and $d_{ij}$,
whereby
the location and clock states are 
characterized by separate messages and, thus, the message calculations can be performed via particle methods for the location states 
and via Gaussian parameter updates for the clock states. To obtain a distributed algorithm in which 
only message parameters have to be communicated between agents, the result of particle-based message multiplication for the location states
is approximated by a Gaussian mixture (see Section \ref{sec:messagerep}).
Next,
we 
present the individual operations used for calculating messages and beliefs.

\vspace{-1mm}

%

\subsection{Prediction} \label{ssec:Msg_tempUpdate}

At time $n \rmv=\rmv 0$, the recursive BP 
algorithm is initialized  by setting 
$b(\bm{\vartheta}_i) \rmv=\rmv f(\bm{\vartheta}_i)$ and $b(\bd{x}_i) \rmv=\rmv f(\bd{x}_i)$, 
where $f(\bm{\vartheta}_i)$ and $f(\bd{x}_i)$ 
are the Gaussian prior pdfs 
in \eqref{eq:prior_theta} and \eqref{eq:prior_x}. The mixture parameters of $\eta_{l_i}(\bd{x}_i) \rmv=\rmv b(\bd{x}_i) \rmv=\rmv f(\bd{x}_i)$ 
are $w_{b_i,1} \!=\! 1$ and $S_{b_i} \!=\! 1$. 
For $n \rmv \ge \rmv 1$, the parameters of the messages $\zeta_{f_i}(\bm{\vartheta}_i)$, $\zeta_{l_i}(\bd{x}_i)$, and $\zeta_{\psi_i}(\tilde{\mathbf{p}}_i)$
are calculated according to \eqref{eq:BP_zeta}, in which the $\eta$ messages are replaced by the respective beliefs $b$ from time $n \rmv-\! 1$ because they are equal.
In the following presentation of these calculations, messages and their parameters that are used from 
time 
$n \rmv-\rmv 1$ are denoted by the superscript ``$-$.''

  
 \subsubsection{Message $\zeta_{f_i}(\bm{\vartheta}_i)$}
  
The parameters of $\zeta_{f_i}(\bm{\vartheta}_i)$ are calculated using the function $f_i$ and the parameters of 
$b^-(\bm{\vartheta}_i)$.
The evaluation
of \eqref{eq:BP_zeta} here simplifies because the function node ${f}_i$ is connected only to two edges \cite{loeliger2007factor}. 
 One obtains 
\vspace{-1mm}
  \begin{align}
    \bm{\mu}_{{f}_i \to {\vartheta}_i} &=\ist \bm{\mu}_{{\vartheta}_i \to {f}_i}^- \, , \label{eq:MU_t1} \\[.5mm]
    \bm{\Sigma}_{{f}_i \to {\vartheta}_i} &=\ist \bm{\Sigma}_{{\vartheta}_i \to {f}_i}^- \rmv+ \bm{\Sigma}_{u_{1,i}}  \, . \label{eq:MU_t2} 
  \end{align}

  \subsubsection{Message $\zeta_{l_i}(\bd{x}_i)$}
  \label{sssec:Msg_locationUpdate}
The parameters of $\zeta_{l_i}(\bd{x}_i)$ are calculated 
using the function $l_i$ and the parameters of 
$b^-(\bd{x}_i)$.
One obtains from \eqref{eq:BP_zeta}
\vspace{-1mm}
  \begin{align*}
    \bm{\mu}_{l_i \to x_i,s} &=\ist \mathbf{G}_1 \bm{\mu}^-_{x_i \to l_i,s} \, , 
    \\[.5mm]
    \bm{\Sigma}_{l_i \to x_i,s} &=\ist \mathbf{G}_1 \bm{\Sigma}^-_{x_i \to l_i,s} \mathbf{G}_1^\text{T} \rmv+ \bm{\Sigma}_{u_{2,i}} \, , 
    \\[-5.5mm]
\nonumber
  \end{align*}
as well as
$w_{x_i,s} \!=\rmv w_{b_i,s}^-$ and 
\vspace{.5mm}
$S_{x_i} \!\rmv=\! S_{b_i}^-$. 

  \subsubsection{Message $\zeta_{\psi_i}(\tilde{\mathbf{p}}_i)$}
  \label{sssec:Msg_tempUpdate}
  Similarly, the parameters of $\zeta_{\psi_i}(\tilde{\mathbf{p}}_i)$ are calculated using the function $\psi_i$ and the parameters of 
  $\eta_{\psi_i}(\mathbf{x}_i)$. (Note that $\eta_{\psi_i}(\mathbf{x}_i)=\zeta_{l_i}(\mathbf{x}_i)$.)
  One obtains
\vspace{-.5mm}
  \begin{align*}
    \bm{\mu}_{\psi_i \to \tilde{p}_i,s} &= \mathbf{P} \bm{\mu}_{l_i \to x_i,s} \, , 
    \\[.5mm]
    \bm{\Sigma}_{\psi_i \to \tilde{p}_i,s} &= \mathbf{P} \bm{\Sigma}_{l_i \to x_i,s} \mathbf{P}^\text{T} . 
    \\[-5.5mm]
    \nn  
  \end{align*}
The $w$ and $S$ parameters are given by $w_{x_i,s}$ and $S_{x_i}$, respectively (see Section \ref{sssec:Msg_locationUpdate}).

\vspace{-1mm}

\subsection{Iterative Message Passing} \label{ssec:Msg_itMsgPass}

Next, we describe the iterative message passing operations performed in iteration $q\rmv\in\rmv\{1,\ldots,Q\}$. 
 The iterations are
initialized by setting $\eta_{{f}_{ij}}^{(0)}({\bm{\vartheta}}_i) \rmv=\rmv \zeta_{{f}_{i}}({\bm{\vartheta}}_i)$,
$\eta_{{\phi}_{ij}}^{(0)}(\tilde{\bd{p}}_i) \rmv=\rmv \zeta_{{\psi}_{i}}(\tilde{\bd{p}}_i)$, and 
$\zeta_{f_{ij}}^{(0)}(d_{ij}) \rmv= \zeta_{\phi_{ij}}^{(0)}(d_{ij}) \rmv=\rmv f(d_{ij})$ for $j \!\in\! \mathcal{T}_i$, 
where $f(d_{ij}) \rmv=\rmv \mathcal{N}(d_{ij};\mu_d,\sigma^2_d)$ with $\mu_d$  
and $\sigma_d^2$ reflecting
prior assumptions on the interagent distances. 
The messages $\eta_{{f}_{ij}}^{(0)}({\bm{\vartheta}}_i)$ and $\eta_{{\phi}_{ij}}^{(0)}(\tilde{\bd{p}}_i)$ are passed to the neighbors $j \!\in\! \mathcal{T}_i$.
For $q \rmv \ge \rmv 1$, the parameters of 
$\zeta_{f_{ij}}^{(q)}(d_{ij})$,
$\zeta_{\phi_{ij}}^{(q)}(d_{ij})$, 
$\zeta^{(q)}_{f_{ij}}(\bm{\vartheta}_i)$, and
$\zeta^{(q)}_{\phi_{ij}}(\tilde{\bd{p}}_i)$
are calculated according to \eqref{eq:BP_zeta}, and the parameters of
$\eta_{{f}_{ij}}^{(q)}({\bm{\vartheta}}_i)$ and 
$\eta_{{\phi}_{ij}}^{(q)}(\tilde{\bd{p}}_i)$
are calculated according to \eqref{eq:BP_eta}, as discussed next.

  
\subsubsection{Message $\zeta_{f_{ij}}^{(q)}(d_{ij})$}
We consider message $\eta_{f_{ij}}^{(q-1)}(\bm{\vartheta}_i)$ (passed
from agent $i$ to neighbor $j$) as \emph{informative} if the trace of its covariance matrix  $\bm{\Sigma}_{\vartheta_i \to f_{ij}}^{(q-1)}$ 
is smaller than a threshold $\tau$ and as \emph{uninformative} otherwise, and we denote by 
$\mathcal{T}_i^{c(q)}$ the set of neighbors $j$ of agent $i$ that provide informative messages 
$\eta_{f_{ji}}^{(q-1)}(\bm{\vartheta}_j)$.
If $\eta_{f_{ij}}^{(q-1)}(\bm{\vartheta}_i)$ is informative, 
then 
the parameters of $\zeta_{f_{ij}}^{(q)}(d_{ij})$, $j \!\in\! \mathcal{T}_i^{c(q)}$ are calculated using the function ${f}_{ij}$ 
and the parameters of
$\eta_{{f}_{ij}}^{(q-1)}(\bm{\vartheta}_{i})$ and $\eta_{{f}_{ji}}^{(q-1)}(\bm{\vartheta}_{j})$. 
Using \eqref{eq:BP_zeta} and standard Gaussian operations \cite{loeliger2007factor}, 
one 
obtains
  \begin{align}
    \sigma^{2\ist (q)}_{{f}_{ij} \to d_{ij}} & \!=\ist \sigma_v^2 \ist \big( \| \bd{a}_d \|^2
    - \bd{q}^{\internode{(q)} \text{T}}_{j \to i, 1} 
    \bd{D}_{ij}^\text{T} \bd{a}_d \big)^{-1} \rmv, \label{eq:IU_d1} \\[1mm]
    \mu_{{f}_{ij} \to d_{ij}}^{\internode{(q)}} & \!=\ist - \sigma^{2\ist (q)}_{{f}_{ij} \to d_{ij}} \bd{q}^{\internode{(q)} \text{T}}_{j \to i, 1} 
    \ist \bm{\Sigma}^{\internode{(q)} \ist -1}_{j \to i, 1} \ist\bm{\mu}^{\internode{(q)}}_{j \to i, 1} \, , \label{eq:IU_d2}
  \end{align}
where $\bd{q}^{\internode{(q)} \text{T}}_{j \to i, 1} \rmv\triangleq \bd{a}_d^{\!\text{T}}  \bd{D}_{ij} \big( \bd{D}_{ij}^\text{T} \bd{D}_{ij} 
      +  \sigma_v^2 \ist {\bm{\Sigma}^{\internode{(q)} \ist -1}_{ j \to i, 1}} \big)^{-1}\!$,
$\bd{D}_{ij} \rmv\triangleq [\bd{A}_{ij} \,\ist \bd{B}_{ij}]$, 
${\bm{\Sigma}^{\internode{(q)}}_{ j \to i, 1}} \!\triangleq\rmv \mathrm{diag}\big\{ \bm{\Sigma}^{\internode{(q-1)}}_{{\vartheta}_i \to {f}_{ij}},\bm{\Sigma}^{\internode{(q-1)}}_{{\vartheta}_j \to {f}_{ji}}\big\}$, 
and $\bm{\mu}^{\internode{(q)}}_{j \to i, 1} 
  \triangleq$\linebreak 
  $\big[\bm{\mu}^{(q-1) \ist \text{T}}_{{\vartheta}_i \to {f}_{ij}} \; \bm{\mu}^{(q-1) \ist \text{T}}_{{\vartheta}_j \to {f}_{ji}} \big]^\text{T}\!$.
Otherwise, i.e., 
if $\eta_{f_{ij}}^{(q-1)}(\bm{\vartheta}_i)$ is 
uninformative or if $j\rmv \notin \rmv \mathcal{T}_i^{c(q)}\!$, we set 
\pagebreak 
$\zeta_{f_{ij}}^{(q)}(d_{ij}) \rmv=\rmv \zeta_{f_{ij}}^{(q-1)}(d_{ij})$.
  
\subsubsection{Message $\zeta_{\phi_{ij}}^{(q)}(d_{ij})$}
\label{sssec:Msg_itMsgPass_zeta-phi-d}
We consider 
$\eta_{\phi_{ij}}^{(q-1)}(\tilde{\bd{p}}_i)$ as
informative if it satisfies a criterion involving two thresholds $\tau_1$ and $\tau_2$ (see Section \ref{sssec:particleMessageMult_1}),
and we denote by $\mathcal{T}_i^{p(q)}$ the set of neighbors $j$ of agent $i$ that provide informative messages $\eta_{\phi_{ji}}^{(q-1)}(\tilde{\bd{p}}_j)$.
If $\eta_{\phi_{ij}}^{(q-1)}(\tilde{\bd{p}}_i)$ is informative, then 
the parameters of $\zeta_{\phi_{ij}}^{(q)}(d_{ij})$, $j \!\in\! \mathcal{T}_i^{p(q)}$ 
are calculated using the function ${\phi}_{ij}$ and the parameters of $\eta_{{\phi}_{ij}}^{(q-1)}(\tilde{\bd{p}}_i)$ 
and $\eta_{{\phi}_{ji}}^{(q-1)}(\tilde{\bd{p}}_j)$. 
Because $\phi_{ij}$ is nonlinear, we use a linearization as discussed
in Appendix \ref{app:gauss_zeta_dp}. This yields a single Gaussian representing $\zeta_{\phi_{ij}}^{(q)}(d_{ij})$,
whose mean and variance are obtained 
as
\begin{align}
      \mu_{\phi_{ij} \to d_{ij}}^{(q)}  &\rmv=  \! \sum_{r = 1}^{S_{i\to j}^{(q-1)}}  \! \sum_{s = 1}^{S_{j\to i}^{(q-1)}} \!\!\rmv 
	w_{i\to j,r}^{(q-1)} \ist w_{j\to i,s}^{(q-1)} \ist 
	\big\|
	\bm{\mu}_{d_{ij},rs}^{(q-1)}
	\big\| \,, 
	\label{eq:IU_d3} \\[1mm] 
      \sigma_{\phi_{ij} \to d_{ij}}^{2\ist(q)}  &\rmv= \! \sum_{r = 1}^{S_{i\to j}^{(q-1)}}  \! \sum_{s = 1}^{S_{j\to i}^{(q-1)}} \!\!\rmv  
	w_{i\to j,r}^{(q-1)} \ist w_{j\to i,s}^{(q-1)} \ist\Big(
	\bar{\bm{\mu}}_{d_{ij},rs}^{(q-1)\text{T}} \bm{\Sigma}_{ij,rs} \ist \bar{\bm{\mu}}_{d_{ij},rs}^{(q-1)}
	\nonumber \\[0mm]
	& \hspace{8mm} + \big( \big\|
	\bm{\mu}_{d_{ij},rs}^{(q-1)}
	\big\| 
	  - \mu_{\phi_{ij} \to d_{ij}}^{(q)} \big)^2 \Big) \ist , 
	  \label{eq:IU_d4}
	  \\[-5.5mm]
    \nonumber
    \end{align}
with 
$\rmv\bm{\mu}_{d_{ij},rs}^{(q-1)} \!\rmv\triangleq\! \bm{\mu}^{\internode{(q-1)}}_{\tilde{p}_i \to \phi_{ij},r} \!- \bm{\mu}^{\internode{(q-1)}}_{\tilde{p}_j \to \phi_{ji},s}$,
$\bm{\Sigma}_{ij,rs} \!\rmv\triangleq\!\mathrm{diag} \big\{ \bm{\Sigma}^{\internode{(q-1)}}_{\tilde{p}_i \to \phi_{ij},r} ,$\linebreak 
$\bm{\Sigma}^{\internode{(q-1)}}_{\tilde{p}_j \to \phi_{ji},s} \big\}$, and 
$\rmv\bar{\bm{\mu}}_{d_{ij},rs}^{(q-1)} \!\triangleq\! 
\big[\bm{\mu}_{d_{ij},rs}^{(q-1)\ist\mathrm{T}} \;\ist {-\bm{\mu}}_{d_{ij},rs}^{(q-1)\ist\mathrm{T}} \big]^{\mathrm{T}} \!/ \big\| \bm{\mu}_{d_{ij},rs}^{(q-1)} \big\|$. \linebreak 
If $\eta_{\phi_{ij}}^{(q-1)}(\tilde{\bd{p}}_i)$ is uninformative or if $j\rmv\notin\rmv\mathcal{T}_i^{p(q)}\!$, we set $\zeta_{\phi_{ij}}^{(q)}(d_{ij}) \rmv=\rmv \zeta_{\phi_{ij}}^{(q-1)}(d_{ij})$.

\subsubsection{Message $\zeta^{(q)}_{f_{ij}}(\bm{\vartheta}_i)$}
\label{sec:zeta-f-theta}
 The parameters of $\zeta^{(q)}_{f_{ij}}(\bm{\vartheta}_i)$, $j{\in}\mathcal{T}_i^{c(q)}$ are calculated using the function ${f}_{ij}$ 
and the parameters of
$\eta_{{f}_{ji}}^{(q-1)}({\bm{\vartheta}}_{j})$ and $\zeta_{{\phi}_{ij}}^{(q)}(d_{ij})$. 
Similarly to \eqref{eq:IU_d1} and \eqref{eq:IU_d2}, one has
  \begin{align}
  \bm{\Sigma}^{\internode{(q)}}_{{f}_{ij}\to {\vartheta}_i} & \rmv=\ist \sigma_v^2 \ist \big( \bd{A}_{ij}^{\!\text{T}}\rmv\bd{A}_{ij} 
    - \bd{Q}^{\internode{(q)}}_{j\to i,2} \rmv \bd{C}_{ij}^\text{T}\rmv\bd{A}_{ij} \big)^{-1} \rmv, \label{eq:IU_t1}\\[.7mm]
    \bm{\mu}^{\internode{(q)}}_{{f}_{ij}\to {\vartheta}_i} & \rmv=\ist - \bm{\Sigma}^{\internode{(q)}}_{{f}_{ij} \to {\vartheta}_i} 
      \rmv\bd{Q}^{\internode{(q)}}_{j \to i, 2} \bm{\Sigma}^{\internode{(q)} \ist -1}_{j \to i, 2} \, \bm{\mu}^{\internode{(q)}}_{j \to i, 2} \, , \label{eq:IU_t2}
  \end{align}
where 
$ \bd{Q}^{\internode{(q)}}_{j \to i, 2} \triangleq \bd{A}_{ij}^{\!\text{T}}  \bd{C}_{ij}\big( \bd{C}_{ij}^\text{T} \bd{C}_{ij} 
      +  \sigma_v^2 \ist \bm{\Sigma}^{\internode{(q)} \ist -1}_{ j \to i, 2} \big)^{-1}\!$,
$\bd{C}_{ij} \triangleq [\bd{B}_{ij} \,\ist \bd{a}_d]$, 
$\bm{\Sigma}^{\internode{(q)} }_{j \to i, 2} \triangleq \mathrm{diag}\big\{ \bm{\Sigma}^{\internode{(q-1)}}_{{\vartheta}_j \to {f}_{ji}},\ist \sigma^{2\ist (q-1)}_{\phi_{ij} \to d_{ij}} \big\}$, 
  and $\bm{\mu}^{\internode{(q)}}_{j \to i, 2} 
  \triangleq  \big[\bm{\mu}^{(q-1) \ist \text{T}}_{{\vartheta}_j \to {f}_{ji}} \;\ist \mu_{\phi_{ij} \to d_{ij}}^{\internode{(q-1)}} \big]^\text{T}\!$.

  
\subsubsection{Message $\zeta^{(q)}_{\phi_{ij}}(\tilde{\bd{p}}_i)$}
\label{sec:zeta-phi-p}
 The parameters of $\zeta^{(q)}_{\phi_{ij}}(\tilde{\bd{p}}_i)$, $j{\in}\mathcal{T}_i^{p(q)}$ (see \eqref{eq:gauss_zeta_px})
are calculated using the function $\phi_{ij}$ 
and the parameters of
$\eta_{{\phi}_{ji}}^{(q-1)}(\tilde{\bd{p}}_{j})$ and $\zeta_{{f}_{ij}}^{(q)}(d_{ij})$.
Again, 
because $\phi_{ij}$ is nonlinear, we
linearize $\| \tilde{\bd{p}}_{i} \!- \rmv \tilde{\bd{p}}_{j} \|$ (considered as a function of $\tilde{\bd{p}}_{i}$, with fixed $\tilde{\bd{p}}_{j} = \bm{\mu}^{(q-1)}_{\tilde{p}_{j}\to \phi_{ji},s}$,
$s \rmv\in\rmv \{1,\ldots,S_{j\to i}^{(q-1)}\}$)
around $\sum_{r=1}^{S_{i\to j}^{(q-1)}} \!\! w_{i\to j,r}^{(q-1)} \ist \bm{\mu}^{(q-1)}_{\tilde{p}_{i}\to \phi_{ij},r}$
and obtain the parameters of \eqref{eq:gauss_zeta_px} as
\cite{sathyan13,meyer_ICASSP13}
\vspace{-1mm}
 \begin{align*}
    r^{\internode{(q)}}_{\phi_{ij}} & =\ist \mu^{(q)}_{f_{ij}\to d_{ij}} \, , 
    \\[1mm]
    \bm{\mu}^{\internode{(q)}}_{\phi_{ij},s} & =\ist \bm{\mu}^{(q-1)}_{\tilde{p}_{j}\to \phi_{ji},s} \, , 
    \\[1mm]
    \sigma_{\phi_{ij},s}^{2 \ist (q)} & =\ist 
    \bar{\bm{\mu}}_{p_{ij},s}^{(q-1)\text{T}} \,\bm{\Sigma}_{\tilde{p}_j \to \phi_{ji},s}^{(q-1)} \, \bar{\bm{\mu}}_{p_{ij},s}^{(q-1)}
    +\ist \sigma_{{f}_{ij} \to d_{ij}}^{2\,(q)} \, , 
  \end{align*}
  where $\bar{\bm{\mu}}_{p_{ij},s}^{(q-1)} \rmv\triangleq\rmv \bm{\mu}_{p_{ij},s}^{(q-1)} / \big\| \bm{\mu}_{p_{ij},s}^{(q-1)} \big\|$ with 
$\bm{\mu}_{p_{ij},s}^{(q-1)} \rmv\triangleq\rmv \bm{\mu}^{(q-1)}_{\tilde{p}_{j}\to \phi_{ji},s}$\linebreak 
$- \sum_{r=1}^{S_{i\to j}^{(q-1)}} \!\! w_{i\to j,r}^{(q-1)} \ist \bm{\mu}^{(q-1)}_{\tilde{p}_{i}\to \phi_{ij},r}$. 
Furthermore,
$w_{j\to i,s}^{(q-1)}$ and $S_{j\to i}^{(q-1)}\rmv$ in \eqref{eq:gauss_zeta_px} equal the respective parameters of 
$\eta_{\phi_{ji}}^{(q-1)}(\tilde{\bd{p}}_j)$ (cf.\ Section \ref{sssec:particleMessageMult_1}).
In this context, 
\pagebreak 
note that 
$\zeta^{\internode{(q)}}_{\phi_{ij}}(\tilde{\mathbf{p}}_i) 
 = \int\!\rmv\int \rmv f(d_{ij}|\tilde{\bd{p}}_i,\tilde{\bd{p}}_j\rmv) \, \eta_{\phi_{ji}}^{\internode{(q-1)}}(\tilde{\bd{p}}_j\rmv) \, \eta_{\phi_{ij}}^{\internode{(q)}}(d_{ij}\rmv) 
 \, \mathrm{d}\tilde{\mathbf{p}}_j \ist \mathrm{d}d_{ij}$ involves only $\eta_{\phi_{ji}}^{\internode{(q-1)}}(\tilde{\bd{p}}_j)$ as Gaussian mixture distribution whereas 
 $\eta_{\phi_{ij}}^{\internode{(q)}}(d_{ij})$ is a 
 Gaussian distribution.

\subsubsection{Message $\eta_{{f}_{ij}}^{(q)}({\bm{\vartheta}}_i)$}
\label{sssec:particleMessageMult}
The parameters of $\eta_{{f}_{ij}}^{(q)}({\bm{\vartheta}}_i)$, $j \rmv \in \rmv \mathcal{T}_i$ 
are calculated from those of $\zeta_{{f}_{i}}({\bm{\vartheta}}_i)$ and $\zeta^{\internode{(q)}}_{{f}_{ij'}} \!({\bm{\vartheta}}_i)$, 
$j' \! \in \rmv \mathcal{T}_i^{c(q)}\setminus$\linebreak 
$\{j\}$ 
according to
\eqref{eq:BP_eta}. Since all involved messages are Gaussian, 
$\eta_{{f}_{ij}}^{(q)}({\bm{\vartheta}}_i)$ is a single Gaussian with parameters \cite{loeliger2007factor} 
    \begin{align}
      \bm{\Sigma}_{{\vartheta}_i \to f_{ij}}^{\internode{(q)}} &\rmv=\ist \Bigg( \rmv\bm{\Sigma}_{{f}_i \to {\vartheta}_i}^{-1} 
	+ \rrmv \sum_{j' \in \mathcal{T}_i^{c(q)}\setminus\{j\}} \!\!\! \bm{\Sigma}_{{f}_{ij'} \to {\vartheta}_i}^{\internode{(q)}\ist-1} \Bigg)^{\!\!-1} \!,
	\label{eq:IU_t3} \\
      \bm{\mu}_{{\vartheta}_i \to f_{ij}}^{\internode{(q)}} &\rmv=\, \bm{\Sigma}_{{\vartheta}_i \to f_{ij}}^{\internode{(q)}} 
	\!\Bigg( \rmv\bm{\Sigma}_{{f}_i \to {\vartheta}_i}^{-1} \bm{\mu}_{{f}_i \to {\vartheta}_i} \nonumber \\[-2.5mm]
      & \hspace{19mm} + \hspace{-2mm} \sum_{j' \in \mathcal{T}_i^{c(q)}\setminus\{j\}} \!\!\! \bm{\Sigma}_{{f}_{ij'} \to {\vartheta}_i}^{\internode{(q)}\ist-1}  
	\bm{\mu}_{{f}_{ij'} \to {\vartheta}_i}^{\internode{(q)}} \Bigg) \ist . \label{eq:IU_t4} \\[-6mm]
	\nn
    \end{align}
 
\subsubsection{Message $\eta_{{\phi}_{ij}}^{(q)}(\tilde{\bd{p}}_i)$}
\label{sssec:particleMessageMult_1}
The parameters of $\eta_{{\phi}_{ij}}^{(q)}(\tilde{\bd{p}}_i)$, $j \rmv \in \rmv \mathcal{T}_i$ 
are calculated from those of $\zeta_{\psi_{i}}(\tilde{\mathbf{p}}_i)$ and 
$\zeta^{(q)}_{\phi_{ij'}}(\tilde{\mathbf{p}}_i)$, $j' \!\in\rmv \mathcal{T}_i^{p(q)}\setminus \{j\}$ via 
\eqref{eq:BP_eta}, which reads
  \be
  \eta^{(q)}_{\phi_{ij}}(\tilde{\mathbf{p}}_i) \ist=\, \zeta_{\psi_{i}}(\tilde{\mathbf{p}}_i) \!\! \prod_{j' \in \mathcal{T}_i^{p(q)}\setminus \{j\}} \!\!\rmv \zeta^{(q)}_{\phi_{ij'}}(\tilde{\mathbf{p}}_i) \,.
    \label{eq:eta_p_mult}
\vspace{-1.5mm}
  \ee
  This product
  involves the annularly shaped messages 
  $\zeta^{\internode{(q)}}_{\phi_{ij'}}(\tilde{\bd{p}}_i)$ (see \eqref{eq:gauss_zeta_px}).
  We 
  use a particle implementation of \eqref{eq:eta_p_mult}
  based on importance sampling \cite{doucet}, which is inspired by an approach proposed for localization in \cite{ihler} and \cite{meyer_ICASSP13}.  
  The resulting particle representation of $\eta^{(q)}_{\phi_{ij}}(\tilde{\mathbf{p}}_i)$ is then approximated by a Gaussian or Gaussian mixture distribution, 
  or the message $\eta^{(q)}_{\phi_{ij}}(\tilde{\mathbf{p}}_i)$ is declared uninformative as explained presently.

The proposal distribution for importance sampling is chosen similarly as in \cite{ihler}, i.e.,
  \be
  p^{(q)}(\tilde{\mathbf{p}}_i) \ist\triangleq\, \zeta_{\psi_{i}}(\tilde{\mathbf{p}}_i) \ist+\!\!\rmv \sum_{j' \in \mathcal{T}_i^{p(q)}\setminus \{j\}} \!\!\rmv\zeta^{(q)}_{\phi_{ij'}}(\tilde{\mathbf{p}}_i) \,.
    \label{eq:proposal}
  \vspace{-.5mm}
  \ee
 To obtain particles representing
  $p^{(q)}(\tilde{\mathbf{p}}_i)$, we first draw particles $\big\{ \tilde{\mathbf{p}}_{\zeta_{i},i}^{(l)} \big\}_{l=1}^{L}$ from the Gaussian or Gaussian mixture 
  message $\zeta_{\psi_{i}}(\tilde{\mathbf{p}}_i)$. 
  Next, for each $j' \!\in\rmv \mathcal{T}_i^{p(q)}\setminus\rmv \{j\}$, we generate particles $\big\{ \tilde{\mathbf{p}}_{\zeta_{ij'},i}^{(l)} \big\}_{l=1}^{L}$ representing 
  $\zeta^{(q)}_{\phi_{ij'}}(\tilde{\mathbf{p}}_i)$ according to \cite{ihler}
  \[
    \tilde{\mathbf{p}}_{\zeta_{ij'},i}^{(l)} =\, \tilde{\mathbf{p}}_{\eta_{ij'},j'}^{(l)} +\ist d_{ij'}^{(l)} \rmv \begin{bmatrix} \sin(\varphi^{(l)}) \\[.4mm] \cos(\varphi^{(l)}) \end{bmatrix}.
  \vspace{-.7mm}
  \]
  This involves particles $\big\{\tilde{\mathbf{p}}_{\eta_{ij'},j'}^{(l)} \big\}_{l=1}^L$ drawn from $\eta^{(q-1)}_{\phi_{ij'}}(\tilde{\mathbf{p}}_{j'})$, 
  particles $\big\{ d_{ij'}^{(l)} \big\}_{l=1}^L$ drawn from $\eta^{(q)}_{{f}_{ij'}}(d_{ij'})$, and 
  particles $\big\{\varphi^{(l)}\big\}_{l=1}^L$ uniformly drawn 
  on $[0,2\pi)$.
  Then, $\big|\mathcal{T}_i^{p(q)}\big| L$ particles $\big\{\tilde{\mathbf{p}}_{i}^{(l)} \big\}_{l=1}^{|\mathcal{T}_i^{p(q)}| L}$
  representing the proposal distribution $p^{(q)}(\tilde{\mathbf{p}}_i)$ in \eqref{eq:proposal} are 
  obtained 
  by fusing the particles $\big\{ \tilde{\mathbf{p}}_{\zeta_i,i}^{(l)} \big\}_{l=1}^{L}$ and 
  $\big\{ \tilde{\mathbf{p}}_{\zeta_{ij'},i}^{(l)} \big\}_{l=1}^{L}$,
  $j' \!\in\rmv \mathcal{T}^{p(q)}_i\setminus\rmv \{j\}$,
  i.e., 
  \[
  \big\{\tilde{\mathbf{p}}_{i}^{(l)} \big\}_{l=1}^{|\mathcal{T}_i^{p(q)}| L} = \big\{ \tilde{\mathbf{p}}_{\zeta_i,i}^{(l)} \big\}_{l=1}^{L}
  \rmv\cup\!\! \bigcup_{j' \!\in\rmv \mathcal{T}_i^{p(q)}\setminus\rmv \{j\}} \!\!\!\rmv \big\{ \tilde{\mathbf{p}}_{\zeta_{ij'},i}^{(l)} \big\}_{l=1}^{L} \ist.
  \]
  The corresponding weights are calculated as 
  \[
  w_i^{(l)} \!=\ist \frac{\eta^{(q)}_{\phi_{ij}}\big(\tilde{\mathbf{p}}_{i}^{(l)}\big)}  {p^{(q)}\big(\tilde{\mathbf{p}}_{i}^{(l)}\big)}
    \ist=\ist \frac{\zeta_{\psi_{i}}\big(\tilde{\mathbf{p}}_{i}^{(l)}\big) \ist \prod_{j' \in \mathcal{T}_i^{p(q)}\setminus \{j\}} \zeta^{(q)}_{\phi_{ij'}}\rmv\big(\tilde{\mathbf{p}}_{i}^{(l)}\big)}
    {\zeta_{\psi_{i}}\big(\tilde{\mathbf{p}}_{i}^{(l)}\big) + \sum_{j' \in \mathcal{T}_i^{p(q)}\setminus \{j\}} \zeta^{(q)}_{\phi_{ij'}}\rmv\big(\tilde{\mathbf{p}}_{i}^{(l)}\big)} \,,
  \]
  for $l \rmv=\rmv 1,\dots,\big|\mathcal{T}_i^{p(q)}\big| L$. 
  This involves an evaluation of the messages $\zeta_{\psi_{i}}(\tilde{\mathbf{p}}_i)$ (cf.\ Section \ref{sssec:Msg_tempUpdate})
  and $\zeta^{(q)}_{\phi_{ij'}}\rmv(\tilde{\mathbf{p}}_i)$, $j' \!\in \mathcal{T}_i^{p(q)}\setminus \{j\}$ in \eqref{eq:gauss_zeta_px} 
  at the particles $\tilde{\mathbf{p}}_{i}^{(l)}\!$, $l \rmv=\rmv 1,\dots,\big|\mathcal{T}_i^{p(q)}\big| L$.
  The complexity of this algorithm for computing the message product \eqref{eq:eta_p_mult}
  scales only linearly in the number of particles. This improves on the quadratic scaling 
  of the particle-based message multiplication method described in \cite{ihler}.

Next, the 
particle representation $\big\{ \big(\tilde{\mathbf{p}}_{i}^{(l)}\rmv, w_i^{(l)} \big) \big\}_{l=1}^{|\mathcal{T}_i^{p(q)}| L}$ of $\eta^{(q)}_{\phi_{ij}}(\tilde{\bd{p}}_i)$ 
is converted into a Gaussian or two-component Gaussian mixture distribution, or the respective message is declared uninformative.
This is done using the procedure described in \cite[Section 4.1]{meyer_ICASSP13}, which involves two thresholds $\tau_1$ and $\tau_2$.
In the informative case, one obtains the Gaussian parameters $\bm{\mu}^{\internode{(q)}}_{\tilde{p}_i \to \phi_{ij},1}$ and $\bm{\Sigma}^{(q)}_{\tilde{p}_i \to \phi_{ij},1}$
(here, $w_{i\to j,1}^{(q)} \rmv=\! 1$ and $S_{i\to j}^{(q)} \rmv=\! 1$) or the Gaussian mixture parameters 
$\bm{\mu}^{\internode{(q)}}_{\tilde{p}_i \to \phi_{ij},s}$, $\bm{\Sigma}^{(q)}_{\tilde{p}_i \to \phi_{ij},s}$, and $w_{i\to j,s}^{(q)}$ for $s \in \{1, 2\}$ (here, $S_{i\to j}^{(q)} \rmv=\rmv 2$).

\vspace{-1mm}

\subsection{Calculation of Messages $\eta_{\psi_{i}}(\tilde{\bd{p}}_i)$ and $\zeta_{\psi_i}(\mathbf{x}_i)$} \label{sssec:Msg_eta-zeta-Q}
The messages $\eta_{\psi_{i}}(\tilde{\bd{p}}_i)$ and $\zeta_{\psi_i}(\mathbf{x}_i)$ are calculated after 
the final message passing iteration ($q\rmv=\rmv Q$).

\subsubsection{Message $\eta_{\psi_{i}}(\tilde{\bd{p}}_i)$}
\label{sssec:MessageMult_eta_zeta}
According to \eqref{eq:BP_eta},\\
  \[
  \eta_{\psi_{i}}(\tilde{\bd{p}}_i) \ist=\! \prod_{j \in \mathcal{T}_i^{p(Q)}} \!\!\zeta^{(Q)}_{\phi_{ij}}(\tilde{\mathbf{p}}_i) \,.
\vspace{-1.5mm}
  \]
If $\mathcal{T}_i^{p(Q)}$ is nonempty, then a 
  Gaussian or Gaussian mixture distribution with 
  parameters $\bm{\mu}_{\tilde{p}_{i} \to \psi_{i},s}$, $\bm{\Sigma}_{\tilde{p}_{i} \to \psi_{i},s}$, $w_{\tilde{p}_{i},s}$, 
  and $S_{\tilde{p}_{i}}\!\in\rmv \{1,2\}$ is obtained by carrying out similar steps as 
  in Section \ref{sssec:particleMessageMult_1},
  using the proposal distribution
  $p(\tilde{\mathbf{p}}_i) \triangleq \sum_{j \in \mathcal{T}_i^{p(Q)}} \zeta^{(Q)}_{\phi_{ij}}(\tilde{\mathbf{p}}_i)$ and replacing 
  $S_{i\to j}^{(q)}$ by $S_{\tilde{p}_i}$. 
If $\mathcal{T}_i^{p(Q)}$ is empty 
or if $\eta_{\psi_{i}}(\tilde{\bd{p}}_i)$ is found to be uninformative,
then $\eta_{\psi_{i}}(\tilde{\bd{p}}_i)$ is set to a constant (i.e., 
$\bm{\Sigma}_{\tilde{p}_i \to \psi_i,s}^{-1}$ is set to the zero matrix). 
   
\subsubsection{Message $\zeta_{\psi_i}(\mathbf{x}_i)$}
\label{sssec:Message_zta_psi}
The parameters of $\zeta_{\psi_i}(\mathbf{x}_i)$ are calculated from those of $\eta_{\psi_{i}}(\tilde{\bd{p}}_i)$ 
based on \eqref{eq:BP_zeta}. One obtains
  \begin{align}
    \bm{\Sigma}_{\psi_i \to x_i,s}^{-1} &= \mathbf{P}^\text{T} \bm{\Sigma}_{\tilde{p}_i \to \psi_i,s}^{-1} \mathbf{P} \, , \nn
    \\[.8mm]
    \bm{\Sigma}_{\psi_i \to x_i,s}^{-1} \ist \bm{\mu}_{\psi_i \to x_i,s} &= \mathbf{P}^\text{T} \bm{\Sigma}_{\tilde{p}_i \to \psi_i,s}^{-1} \ist 
      \bm{\mu}_{\tilde{p}_i \to \psi_i,s} \, . 
      \label{eq:MU_b2}
  \end{align}
  Note that \eqref{eq:MU_b2} yields $\bm{\Sigma}_{\psi_i \to x_i,s}^{-1} \ist \bm{\mu}_{\psi_i \to x_i,s}$ (instead of $\bm{\mu}_{\psi_i \to x_i,s}$) because that 
  product will be used in \eqref{eq:MU_b4}. The $w$ and $S$ parameters are $w_{\tilde{p}_{i},s}$ and $S_{\tilde{p}_{i}}$
  (see Section \ref{sssec:MessageMult_eta_zeta}).

\vspace{-1mm}

\subsection{Calculation of Beliefs} \label{sssec:Msg_bCalc}

Once the parameters of $\eta_{\psi_{i}}(\tilde{\bd{p}}_i)$ and $\zeta_{\psi_i}(\mathbf{x}_i)$ are available,
the beliefs $b(\bm{\vartheta}_i)$ and $b(\bd{x}_i)$ are calculated according to \eqref{eq:BP_be}. 
 

\subsubsection{Belief $b(\bm{\vartheta}_i)$}
\label{sec:b_theta}
  The parameters $\bm{\Sigma}_{\vartheta_i \to f_i}$ and $\bm{\mu}_{\vartheta_i \to f_i}$
  of belief $b(\bm{\vartheta}_i)$ are calculated from those of $\zeta_{f_i}(\bm{\vartheta}_i)$ and 
  $\zeta_{f_{ij}}^{(Q)}(\bm{\vartheta}_i)$, $j \!\in\! \mathcal{T}_i^{c(Q)}\!$. This is done by calculating the expressions in \eqref{eq:IU_t3} and \eqref{eq:IU_t4}, respectively,
  in which $q$ is replaced by $Q$, the summation index set $\mathcal{T}_i^{c(q)}\setminus \{j\}$ is replaced by $\mathcal{T}_i^{c(Q)}\!$, and all the terms involving 
  $\bm{\Sigma}_{{f}_i \to {\vartheta}_i}^{-1}$ are suppressed.  

\subsubsection{Belief $b(\bd{x}_i)$}
\label{sec:b_x}
The parameters $\bm{\mu}_{{x}_{i} \to l_{i},s}$, $\bm{\Sigma}_{{x}_{i} \to l_{i},s}$, $w_{{b}_{i},s}$, and $S_{{b}_{i}} \!\in\rmv \{1,2\}$ of belief $b(\mathbf{x}_i)$ 
are obtained by multiplying $\zeta_{\psi_i}(\mathbf{x}_i)$ and $\zeta_{l_i}(\mathbf{x}_i)$.
These messages are mixtures of, respectively, $S_{\tilde{p}_{i}}$ and $S_{{x}_{i}}$ components. This results in $S_{\tilde{p}_{i}} S_{{x}_{i}}$ mixture components 
for $b(\mathbf{x}_i)$, with parameters
  \begin{align}
    \bm{\Sigma}_{x_i \to l_i,(r,s)}  &= \big(\bm{\Sigma}_{\psi_i \to x_i,r}^{-1} + \bm{\Sigma}_{l_i \to x_i,s}^{-1}\big)^{-1} \rmv, \label{eq:MU_b3}  \\[.5mm]
    \bm{\mu}_{x_i \to l_i,(r,s)}  &=\ist \bm{\Sigma}_{x_i \to l_i,(r,s)} \big( \bm{\Sigma}_{\psi_i \to x_i,r}^{-1} \ist \bm{\mu}_{\psi_i \to x_i,r} \nonumber \\
    & \hspace{22mm} + \bm{\Sigma}_{l_i \to x_i,s}^{-1} \ist\bm{\mu}_{l_i \to x_i,s} \big) \label{eq:MU_b4}\\[-6mm]
  \nonumber
  \end{align}
  and weights (before normalization)
  \begin{align*}
    \tilde{w}_{b_i,(r,s)} &=\ist w_{\tilde{p}_{i},r} w_{{x}_{i},s} \\
      & \hspace{5mm} \times \mathrm{exp}\big( {-\ist h_{\psi_i \to x_i,r}} - h_{l_i \to x_i,s} + h_{x_i \to l_i,(r,s)} \big) \, ,
  \end{align*}
  where 
  $h_{\psi_i \to x_i,r} \triangleq \bm{\mu}_{\psi_i \to x_i,r}^{\text{T}} \ist   \bm{\Sigma}_{\psi_i \to x_i,r}^{-1} \ist   \bm{\mu}_{\psi_i \to x_i,r}$,
  $h_{l_i \to x_i,s} \triangleq$\linebreak 
  $\bm{\mu}_{l_i \to x_i,s}^{\text{T}} \ist  \bm{\Sigma}_{l_i \to x_i,s}^{-1} \ist \bm{\mu}_{l_i \to x_i,s}$, and
  $h_{x_i \to l_i,(r,s)} \triangleq \bm{\mu}_{x_i \to l_i,(r,s)}^{\text{T}}$\linebreak 
  $\times \bm{\Sigma}_{x_i \to l_i,(r,s)}^{-1} \ist \bm{\mu}_{x_i \to l_i,(r,s)}$.
  Note that $S_{\tilde{p}_{i}}S_{{x}_{i}}$ may be $1$, $2$, or $4$.
  If $S_{\tilde{p}_{i}} S_{{x}_{i}}$ is $1$ or $2$, we use all the 
  mixture components to represent the product message $b(\mathbf{x}_i)$, i.e., $S_{{b}_{i}} \!=\rmv S_{\tilde{p}_{i}} S_{{x}_{i}}$,
  and the final weights $w_{b_i,(r,s)}$ are obtained by normalizing the $\tilde{w}_{b_i,(r,s)}$.
  However, if $S_{\tilde{p}_{i}} S_{{x}_{i}} \!=\rmv 4$, we set $S_{{b}_{i}} \!=\rmv 2$ and use only the two strongest mixture components,
  corresponding to the two 
  index tuples $(r,s)$ whose weights $\tilde{w}_{b_i,(r,s)}$ are largest. These weights are then normalized. The parameters and weights obtained in this way are then
  assigned to $\bm{\Sigma}_{x_i \to l_i,s'}$, $\bm{\mu}_{x_i \to l_i,s'}$, and $w_{b_{i},s'}$ with $s' \in \{1,\ldots, S_{{b}_{i}} \}$.

%
  \begin{table}[t!]
\vspace{-.1mm}
      \small    
      \caption{CoSLAS 
       BP Algorithm---Operations Performed by Agent $i$} 
	  \label{alg:BP_ex}
      \vspace{-2mm}

	  {\hrule height .5pt} 
      \vspace{2.5mm}
	  
\textbf{Initialization at time $n \!=\! 0$:}\,
      
      \vspace{1mm}
      The temporal recursion is initialized by setting 
\vspace{-.3mm}
      $b({\bm{\vartheta}}_i) \rmv=\rmv f\big({\bm{\vartheta}}_i^{(0)}\big)$ and 
      $b(\bd{x}_i) \rmv=\rmv f\big(\bd{x}_i^{(0)}\big)$
      (see \eqref{eq:prior_theta} and \eqref{eq:prior_x}, respectively). 

      \vspace{2mm}
		      
      \textbf{Temporal recursion at times $n \rmv\ge\rmv 1$:}\, 

      \vspace{1.5mm}
      
      \emph{Step 1 -- Prediction}: 

	  \begin{enumerate} 
	    \vspace{.5mm}
	    
	    \item[1.1)] 
		The clock message
		$\zeta_{{f}_i}({\bm{\vartheta}}_i)$ is calculated from 
		$b^-({\bm{\vartheta}}_i)$ 
		(which was calculated at time $n\!-\rmv\!1$)
		according to \eqref{eq:MU_t1} and  \eqref{eq:MU_t2}. 

		\vspace{1mm}
 
 	    \item[1.2)] 
		The location message
		$\zeta_{l_i}(\bd{x}_i)$ is calculated from 
		$b^-(\bd{x}_i)$ 
		(which was calculated at time $n\!-\rmv\!1$) according to Section \ref{sssec:Msg_locationUpdate}.

		\vspace{1mm}
      
	    \item[1.3)] The location 
		message $\zeta_{\psi_i}(\tilde{\bd{p}}_i)$ is calculated from $\eta_{\psi_i}(\bd{x}_i) = \zeta_{l_i}(\bd{x}_i)$
		according to Section \ref{sssec:Msg_tempUpdate}.

	    \vspace{1mm}

	  \end{enumerate} 
      
      \emph{Step 2 -- Iterative message passing}:\, 
      The message passing iteration is initialized by setting 
\vspace{-.3mm}
      $\eta_{{f}_{ij}}^{(0)}({\bm{\vartheta}}_i) \rmv=\rmv \zeta_{{f}_{i}}({\bm{\vartheta}}_i)$,
      $\eta_{\phi_{ij}}^{(0)}(\tilde{\bd{p}}_i) \rmv= \zeta_{\psi_{i}}(\tilde{\bd{p}}_i)$, and
      $\zeta_{f_{ij}}^{(0)}(d_{ij}) = \zeta_{\phi_{ij}}^{(0)}(d_{ij}) = f(d_{ij})$ 
      for all $j \!\in\! \cl{T}_i$. 
\vspace{-.3mm}
      Furthermore,       
      $\eta_{{f}_{ij}}^{(0)}({\bm{\vartheta}}_i)$ and 
      $\eta_{\phi_{ij}}^{(0)}(\tilde{\bd{p}}_i)$ are transmitted to the respective neighbors $j \!\in\! \cl{T}_i$.
      Then, for $q=1,\dots,Q$:
			
	  \begin{enumerate} 
	    \vspace{.5mm}
			
	    \item[2.1)] 
	    The 
	    messages
	    $\eta_{f_{ji}}^{(q-1)}({\bm{\vartheta}}_j)$ and $\eta_{\phi_{ji}}^{(q-1)}(\tilde{\bd{p}}_j)$ (calculated at the previous iteration) 
	    are received from the respective neighbors $j \!\in\! \mathcal{T}_i$. The sets 
	    $\mathcal{T}_i^{c(q)} \!= \rmv \big\{j \big| \eta_{f_{ji}}^{(q-1)}({\bm{\vartheta}}_j) \text{ is informative}\big\}$ 
\vspace{-.5mm}
and $\mathcal{T}_i^{p(q)} \!=\rmv \big\{j \big| \eta_{\phi_{ji}}^{(q-1)}(\tilde{\bd{p}}_j) \text{ is informative}\big\}$ are determined.

	    \vspace{.5mm}
	    
	    \item[2.2)] 
	    If 
	    $\eta_{f_{ij}}^{(q-1)}({\bm{\vartheta}}_i)$
	    is informative, then for all $j \!\in\! \mathcal{T}_i^{c(q)}\!$, 
\vspace{-.5mm}
	    the 
	   messages $\zeta_{f_{ij}}^{(q)}(d_{ij})$ are calculated from 
\vspace{-.3mm}
	    $\eta_{f_{ij}}^{(q-1)}({\bm{\vartheta}}_i)$ and $\eta_{f_{ji}}^{(q-1)}({\bm{\vartheta}}_j)$
	    according to \eqref{eq:IU_d1} and \eqref{eq:IU_d2}.
	    Otherwise $\zeta_{f_{ij}}^{(q)}(d_{ij}) \rmv= \zeta_{f_{ij}}^{(q-1)}(d_{ij})$.
	    
	    \vspace{.7mm}
	   
	    \item[2.3)]  
	    If 
	    $\eta_{\phi_{ij}}^{(q-1)}(\tilde{\bd{p}}_i)$ 
	    is informative, then 
\vspace{-.5mm}
for all $j \!\in\! \mathcal{T}_i^{p(q)}\!$, 
	    the 
	    messages $\zeta_{\phi_{ij}}^{(q)}(d_{ij})$ are 
\vspace{-.5mm}
calculated from 
	    $\eta_{\phi_{ij}}^{(q-1)}(\tilde{\bd{p}}_i)$ and $\eta_{\phi_{ji}}^{(q-1)}(\tilde{\bd{p}}_j)$
	    according to to \eqref{eq:IU_d3} and \eqref{eq:IU_d4}. 
	    Otherwise $\zeta_{\phi_{ij}}^{(q)}(d_{ij})\rmv= \zeta_{\phi_{ij}}^{(q-1)}(d_{ij})$.
	    
	    \vspace{.5mm}
	    
	    \item[2.4)] 
	    For 
	    $j \!\in\! \mathcal{T}_i^{c(q)}\!$,
	    the 
\vspace{-.5mm}
	    messages
	    $\zeta_{{f}_{ij}}^{(q)}({\bm{\vartheta}}_i)$ are calculated
	    \vspace{-.3mm}
 from $\eta_{{f}_{ji}}^{(q-1)}({\bm{\vartheta}}_j)$ and $\zeta_{{\phi}_{ij}}^{(q)}(d_{ij})$ according to \eqref{eq:IU_t1} and \eqref{eq:IU_t2}.
	    
	    \vspace{1mm}
	     	    
	    \item[2.5)]
	    For 
	    $j \!\in\! \mathcal{T}_i^{p(q)}\!$,
	    the 
	    messages $\zeta_{\phi_{ij}}^{(q)}(\tilde{\bd{p}}_i)$ 
\vspace{-.5mm}
	    are calculated from $\eta_{\phi_{ji}}^{(q-1)}(\tilde{\bd{p}}_j)$ and $\zeta_{{f}_{ij}}^{(q)}(d_{ij})$ according to Section \ref{sec:zeta-phi-p}.
	    
	    \vspace{.7mm}

	    \item[2.6)] 
	    For 
	    $j\!\in\!\mathcal{T}_i$,
	    the 
	    messages
\vspace{-.5mm}
	    $\eta_{f_{ij}}^{(q)}({\bm{\vartheta}}_i)$ are calculated 
	    from $\zeta_{{f}_i}({\bm{\vartheta}}_i)$ and $\zeta_{{f}_{ij'}}^{(q)}({\bm{\vartheta}}_i)$, $j'\!\rmv\in\! \mathcal{T}_i^{c(q)}\!\setminus\!\{j\}$ 
	    according to \eqref{eq:IU_t3} and \eqref{eq:IU_t4}.
	    
	    \vspace{.7mm}

	    \item[2.7)] 
	    For 
	    $j\!\in\!\mathcal{T}_i$,
	    the 
	    messages $\eta_{\phi_{ij}}^{(q)}(\tilde{\bd{p}}_i)$ are calculated from $\zeta_{\psi_i}(\tilde{\bd{p}}_i)$ and 
	    $\zeta_{\phi_{ij'}}^{(q)}(\tilde{\bd{p}}_i)$, $j'\!\rmv\in\! \mathcal{T}_i^{p(q)}\!\setminus\!\{j\}$ according to Section \ref{sssec:particleMessageMult_1}. 
	    
	    \vspace{1mm}

	    \item[2.8)] 
	    The 
	    (parameters of) the messages $\eta_{f_{ij}}^{(q)}({\bm{\vartheta}}_i)$ and
	    $\eta_{\phi_{ij}}^{(q)}(\tilde{\bd{p}}_i)$ are transmitted to the respective neighbors $j \!\in\! \mathcal{T}_i$.

      \vspace{1mm}
	    
	  \end{enumerate}

      \emph{Step 3 -- Belief calculation}: 
      			
	  \begin{enumerate} 
	    \vspace{.5mm}

	    \item[3.1)] 
	    The belief $b({\bm{\vartheta}}_i) = \eta_{f_i}({\bm{\vartheta}}_i)$ is calculated from $\zeta_{f_i}(\bm{\vartheta}_i)$ and $\zeta_{f_{ij}}^{(Q)}(\bm{\vartheta}_i)$, 
	    $j \!\in\! \mathcal{T}_i^{c(Q)}$ according to \eqref{eq:IU_t3} and \eqref{eq:IU_t4}
	    in which $q$ is replaced by $Q$, the summation index set $\mathcal{T}_i^{c(q)}\setminus \{j\}$ is replaced by $\mathcal{T}_i^{c(Q)}\!$, and all terms involving 
	    $\bm{\Sigma}_{{f}_i \to {\vartheta}_i}^{-1}$ are suppressed.

	    \vspace{1mm}
	    
	    \item[3.2)]
	    The message $\eta_{\psi_i}(\tilde{\bd{p}}_i)$ is calculated from $\zeta_{\phi_{ij}}^{(Q)}(\tilde{\bd{p}}_i)$, $j \!\in\! \mathcal{T}_i^{p(Q)}$ according to 
	    Section \ref{sssec:MessageMult_eta_zeta}. Next, the message $\zeta_{\psi_i}({\bd{x}}_i)$ is calculated from $\eta_{\psi_i}(\tilde{\bd{p}}_i)$ according to 
	    Section \ref{sssec:Message_zta_psi}.
	    Finally, the belief $b(\bd{x}_i) = \eta_{l_i}(\bd{x}_i)$ is calculated from $\zeta_{\psi_i}({\bd{x}}_i)$ and $\zeta_{l_i}({\bd{x}}_i)$ according to 
	    Section \ref{sec:b_x}.

	    \vspace{1mm}

	  \end{enumerate} 
	  
      \emph{Step 4 -- Estimation}: 
      The clock estimates $\hat{\alpha}_i$ and $\hat{\beta}_i$ and the location-related estimates $\hat{\bd{x}}_i$ are obtained from the parameters
      of $b({\bm{\vartheta}}_i)$ and $b(\bd{x}_i)$, respectively as described in Section \ref{sssec:Msg_Est}.\\[-1mm]
	  
	{\hrule height .5pt}
  \end{table}
%

\vspace{-1mm}

\subsection{Estimation} \label{sssec:Msg_Est}
Approximations $\hat{{\bm{\vartheta}}}^{(n)}_{i}$ and $\hat{{\bd{x}}}^{(n)}_{i}$ of the MMSE estimates 
$\hat{{\bm{\vartheta}}}^{(n)}_{i,\text{MMSE}}$ and $\hat{{\bd{x}}}^{(n)}_{i,\text{MMSE}}$ are obtained 
by replacing in \eqref{eq:mmse_theta} and \eqref{eq:mmse_x} the marginal posterior pdfs $f\big({\bm{\vartheta}}^{\internode{(n)}}_{i} \big|\bd{y}^{\internode{(1:n)}}\big)$ 
and $f\big(\bd{x}^{\internode{(n)}}_{i} \big|\bd{y}^{\internode{(1:n)}}\big)$ by the beliefs $b(\bm{\vartheta}_i)$ and $b(\bd{x}_i)$, respectively.
Using the parametric representations of $b(\bm{\vartheta}_i)$ and $b(\bd{x}_i)$ discussed in Sections \ref{sec:messagerep} and \ref{sssec:Msg_bCalc},
$\hat{{\bm{\vartheta}}}^{(n)}_{i}$ is directly given by $\bm{\mu}_{\vartheta_i \to f_i}$, and $\hat{{\bd{x}}}^{(n)}_{i}$
by $\sum_{s \in S_{b_i}} \!\! w_{b_i,s} \ist \bm{\mu}_{x_i \to l_i,s}$.
Finally, estimates of the primary clock parameters $\alpha_i^{(n)}$ and $\beta_i^{(n)}$ (see Section \ref{sec:systemModel-1}) 
are obtained as
$\hat{\alpha}_i^{(n)} \!=\rmv 1/{[\hat{{\bm{\vartheta}}}^{(n)}_{i}]}_2$ and 
$\hat{\beta}_i^{(n)} \!=\rmv \hat{\alpha}_i^{(n)} {[\hat{{\bm{\vartheta}}}^{(n)}_{i}]}_1$, where ${[\cdot]}_l$ denotes the $l$th element of a vector.

\vspace{-1mm}

\subsection{Algorithm Summary and Communication Requirements} \label{ssec:alg-summary}
  
  A summary of the overall algorithm is provided in Table~\ref{alg:BP_ex}. The communication requirements 
  are as follows.
  At any time $n$, in any message passing iteration $q$, the parameters of the two-dimensional messages $\eta_{\phi_{ij}}^{(q)}(\tilde{\bd{p}}_i)$ and $\eta_{f_{ij}}^{(q)}({\bm{\vartheta}}_i)$
  have to be transmitted from agent $i$ to agent $j \rmv\in\rmv \mathcal{T}_i^{(n)}\!$. According to Section \ref{sssec:particleMessageMult_1},
  $\eta_{\phi_{ij}}^{(q)}(\tilde{\bd{p}}_i)$ is either uninformative or represented by a Gaussian or two-component Gaussian mixture distribution.
  In the last case, which corresponds to maximum communication requirements,
the parameters of $\eta_{\phi_{ij}}^{(q)}(\tilde{\bd{p}}_i)$ are two mean vectors, 
  two covariance matrices, and one weight 
(as the two weights are normalized, only one of them has to be known). 
Furthermore, according to Section \ref{sssec:particleMessageMult},
  $\eta_{{f}_{ij}}^{(q)}({\bm{\vartheta}}_i)$ is represented by a single Gaussian, i.e., by one mean vector and one covariance matrix.
Hence, the total number of real values that have to be transmitted from agent $i \rmv\in\rmv \mathcal{I}$ to agent $j \rmv\in\rmv \mathcal{T}_i^{(n)}$ per iteration $q$ 
is maximally $(2 \rmv+\! 1)\,(2 \rmv+\rmv 3)+1=16$.

\section{Numerical Study} 
\label{sec:simulations}

In this section, we analyze the performance of the proposed CoSLAS algorithm 
and compare it with that of two variants with
perfect clock or location-velocity information.

\vspace{-1mm}


\subsection{Simulation Setting} 
\label{ssec:simu-set}

\begin{figure}
    \begin{center}
      \begin{psfrags}
      \begin{tabular}{cc}
	  \psfragscanon%
	  \psfrag{s01}[][]{\color[rgb]{0,0,0}\setlength{\tabcolsep}{0pt}\begin{tabular}{c}\footnotesize$ $\end{tabular}} %
	  \psfrag{x01}[t][t]{\footnotesize0} \psfrag{x02}[t][t]{\footnotesize10} \psfrag{x03}[t][t]{\footnotesize20} \psfrag{x04}[t][t]{\footnotesize30} \psfrag{x05}[t][t]{\footnotesize40} \psfrag{x06}[t][t]{\footnotesize50}%
	  \psfrag{v01}[r][r]{\footnotesize0} \psfrag{v02}[r][r]{\footnotesize10} \psfrag{v03}[r][r]{\footnotesize20} \psfrag{v04}[r][r]{\footnotesize30} \psfrag{v05}[r][r]{\footnotesize40} \psfrag{v06}[r][r]{\footnotesize50}%
	  \includegraphics[width=3.8cm]{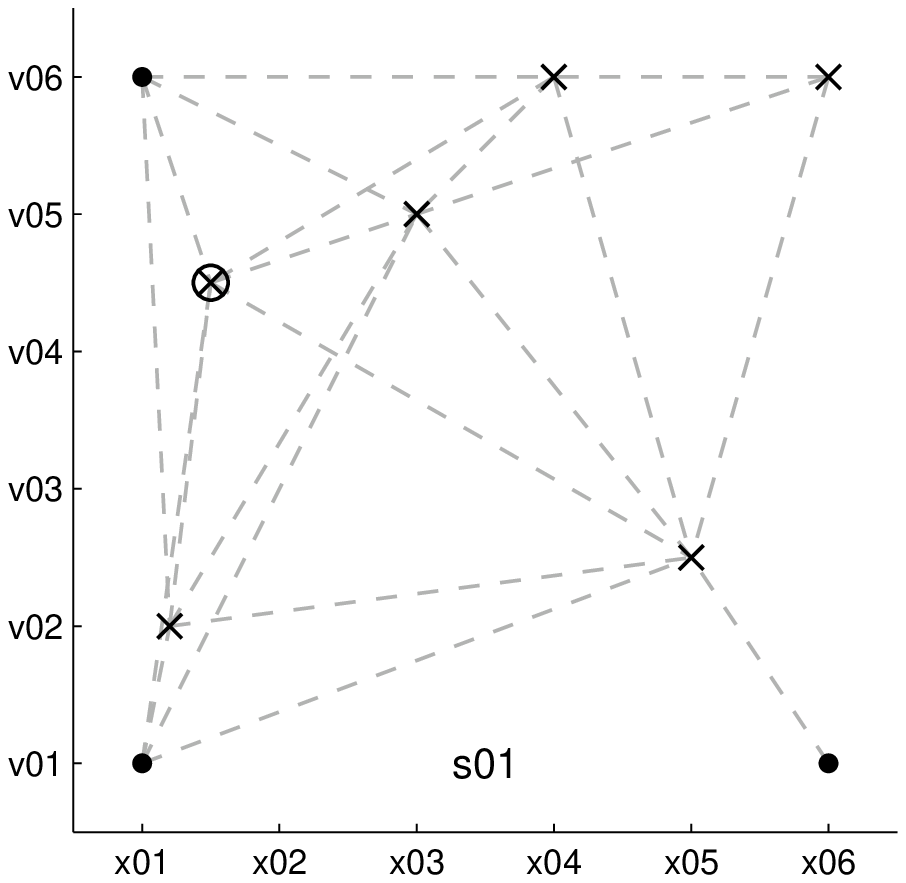}
      & \hspace{-0.5cm} 
	  \psfragscanon%
	  \psfrag{s01}[][]{\color[rgb]{0,0,0}\setlength{\tabcolsep}{0pt}\begin{tabular}{c}\footnotesize$ $\end{tabular}} %
	  \psfrag{x01}[t][t]{\footnotesize0} \psfrag{x02}[t][t]{\footnotesize10} \psfrag{x03}[t][t]{\footnotesize20} \psfrag{x04}[t][t]{\footnotesize30} \psfrag{x05}[t][t]{\footnotesize40} \psfrag{x06}[t][t]{\footnotesize50}%
	  \psfrag{v01}[r][r]{\footnotesize0} \psfrag{v02}[r][r]{\footnotesize10} \psfrag{v03}[r][r]{\footnotesize20} \psfrag{v04}[r][r]{\footnotesize30} \psfrag{v05}[r][r]{\footnotesize40} \psfrag{v06}[r][r]{\footnotesize50}%
	  \includegraphics[width=3.8cm]{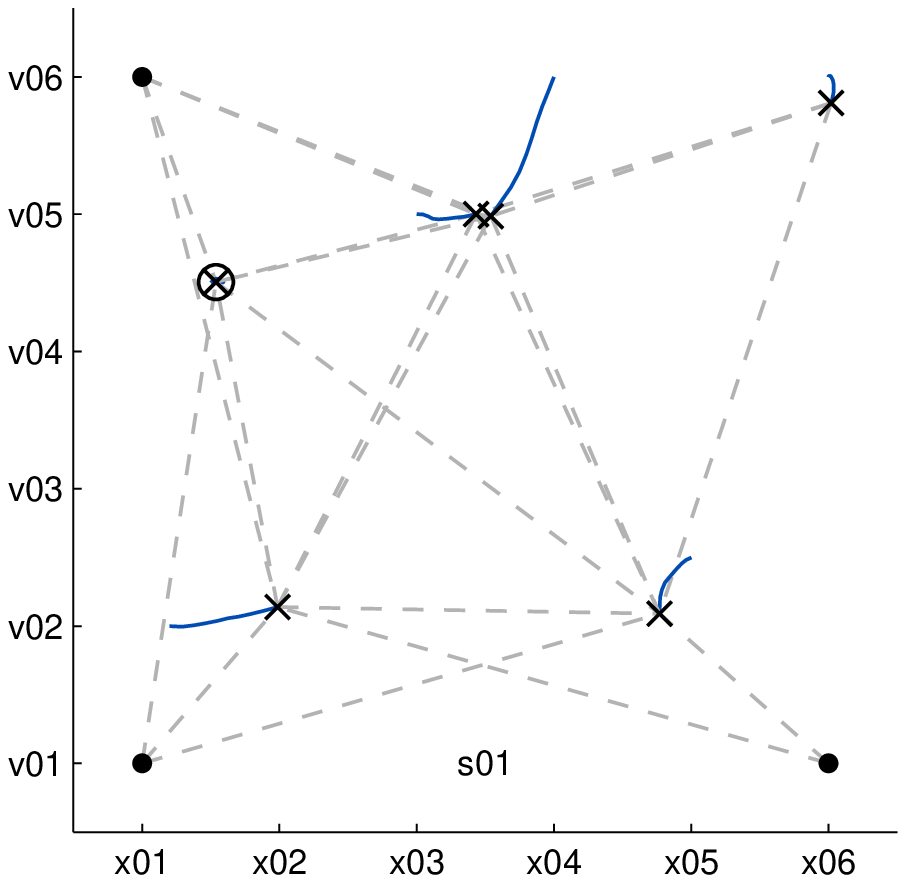}
      \\[1mm]
	  \psfragscanon%
	  \psfrag{s01}[][]{\color[rgb]{0,0,0}\setlength{\tabcolsep}{0pt}\begin{tabular}{c}\footnotesize$ $\end{tabular}} %
	  \psfrag{x01}[t][t]{\footnotesize0} \psfrag{x02}[t][t]{\footnotesize10} \psfrag{x03}[t][t]{\footnotesize20} \psfrag{x04}[t][t]{\footnotesize30} \psfrag{x05}[t][t]{\footnotesize40} \psfrag{x06}[t][t]{\footnotesize50}%
	  \psfrag{v01}[r][r]{\footnotesize0} \psfrag{v02}[r][r]{\footnotesize10} \psfrag{v03}[r][r]{\footnotesize20} \psfrag{v04}[r][r]{\footnotesize30} \psfrag{v05}[r][r]{\footnotesize40} \psfrag{v06}[r][r]{\footnotesize50}%
	  \includegraphics[width=3.8cm]{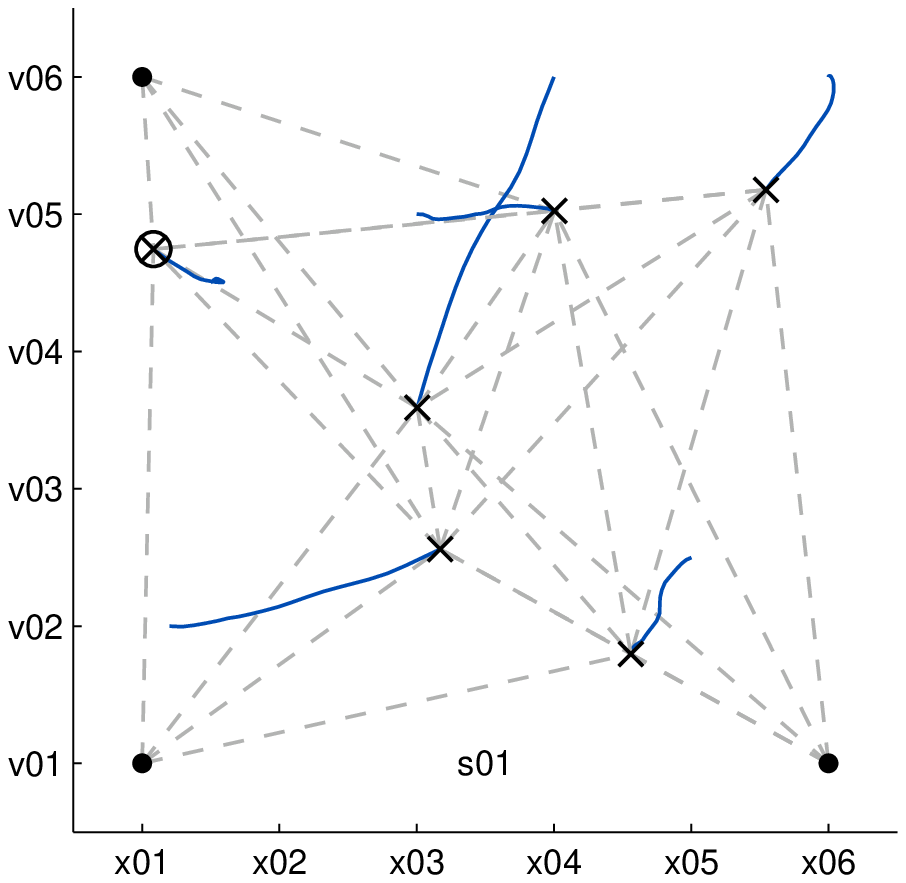}
      & \hspace{-0.5cm} 
	  \psfragscanon%
	  \psfrag{s01}[][]{\color[rgb]{0,0,0}\setlength{\tabcolsep}{0pt}\begin{tabular}{c}\footnotesize$ \quad$\end{tabular}} %
	  \psfrag{x01}[t][t]{\footnotesize0} \psfrag{x02}[t][t]{\footnotesize10} \psfrag{x03}[t][t]{\footnotesize20} \psfrag{x04}[t][t]{\footnotesize30} \psfrag{x05}[t][t]{\footnotesize40} \psfrag{x06}[t][t]{\footnotesize50}%
	  \psfrag{v01}[r][r]{\footnotesize0} \psfrag{v02}[r][r]{\footnotesize10} \psfrag{v03}[r][r]{\footnotesize20} \psfrag{v04}[r][r]{\footnotesize30} \psfrag{v05}[r][r]{\footnotesize40} \psfrag{v06}[r][r]{\footnotesize50}%
	  \includegraphics[width=3.8cm]{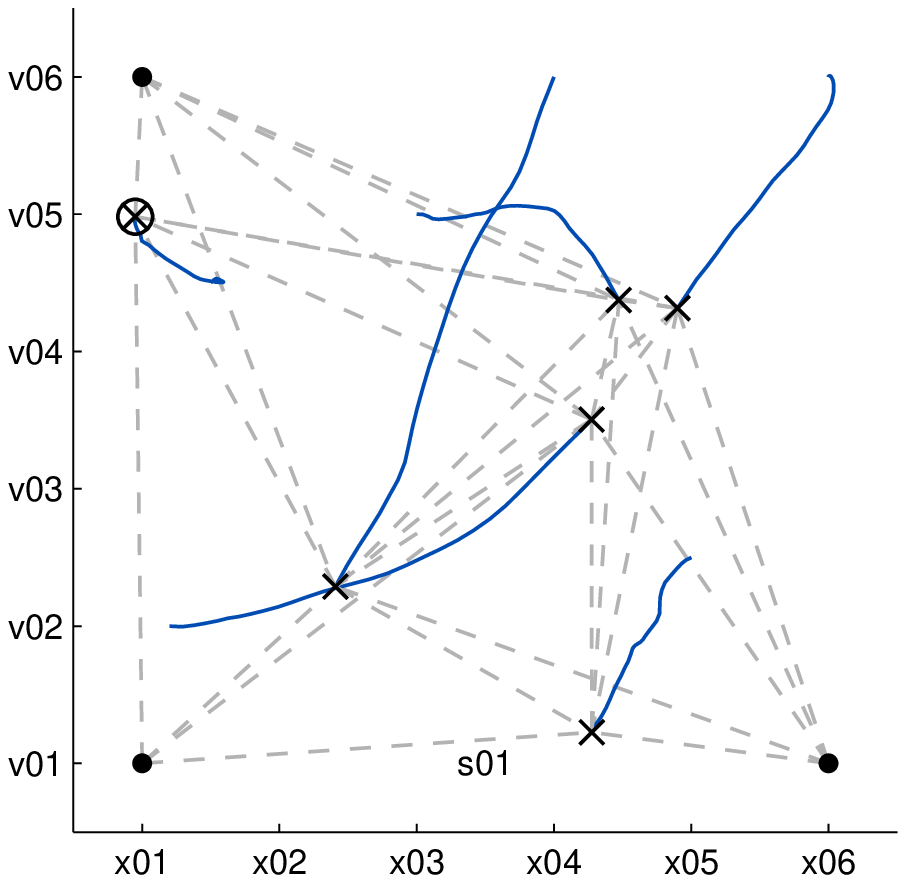}
      \end{tabular}
      \end{psfrags}
    \end{center}
    
    \vspace{-6.7cm}
      \begin{tikzpicture}[scale=1]
      \node[rotate=90] (A) at (-0.3,0) {\footnotesize$y\,$[m]};
      \node[rotate=90] (B) at (-0.3,-4.0) {\footnotesize$y\,$[m]};
      \node (C) at (2.2,-6.3) {\footnotesize$x\,$[m]};
      \node (D) at (6.05,-6.3) {\footnotesize$x\,$[m]};
      \node[rectangle,draw=black,very thin,fill=white, inner sep=0.8mm] at (2.05,-1.5) {\footnotesize$n = 0$};
      \node[rectangle,draw=black,very thin,fill=white, inner sep=0.8mm] at (5.7,-1.5) {\footnotesize$n = 10$};
      \node[rectangle,draw=black,very thin,fill=white, inner sep=0.8mm] at (2.05,-5.45) {\footnotesize$n = 20$};
      \node[rectangle,draw=black,very thin,fill=white, inner sep=0.8mm] at (5.7,-5.45) {\footnotesize$n = 30$};
    \end{tikzpicture}
    \vspace{-2mm}
  \caption{\label{sim:scenario} Agent locations at times
  $n=0$, $10$, $20$, and $30$. 
  Dots indicate the locations of the spatial reference agents, crosses indicate 
  the locations of the mobile agents, 
  the circle indicates 
  the 
  location of the temporal reference agent (one of the mobile agents), blue solid lines indicate 
  the agent trajectories, and dashed gray lines indicate 
  the measurement/communication links.}
\vspace{-1mm}
\end{figure}

We consider a network of $I \rmv=\rmv 9$ agents located in a square area of size $50\ist\text{m}\times50\ist\text{m}$, as shown 
in Fig.~\ref{sim:scenario}. 
The time interval length is $T \rmv=\rmv 1\ist$s.
Three of the agents ($i\rmv\rmv \in \rmv\rmv\{1,2,3\}$) are nonmobile spatial references located in three corners of the square area, and the remaining six agents
($i \rmv\rmv \in \rmv\rmv \{4,\ldots,9\}$) are mobile. 
Mobile agent $i \!=\! 7$ is a clock reference with known clock states $\bm{\vartheta}_7^{(n)} \!\rmv=\rmv [0 \;\, 1]^\text{T}$ for all $n$.
For $i \rmv\not=\rmv 7$, the clock states $\bm{\vartheta}_{i}^{(n)}$ evolve according to \eqref{eq:clockEvol} with
process noise standard deviations 
$\sigma_{1,i} \rmv=\rmv 1\ist\mu$s and $\sigma_{2,i} \rmv=\rmv 10\,$ppm, and 
with the initial clock states $\bm{\vartheta}^{\internode{(0)}}_{i}\!$, $i \rmv\not=\rmv 7$ randomly drawn according to \eqref{eq:prior_theta}
with $\sigma_{\nu_i} \!\rmv=\! 1\ist$s, $\sigma_{\lambda_i} \!\rmv=\rmv 150\,$ppm, and $\bm{\mu}_{f_i\to \vartheta_i}^{(0)} = [0 \;\, 1]^\text{T}\rmv$.
The location-related states $\bd{x}_{i}^{(n)}$ of the mobile agents evolve according to \eqref{eq:posEvol} with process noise standard deviation $\sigma_{u_2,i} \rmv=\rmv 2\ist$m, 
and with the initial values
$\bd{x}_{i}^{(0)}$ 
chosen as shown in Fig.~\ref{sim:scenario}.
A realization of the states $\bm{\vartheta}_{i}^{(n)}$ and $\bd{x}_{i}^{(n)}\!$, $n \rmv=\rmv 0,1,\ldots$ was generated as described above and used for all simulation runs.
Fig.~\ref{sim:scenario} shows the locations of the agents at four different times $n$.

Each agent communicates with other agents within a radius of $40\,$m, i.e., 
$\mathcal{T}_i^{(n)} \!=\rmv \big\{j \rmv\in\rmv \mathcal{I} \ist \big| \ist \big\|\bd{p}_i^{(n)} \!\rmv-\rmv\bd{p}_j^{(n)} \big\| \rmv\leq\rmv 40\,\text{m} \big\}$.  \nolinebreak 
The \nolinebreak 
net\-work \nolinebreak 
connectivity is time-varying (cf.\ Fig.~\ref{sim:scenario}) but the network is always connected, as required by our initialization protocol 
in Section \ref{ssec:asymTimeStamp}.
The agents perform 
$K_{ij} \rmv=\rmv K_{ji} \rmv=\! 10$ noisy\linebreak 
measurements relative to each neighbor according to \eqref{eq:mess_rel}. 
In each of the 100 simulation runs we performed, the \nolinebreak 
mea\-surement \nolinebreak 
noises $v_{ij}^{(n,k)}$ in \eqref{eq:mess_rel} were 
drawn independently for all $(i,j)\in \mathcal{C}^{(n)}\rmv$, $n$,
and $k \rmv\in\rmv \{1,\ldots,K_{ij} \!=\! 10\}$, with a noise standard deviation of $\sigma_v \rmv=\rmv 10\ist$ns.

In the simulated algorithms, the parameters 
used to initialize the distance messages $\zeta_{f_{ij}}^{(0)}(d_{ij}) \rmv= \zeta_{\phi_{ij}}^{(0)}(d_{ij})$ (see Section \ref{ssec:Msg_itMsgPass})
are $\mu_d \rmv=\rmv 27\,$m and $\sigma_d \rmv=\rmv 10\,$m. The process noise parameters and the parameters $\sigma_{\nu_i}$ and $\sigma_{\lambda_i}$ 
are as stated earlier.
The number of particles used for message multiplication (see Section \ref{sssec:particleMessageMult_1})
is $\big|\cl{T}_i^{p(q)}\big| L =\rmv 1000$. 
The threshold parameters 
(see Section \ref{ssec:Msg_itMsgPass}) are $\tau \rmv=\rmv 2$, $\tau_1 \rmv=\rmv 15$, and $\tau_2 \rmv=\rmv 40$. 
The initial covariance matrix of $\bd{x}_i^{(0)}\!$,
$\bm{\Sigma}_{l_i\to x_i}^{(0)}\rmv$ (see \eqref{eq:prior_x}),
is defined by $\sigma_{x_i} \!\rmv=\rmv 5\,$m and $\sigma_{\dot{x}_i} \!\rmv=\rmv 2\,$m/s,
and the initial mean 
is modeled randomly as $\bm{\mu}_{l_i \to x_i}^{(0)} \!\rmv=\rmv \bd{x}_i^{(0)} \rmv+ \bm{\varepsilon}_i$, 
where $\bd{x}_i^{(0)}$ is the actual initial location-related state and $\bm{\varepsilon}_i \rmv\sim\rmv \mathcal{N}\big(\bm{\varepsilon}_i;\bd{0},\bm{\Sigma}_{l_i \to x_i}^{(0)} \big)$
was drawn independently for all $i$ and all simulation runs.

\vspace{-1mm}

\subsection{Simulation Results} 
\label{ssec:simu-res}

We consider the proposed CoSLAS algorithm (briefly referred to as CoSLAS) and two variants performing only 
localization or 
synchronization. 
In the first variant, dubbed
ClkRef, 
all agents know
their clock parameters, and
in the second variant, 
LocRef, 
all agents know
their location and velocity. We 
are not able to present a comparison with other
methods because, to the best of our knowledge, there 
are no other SLAS methods for time-varying clock skew and clock offset. 
Our measure of performance is the root mean square error (RMSE) of the various parameters averaged over 
100 simulation runs and 
those agents that are not reference agents.

\begin{figure}[t!]
    \begin{center}
    \begin{psfrags}
      \begin{tabular}{ccc}
      & & \\[-3mm]
      \hspace{0.2cm} 
	\psfragscanon%
	\psfrag{x01}[t][t]{\footnotesize0} \psfrag{x02}[t][t]{\footnotesize1} \psfrag{x03}[t][t]{\footnotesize2} \psfrag{x04}[t][t]{\footnotesize3} \psfrag{x05}[t][t]{\footnotesize4} \psfrag{x06}[t][t]{\footnotesize5}%
	\psfrag{v01}[r][r]{\footnotesize 0} \psfrag{v02}[r][r]{\footnotesize 2} \psfrag{v03}[r][r]{\footnotesize 4} \psfrag{v04}[r][r]{\footnotesize 6} \psfrag{v05}[r][r]{\footnotesize 8} \psfrag{v06}[r][r]{\footnotesize 10} 
	\includegraphics[width=2.8cm]{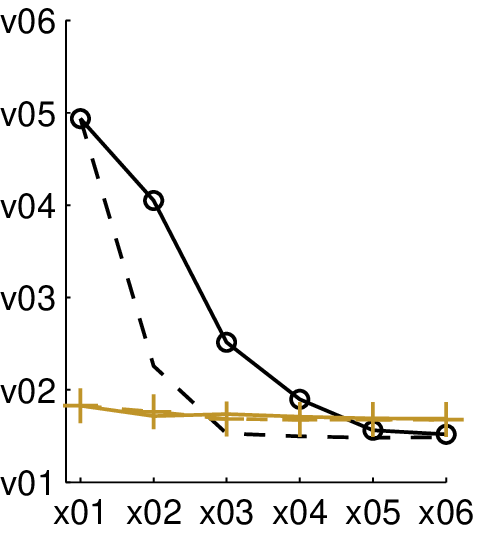}
      & \hspace{-0.5cm} 
	\psfrag{x01}[t][t]{\footnotesize0} \psfrag{x02}[t][t]{\footnotesize1} \psfrag{x03}[t][t]{\footnotesize2} \psfrag{x04}[t][t]{\footnotesize3} \psfrag{x05}[t][t]{\footnotesize4} \psfrag{x06}[t][t]{\footnotesize5}%
	\psfrag{v01}[r][r]{\footnotesize 0} \psfrag{v02}[r][r]{\footnotesize } \psfrag{v03}[r][r]{\footnotesize 1} \psfrag{v04}[r][r]{\footnotesize } \psfrag{v05}[r][r]{\footnotesize 2} \psfrag{v06}[r][r]{\footnotesize } \psfrag{v07}[r][r]{\footnotesize 3} \psfrag{v08}[r][r]{\footnotesize } \psfrag{v09}[r][r]{\footnotesize 4} \psfrag{v10}[r][r]{\footnotesize } \psfrag{v11}[r][r]{\footnotesize 5}
	\includegraphics[width=2.8cm]{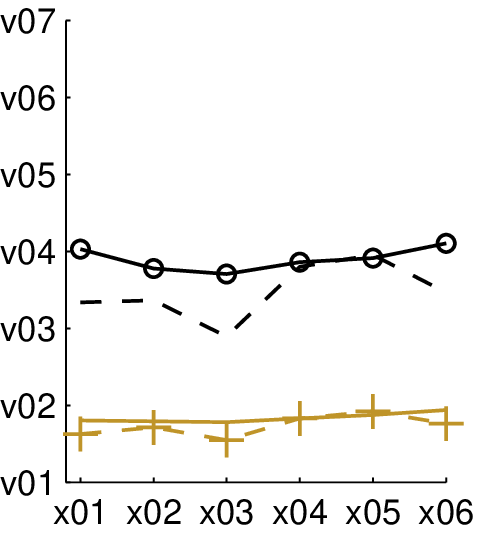}
      & \hspace{-0.5cm} 
	\psfrag{x01}[t][t]{\footnotesize0} \psfrag{x02}[t][t]{\footnotesize1} \psfrag{x03}[t][t]{\footnotesize2} \psfrag{x04}[t][t]{\footnotesize3} \psfrag{x05}[t][t]{\footnotesize4} \psfrag{x06}[t][t]{\footnotesize5}%
	\psfrag{v01}[r][r]{\footnotesize 0} \psfrag{v02}[r][r]{\footnotesize } \psfrag{v03}[r][r]{\footnotesize 1} \psfrag{v04}[r][r]{\footnotesize } \psfrag{v05}[r][r]{\footnotesize 2} \psfrag{v06}[r][r]{\footnotesize } \psfrag{v07}[r][r]{\footnotesize 3}
	    \psfrag{s06}[l][l]{\color[rgb]{0,0,0}\scriptsize $\bd{p}_i$ CoSLAS} \psfrag{s07}[l][l]{\color[rgb]{0,0,0}\scriptsize $\dot{\bd{p}}_i$ CoSLAS}%
	    \psfrag{s08}[l][l]{\color[rgb]{0,0,0}\scriptsize $\bd{p}_i$ ClkRef} \psfrag{s09}[l][l]{\color[rgb]{0,0,0}\scriptsize $\dot{\bd{p}}_i$ ClkRef}%
	    \psfrag{s11}[][]{} \psfrag{s12}[][]{}%
	\includegraphics[width=2.8cm]{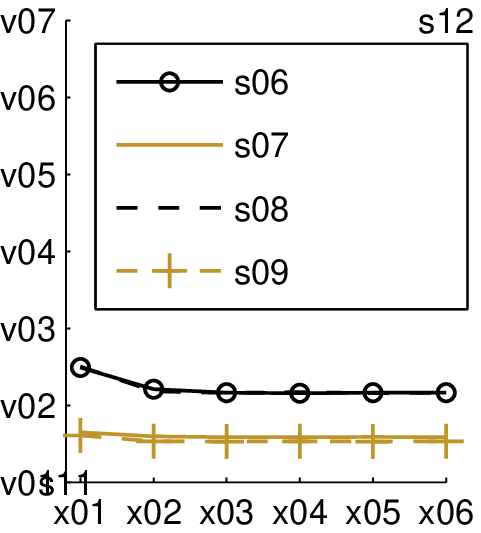}
      \\ \vspace{-4mm}  \\ \hspace{0.2cm} 
	\psfragscanon
	\psfrag{x01}[t][t]{\footnotesize0} \psfrag{x02}[t][t]{\footnotesize1} \psfrag{x03}[t][t]{\footnotesize2} \psfrag{x04}[t][t]{\footnotesize3} \psfrag{x05}[t][t]{\footnotesize4} \psfrag{x06}[t][t]{\footnotesize5}%
	\psfrag{v01}[r][r]{\tiny$0.01\!\!$} \psfrag{v02}[r][r]{\tiny$1$} \psfrag{v03}[r][r]{\tiny$100$} \psfrag{v04}[r][r]{\tiny$10^4$} \psfrag{v05}[r][r]{\tiny$10^6$} \psfrag{v06}[r][r]{\tiny$10^8$}%
	\includegraphics[width=2.8cm]{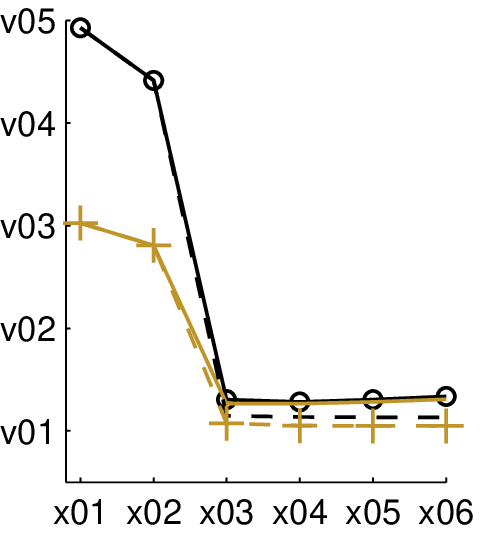} 
      & \hspace{-0.5cm}
	\psfrag{x01}[t][t]{\footnotesize0} \psfrag{x02}[t][t]{\footnotesize1} \psfrag{x03}[t][t]{\footnotesize2} \psfrag{x04}[t][t]{\footnotesize3} \psfrag{x05}[t][t]{\footnotesize4} \psfrag{x06}[t][t]{\footnotesize5}%
	\psfrag{v01}[r][r]{\tiny$0.01\!\!$} \psfrag{v02}[r][r]{\tiny$1$} \psfrag{v03}[r][r]{\tiny$100$} \psfrag{v04}[r][r]{\tiny$10^4$} \psfrag{v05}[r][r]{\tiny$10^6$} \psfrag{v06}[r][r]{\tiny$10^8$}%
	\includegraphics[width=2.8cm]{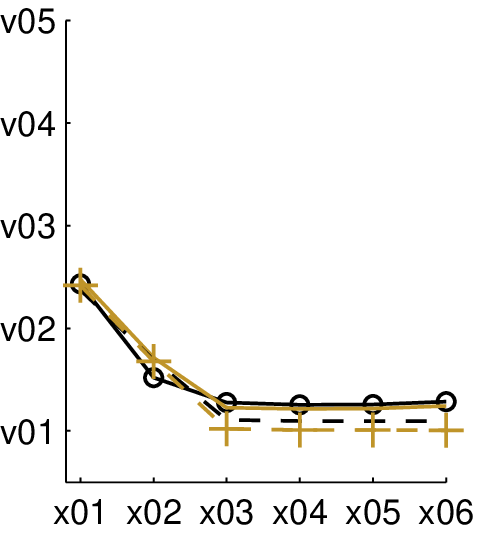}
      & \hspace{-0.5cm}
	\psfrag{x01}[t][t]{\footnotesize0} \psfrag{x02}[t][t]{\footnotesize1} \psfrag{x03}[t][t]{\footnotesize2} \psfrag{x04}[t][t]{\footnotesize3} \psfrag{x05}[t][t]{\footnotesize4} \psfrag{x06}[t][t]{\footnotesize5}%
	\psfrag{v01}[r][r]{\tiny$0.01\!\!$} \psfrag{v02}[r][r]{\tiny$1$} \psfrag{v03}[r][r]{\tiny$100$} \psfrag{v04}[r][r]{\tiny$10^4$} \psfrag{v05}[r][r]{\tiny$10^6$} \psfrag{v06}[r][r]{\tiny$10^8$}%
	    \psfrag{s06}[l][l]{\color[rgb]{0,0,0}\scriptsize $\beta_i$ CoSLAS} \psfrag{s07}[l][l]{\color[rgb]{0,0,0}\scriptsize $\alpha_i$ CoSLAS}%
	    \psfrag{s08}[l][l]{\color[rgb]{0,0,0}\scriptsize $\beta_i$ LocRef} \psfrag{s09}[l][l]{\color[rgb]{0,0,0}\scriptsize $\alpha_i$ LocRef}%
	    \psfrag{s11}[][]{} \psfrag{s12}[][]{}%
	\includegraphics[width=2.8cm]{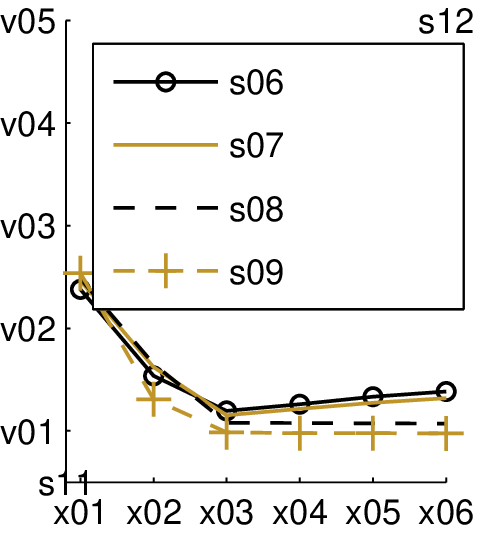}
      \end{tabular}
    \end{psfrags}
    \end{center}
    
    \vspace{-6.9cm}
    \begin{tikzpicture}[scale=1]
      \node at (0,1.5) {};
    
      \node[rotate=90] (A) at (0,0) {\footnotesize RMSE\;[m,\,m/s]};
      \node[rotate=90] (B) at (0,-3.3) {\footnotesize RMSE\;[$\mu$s,\,ppm]};
      \node (C) at (1.8,-5.15) {\footnotesize Iteration index $q$};
      \node (D) at (4.7,-5.15) {\footnotesize Iteration index $q$};
      \node (E) at (7.5,-5.15) {\footnotesize Iteration index $q$};
      
      \node[rectangle,draw=black,very thin,fill=white, inner sep=0.8mm] at (1.8,1.35) {\footnotesize$n = 0$};
      \node[rectangle,draw=black,very thin,fill=white, inner sep=0.8mm] at (4.7,1.35) {\footnotesize$n = 10$};
      \node[rectangle,draw=black,very thin,fill=white, inner sep=0.8mm] at (7.5,1.35) {\footnotesize$n = 20$};
    \end{tikzpicture}
    \vspace{-5.5mm}
  \caption{\label{sim:iterations} RMSEs versus message passing iteration index $q$ at times $n \rmv=\rmv 1$, $10$, and $20$.
 Top: location-related parameters, bottom: clock parameters.} 
\vspace{-.3mm}
\end{figure}

For times $n \!=\! 1$, $10$, and $20$, Fig.~\ref{sim:iterations} shows the \nolinebreak 
depend\-ence \nolinebreak 
of the RMSEs of 
location, velocity, clock phase, and clock skew (cf. Section \ref{sssec:Msg_Est}) 
on the message passing iteration index $q$. Here, differently from Section \ref{sssec:Msg_Est} and
Table \ref{alg:BP_ex}, the belief calculation and estimation steps were performed in each iteration $q$, for a total of $Q \rmv=\rmv 5$ iterations.
At $n \rmv=\rmv 1$, the RMSE of the locations $\bd{p}_i$ is seen to converge to a minimum after $q \rmv=\rmv 4$ 
iterations for CoSLAS and after $q \rmv=\rmv 2$ iterations for ClkRef. 
This difference can be explained by the fact that in ClkRef, all agents know their clocks
whereas in CoSLAS, distance messages can only be calculated when the agents possess informative clock messages 
(cf.\ Step 2.2 in Table~\ref{alg:BP_ex}). \linebreak 
Furthermore, the RMSE of $\dot{\bd{p}}_i$ 
does not decrease
with increasing $q$. This 
can be explained as follows. 
Via
\eqref{eq:MU_b3} and \eqref{eq:MU_b4}, the location accuracy expressed by $\zeta_{\psi_i}(\bd{x}_i)$ and $\zeta_{l_i}(\bd{x}_i)$---or, more specifically, 
by the first two (block) entries of the corresponding parameters $\bm{\mu}_{\psi_i \to x_i,r}$, $\bm{\Sigma}_{\psi_i \to x_i,r}$ and $\bm{\mu}_{l_i \to x_i,s}$, 
$\bm{\Sigma}_{l_i \to x_i,s}$, respectively---strongly influences the velocity accuracy expressed by $b(\bd{x}_i)$---or, more specifically, by the second two (block) entries of 
$\bm{\mu}_{x_i \to l_i,s}$, $\bm{\Sigma}_{x_i \to l_i,s}$.
%
But at $n \rmv=\rmv 1$,
$\zeta_{l_i}(\bd{x}_i)$ still contains large uncertainties inherited from the initial prior $f\big(\bd{x}_i^{(0)}\big)$. 
Therefore, $\dot{\bd{p}}_i$ cannot be estimated accurately at time $n \rmv=\rmv 1$. 
The RMSEs of the clock parameters $\alpha_i$ and $\beta_i$ converge to a minimum after
$q \rmv=\rmv 2$ iterations for both CoSLAS and LocRef. We note that $q \rmv=\rmv 2$ iterations 
correspond to the maximum hop distance from any nonreference agent to a spatial/temporal reference agent 
(in each 
iteration, the clock and location information is propagated by one hop).

At $n\rmv=\rmv 10$ and $n\rmv=\rmv 20$, the RMSEs of 
$\alpha_i$ and $\beta_i$ converge to a minimum 
in $q \rmv=\rmv 2$ iterations. 
At $n \rmv=\rmv 10$, the RMSE of $\bd{p}_i$ is rather high for all $q$.
This is because 
the top right agent in the ``$n \rmv=\rmv 10$'' part of Fig.~\ref{sim:scenario} has
two of its three neighbors effectively located in the same direction. This is no longer the case at $n\rmv=\rmv 20$,
and indeed the RMSE of $\bd{p}_i$ here converges approximately to a minimum in only $q \rmv=\rmv 1$ iteration.
Thus, one can obtain low communication cost without compromising the convergence of $\bd{p}_i$ by performing
only one message passing iteration per time step ($Q \rmv=\rmv 1$, which is sometimes referred to as ``real-time BP'' \cite{savic13simul}). 
We also see that at $n\rmv=\rmv 10$ and $n\rmv=\rmv 20$, remarkably, the RMSEs of CoSLAS are similar to or only slightly higher than those of ClkRef and LocRef. Thus, we can conclude that after a moderate number of time intervals,
CoSLAS compensates for the lack of perfect knowledge of 
the clock or location-related parameters. 

\begin{figure*}
    \begin{center}
      \begin{psfrags}
	\begin{tabular}{ccc}
	    \hspace{1cm}
	    \psfrag{x01}[t][t]{\footnotesize0} \psfrag{x02}[t][t]{\footnotesize10} \psfrag{x03}[t][t]{\footnotesize20} \psfrag{x04}[t][t]{\footnotesize30} \psfrag{x05}[t][t]{\footnotesize40} \psfrag{x06}[t][t]{\footnotesize50}%
	    \psfrag{v01}[r][r]{\footnotesize0} \psfrag{v02}[r][r]{\footnotesize10} \psfrag{v03}[r][r]{\footnotesize20} \psfrag{v04}[r][r]{\footnotesize30} \psfrag{v05}[r][r]{\footnotesize40} \psfrag{v06}[r][r]{\footnotesize50}%
	    \begin{tabular}{c}
	    \\[-3.82cm]
	    \includegraphics[width=3.65cm]{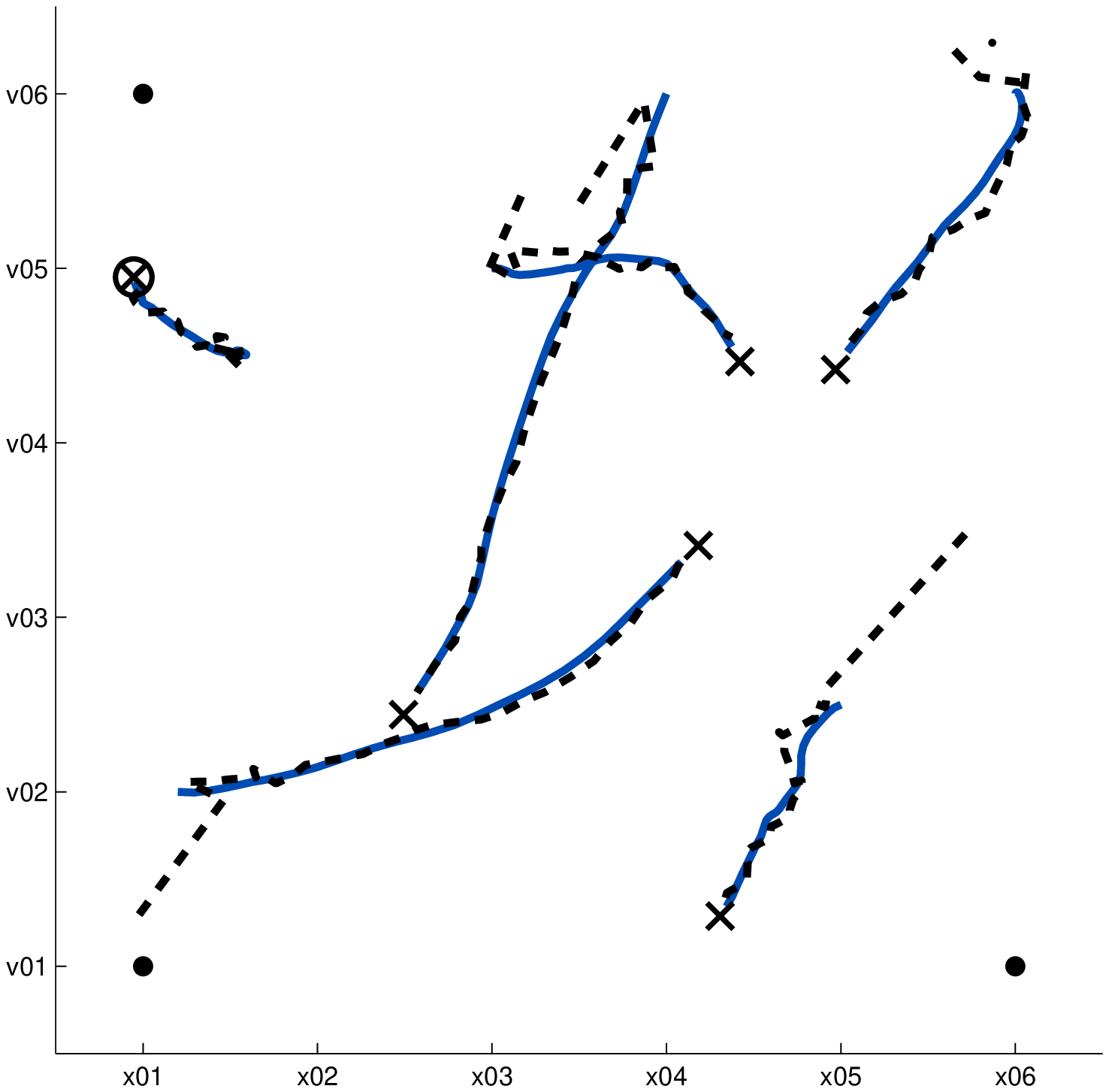}
	    \end{tabular}
	  & \hspace{0.2cm} 
	    \psfrag{s06}[l][l]{\color[rgb]{0,0,0}\scriptsize $\bd{p}_i$ CoSLAS} \psfrag{s07}[l][l]{\color[rgb]{0,0,0}\scriptsize $\dot{\bd{p}}_i$ CoSLAS}%
	    \psfrag{s08}[l][l]{\color[rgb]{0,0,0}\scriptsize $\bd{p}_i$ ClkRef} \psfrag{s09}[l][l]{\color[rgb]{0,0,0}\scriptsize $\dot{\bd{p}}_i$ ClkRef}%
	    \psfrag{s11}[][]{} \psfrag{s12}[][]{}%
	    \psfrag{x01}[t][t]{\footnotesize0} \psfrag{x02}[t][t]{\footnotesize5} \psfrag{x03}[t][t]{\footnotesize10} \psfrag{x04}[t][t]{\footnotesize15} \psfrag{x05}[t][t]{\footnotesize20} \psfrag{x06}[t][t]{\footnotesize25} \psfrag{x07}[t][t]{\footnotesize30}
	    \psfrag{v01}[r][r]{\footnotesize0} \psfrag{v02}[r][r]{\footnotesize} \psfrag{v03}[r][r]{\footnotesize1} \psfrag{v04}[r][r]{\footnotesize} \psfrag{v05}[r][r]{\footnotesize2} \psfrag{v06}[r][r]{\footnotesize} \psfrag{v07}[r][r]{\footnotesize3} \psfrag{v08}[r][r]{\footnotesize} \psfrag{v09}[r][r]{\footnotesize4} \psfrag{v10}[r][r]{\footnotesize} \psfrag{v11}[r][r]{\footnotesize5}%
	    \includegraphics[width=4.8cm]{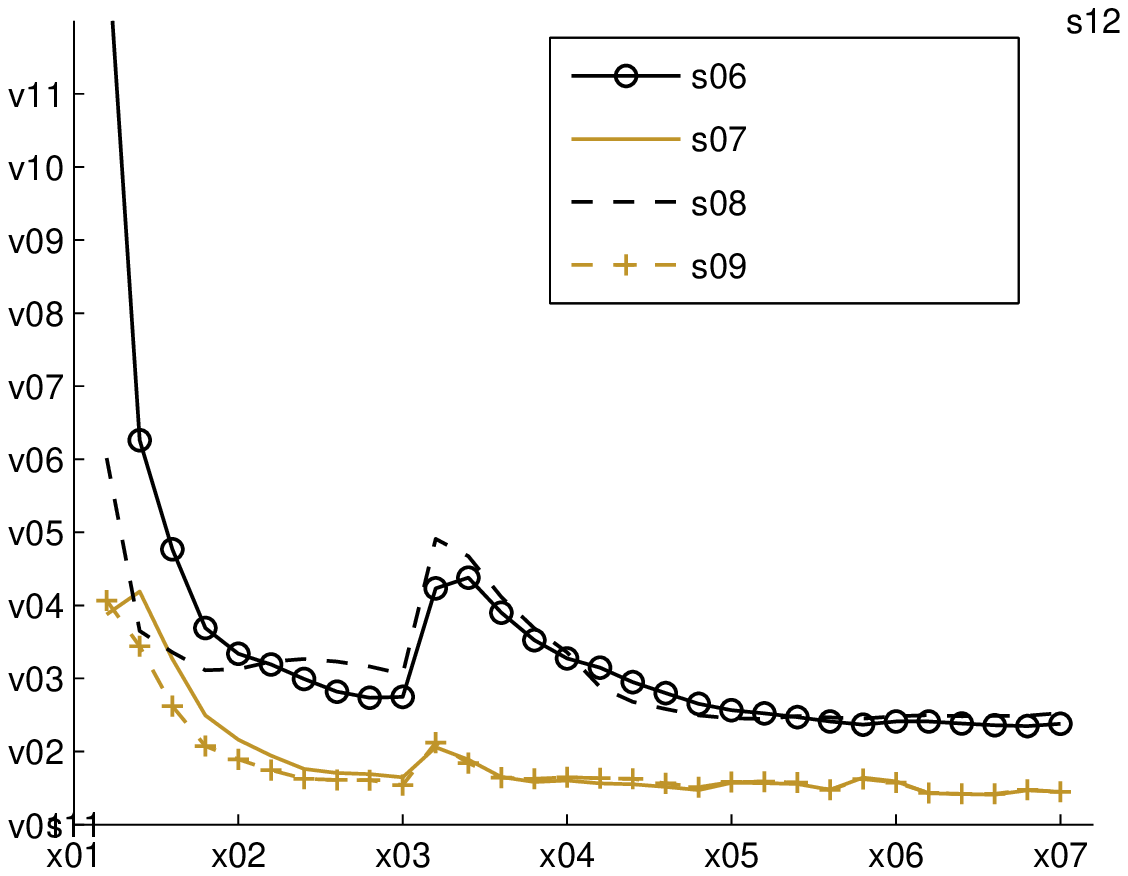}
	  & \hspace{0.2cm} 
	    \psfrag{s06}[l][l]{\color[rgb]{0,0,0}\scriptsize $\beta_i$ CoSLAS} \psfrag{s07}[l][l]{\color[rgb]{0,0,0}\scriptsize $\alpha_i$ CoSLAS}%
	    \psfrag{s08}[l][l]{\color[rgb]{0,0,0}\scriptsize $\beta_i$ LocRef} \psfrag{s09}[l][l]{\color[rgb]{0,0,0}\scriptsize $\alpha_i$ LocRef}%
	    \psfrag{s11}[][]{} \psfrag{s12}[][]{}%
	    \psfrag{x01}[t][t]{\footnotesize0} \psfrag{x02}[t][t]{\footnotesize5} \psfrag{x03}[t][t]{\footnotesize10} \psfrag{x04}[t][t]{\footnotesize15} \psfrag{x05}[t][t]{\footnotesize20} \psfrag{x06}[t][t]{\footnotesize25} \psfrag{x07}[t][t]{\footnotesize30}
	    \psfrag{v01}[r][r]{\footnotesize0} \psfrag{v02}[r][r]{\footnotesize } \psfrag{v03}[r][r]{\footnotesize 0.4} \psfrag{v04}[r][r]{\footnotesize } \psfrag{v05}[r][r]{\footnotesize 0.8} \psfrag{v06}[r][r]{\footnotesize} \psfrag{v07}[r][r]{\footnotesize 1.2} \psfrag{v08}[r][r]{\footnotesize } \psfrag{v09}[r][r]{\footnotesize 1.6} \psfrag{v10}[r][r]{\footnotesize } \psfrag{v11}[r][r]{\footnotesize2}%
	    \includegraphics[width=4.8cm]{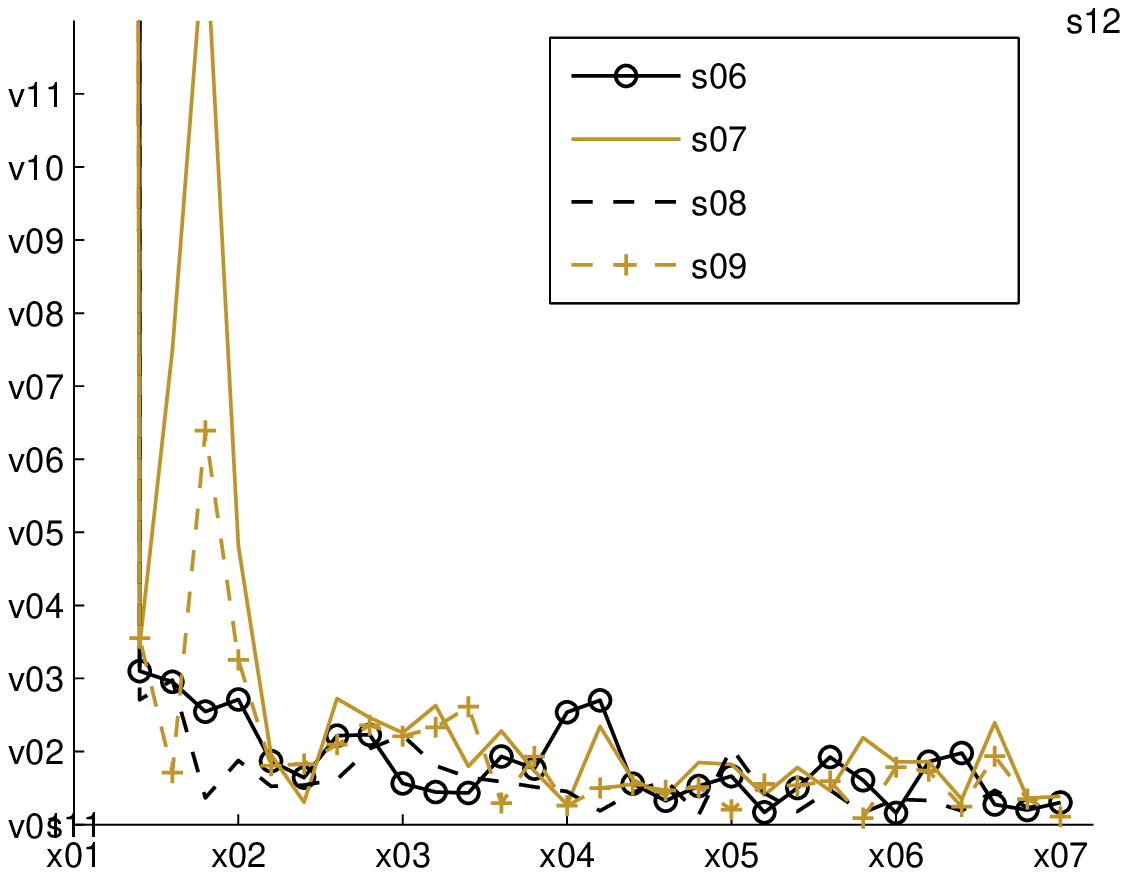}
	\end{tabular}
      \end{psfrags}
    \end{center}
    
    \vspace{-5.15cm}
      \begin{tikzpicture}[scale=1]
      \node(dum1) at (-0.5,0) {};
      \node(dum2) at (-0.5,5) {};
      \node(dum3) at (20,0) {};
      \node[rectangle,draw=black] at (0.2,2.3) {{\footnotesize $Q\rmv=\rmv1$}};
      \node[rotate=90] (A1) at (1.2,2.3) {\footnotesize$y\,$[m]};
      \node (B1) at (3.5,0) {\footnotesize$x\,$[m]};
      \node[rotate=90] (A2) at (6.0,2.3) {\footnotesize RMSE\;[m,\,m/s]};
      \node (B2) at (8.7,0) {\footnotesize Time index $n$};
      \node[rotate=90] (A3) at (11.4,2.3) {\footnotesize RMSE\;[$\mu$s,\,ppm]};
      \node (B3) at (14.3,0) {\footnotesize Time index $n$};
    \end{tikzpicture}
    
    \vspace{-2mm}
    \begin{center}
    \begin{psfrags}
      \begin{tabular}{ccc}
	  \hspace{1cm}
	    \psfrag{x01}[t][t]{\footnotesize0} \psfrag{x02}[t][t]{\footnotesize10} \psfrag{x03}[t][t]{\footnotesize20} \psfrag{x04}[t][t]{\footnotesize30} \psfrag{x05}[t][t]{\footnotesize40} \psfrag{x06}[t][t]{\footnotesize50}%
	    \psfrag{v01}[r][r]{\footnotesize0} \psfrag{v02}[r][r]{\footnotesize10} \psfrag{v03}[r][r]{\footnotesize20} \psfrag{v04}[r][r]{\footnotesize30} \psfrag{v05}[r][r]{\footnotesize40} \psfrag{v06}[r][r]{\footnotesize50}%
	    \begin{tabular}{c}
	    \\[-3.82cm]
	    \includegraphics[width=3.65cm]{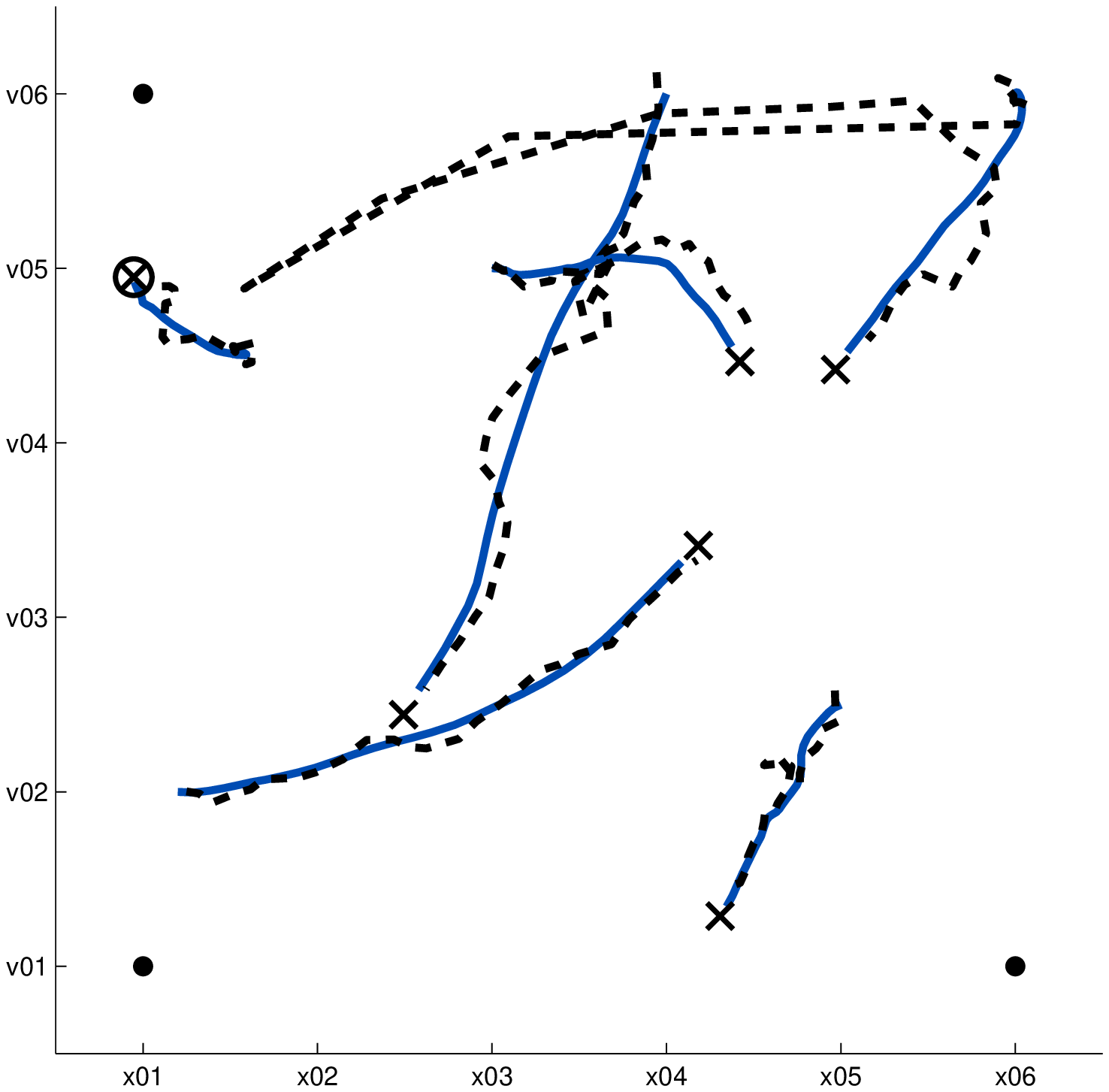}
	    \end{tabular}
	& \hspace{0.2cm} 
	    \psfrag{s06}[l][l]{\color[rgb]{0,0,0}\scriptsize $\bd{p}_i$ CoSLAS} \psfrag{s07}[l][l]{\color[rgb]{0,0,0}\scriptsize $\dot{\bd{p}}_i$ CoSLAS}%
	    \psfrag{s08}[l][l]{\color[rgb]{0,0,0}\scriptsize $\bd{p}_i$ ClkRef} \psfrag{s09}[l][l]{\color[rgb]{0,0,0}\scriptsize $\dot{\bd{p}}_i$ ClkRef}%
	    \psfrag{s11}[][]{} \psfrag{s12}[][]{}%
	    \psfrag{x01}[t][t]{\footnotesize0} \psfrag{x02}[t][t]{\footnotesize5} \psfrag{x03}[t][t]{\footnotesize10} \psfrag{x04}[t][t]{\footnotesize15} \psfrag{x05}[t][t]{\footnotesize20} \psfrag{x06}[t][t]{\footnotesize25} \psfrag{x07}[t][t]{\footnotesize30}
	    \psfrag{v01}[r][r]{\footnotesize0} \psfrag{v02}[r][r]{\footnotesize} \psfrag{v03}[r][r]{\footnotesize1} \psfrag{v04}[r][r]{\footnotesize} \psfrag{v05}[r][r]{\footnotesize2} \psfrag{v06}[r][r]{\footnotesize} \psfrag{v07}[r][r]{\footnotesize3} \psfrag{v08}[r][r]{\footnotesize} \psfrag{v09}[r][r]{\footnotesize4} \psfrag{v10}[r][r]{\footnotesize} \psfrag{v11}[r][r]{\footnotesize5}%
	    \includegraphics[width=4.8cm]{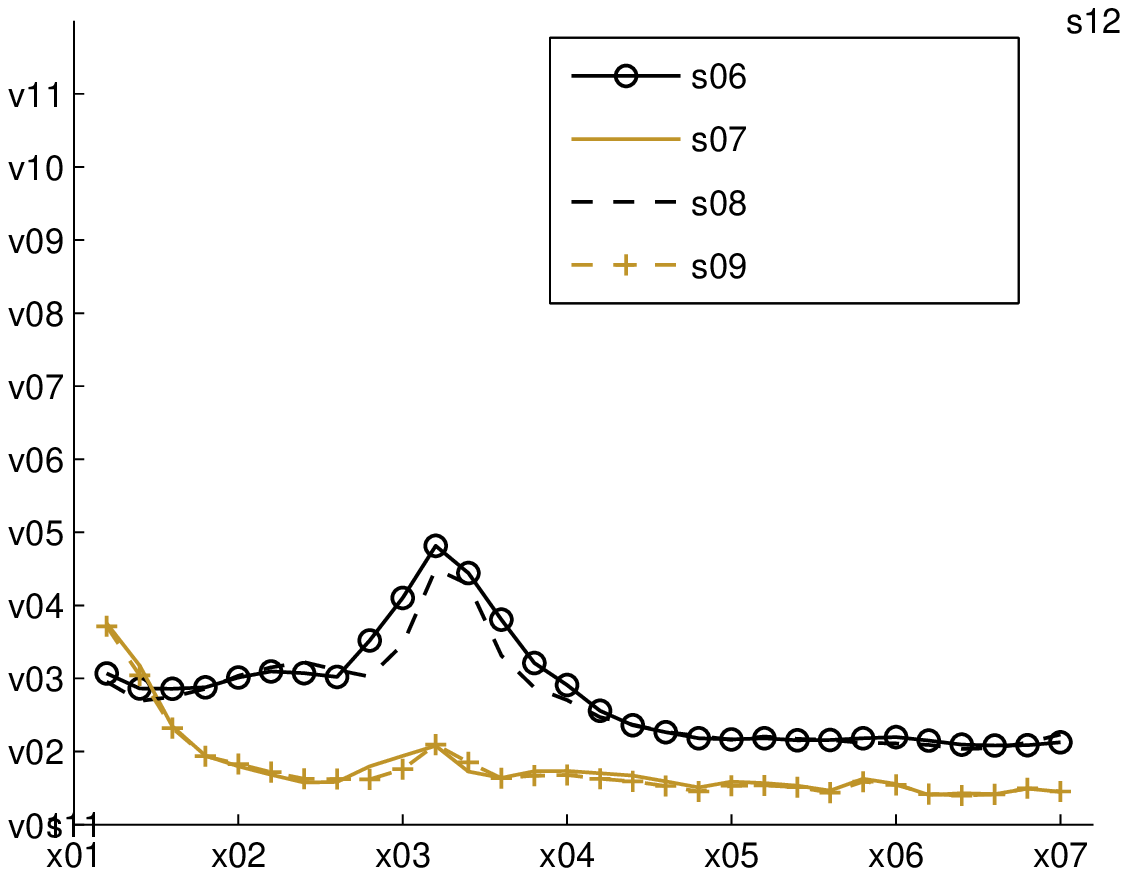}
	& \hspace{0.2cm} 
	    \psfrag{s06}[l][l]{\color[rgb]{0,0,0}\scriptsize $\beta_i$ CoSLAS} \psfrag{s07}[l][l]{\color[rgb]{0,0,0}\scriptsize $\alpha_i$ CoSLAS}%
	    \psfrag{s08}[l][l]{\color[rgb]{0,0,0}\scriptsize $\beta_i$ LocRef} \psfrag{s09}[l][l]{\color[rgb]{0,0,0}\scriptsize $\alpha_i$ LocRef}%
	    \psfrag{s11}[][]{} \psfrag{s12}[][]{}%
	    \psfrag{x01}[t][t]{\footnotesize0} \psfrag{x02}[t][t]{\footnotesize5} \psfrag{x03}[t][t]{\footnotesize10} \psfrag{x04}[t][t]{\footnotesize15} \psfrag{x05}[t][t]{\footnotesize20} \psfrag{x06}[t][t]{\footnotesize25} \psfrag{x07}[t][t]{\footnotesize30}
	    \psfrag{v01}[r][r]{\footnotesize0} \psfrag{v02}[r][r]{\footnotesize 0.02} \psfrag{v03}[r][r]{\footnotesize 0.04} \psfrag{v04}[r][r]{\footnotesize 0.06} \psfrag{v05}[r][r]{\footnotesize0.08} \psfrag{v06}[r][r]{\footnotesize0.1} \psfrag{v07}[r][r]{\footnotesize} 
	    \includegraphics[width=4.8cm]{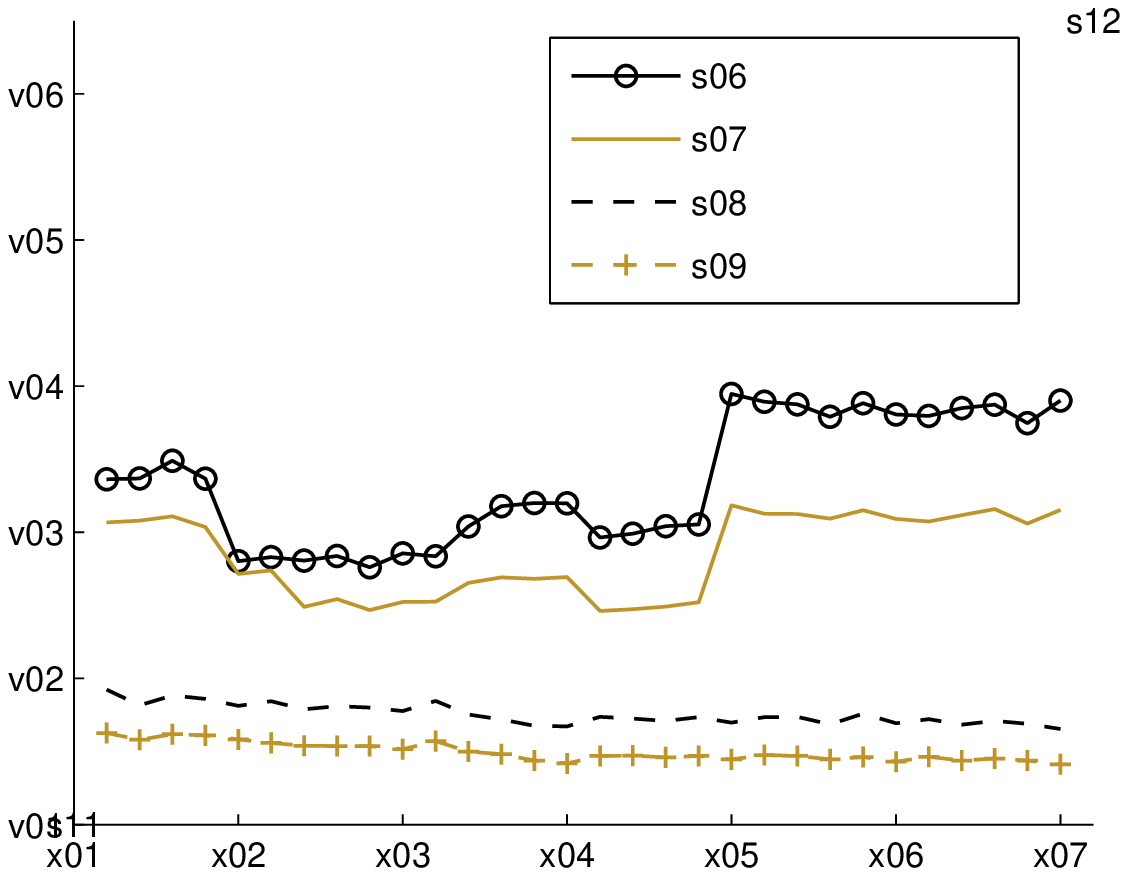}
      \end{tabular}
    \end{psfrags}
    \end{center}
    
    \vspace{-5.15cm}
      \begin{tikzpicture}[scale=1]
      \node(dum1) at (-0.5,0) {};
      \node(dum2) at (-0.5,5) {};
      \node(dum3) at (20,0) {};
      \node[rectangle,draw=black] at (0.2,2.3) {{\footnotesize$Q\rmv=\rmv5$}};
      \node[rotate=90] (A1) at (1.2,2.3) {\footnotesize$y\,$[m]};
      \node (B1) at (3.5,0) {\footnotesize$x\,$[m]};
      \node[rotate=90] (A2) at (6,2.3) {\footnotesize RMSE\;[m,\,m/s]};
      \node (B2) at (8.7,0) {\footnotesize Time index $n$};
      \node[rotate=90] (A3) at (11.15,2.3) {\footnotesize RMSE\;[$\mu$s,\,ppm]};
      \node (B3) at (14.3,0) {\footnotesize Time index $n$};
    \end{tikzpicture}
    \vspace{-6mm}
  \caption{\label{sim:timeevolution} Trajectories and RMSEs
 for $Q \rmv=\rmv 1$ (top) and $Q \rmv=\rmv 5$ (bottom). 
  In the leftmost figures, solid blue lines indicate the true trajectories and dashed black lines 
  the estimated trajectories.} 
\vspace{-1.5mm}
\end{figure*}


In Fig.~\ref{sim:timeevolution}, we show the estimated and true 
trajectories and the RMSEs versus time $n$ for 
$Q \rmv=\rmv 1$ and $Q \rmv=\rmv 5$.
It is seen
that at early times, the location RMSE is higher for $Q \rmv=\rmv 1$
than for $Q \rmv=\rmv 5$.
The increased location RMSE
around time $n \rmv=\rmv 10$ can be explained as before. 
The clock RMSE is generally higher for $Q \rmv=\rmv 1$ since the clock information provided by the temporal 
reference agents cannot be disseminated throughout the network during one 
message passing iteration, and hence (because $Q \rmv=\rmv 1$) during one time step.
However, the location-related RMSEs 
suggest that the local synchronicity between neighboring agents is sufficient for obtaining accurate location-related estimates.
The fluctuation of the clock RMSEs
is caused by the time-varying network connectivity and the random-walk evolution model \eqref{eq:clockEvol}. 
Finally, 
the performance of CoSLAS is again generally close to that of ClkRef and 
\vspace{-1.3mm}
LocRef. 


\section{Conclusion}

We presented a distributed, sequential belief propagation (BP) 
algorithm for cooperative simultaneous localization and synchronization (CoSLAS)
in mobile, decentralized agent networks with time-varying clocks. The agents 
acquire 
interagent distance estimates from time-of-flight measurements.
We exploited the resulting close relation between localization and synchronization to establish a common statistical formulation that
features a 
conditional independence of 
time measurements and 
location-related parameters given the interagent distances.
This independence is leveraged by the proposed BP algorithm 
to obtain reduced dimensions of the messages and thus a reduced complexity.
The combined use of particle representations and parametric 
representations leads to high accuracy at low communication cost,
and a judiciously chosen message schedule 
allows for real-time operation in networks with rapidly changing connectivity.
Simulation results demonstrated the good performance of the proposed algorithm in a challenging scenario with only one temporal reference agent
and time-varying network connectivity.

\appendices

\section{}
\label{app:gauss_zeta_dp}


We derive the Gaussian approximation of $\zeta^{\internode{(q)}}_{\phi_{ij}}(d_{ij})$ presented
in Section \ref{sssec:Msg_itMsgPass_zeta-phi-d}.
According to \eqref{eq:BP_zeta}, we have
\[
\zeta^{\internode{(q)}}_{\phi_{ij}}(d_{ij})  \ist=\rmv \int\!\!\rmv\int \rmv \phi_{ij} \, \eta^{\internode{(q-1)}}_{\phi_{ij}}(\tilde{\bd{p}}_i) \, \eta^{\internode{(q-1)}}_{\phi_{ji}}(\tilde{\bd{p}}_j) \,  
  \mathrm{d}\tilde{\bd{p}}_i \ist \mathrm{d}\tilde{\bd{p}}_j \ist,
\]
with ${\phi}_{ij} = \delta\big( \| \tilde{\bd{p}}_{i} \!-\rmv \tilde{\bd{p}}_{j} \| \rmv-\rmv d_{ij} \big)$. Inserting 
the Gaussian mixture representations of $\eta^{\internode{(q-1)}}_{\phi_{ij}}(\tilde{\bd{p}}_i)$ and $\eta^{\internode{(q-1)}}_{\phi_{ji}}(\tilde{\bd{p}}_j)$  
(cf.\ Table \ref{tab:gausspar-2}) gives
\vspace{-3mm}
\be
\zeta^{\internode{(q)}}_{\phi_{ij}}(d_{ij})  \ist=\rmv \sum_{r=1}^{S_{i\to j}^{(q-1)}} \rmv \sum_{s=1}^{S_{j\to i}^{(q-1)}} \!\!w_{i\to j,r}^{(q-1)} \ist w_{j\to i,s}^{(q-1)} 
  \ist \Psi_{ij,rs}(d_{ij}) \ist,
\label{eq:app_zeta_df_mixture_sum}
\vspace{-2mm}
\ee
where 
\begin{align}
&\hspace{-1mm}\Psi_{ij,rs}(d_{ij})\nn \\[0mm]
&\,\triangleq \! \int\!\!\rmv \int \rmv \delta\big( \| \tilde{\bd{p}}_{i,r} \!-\rmv \tilde{\bd{p}}_{j,s} \| \rmv-\rmv d_{ij} \big) \, 
  \mathcal{N}\big( \tilde{\bd{p}}_{i,r}; \bm{\mu}^{\internode{(q-1)}}_{\tilde{p}_{i} \to \phi_{ij},r},  \bm{\Sigma}^{\internode{(q-1)}}_{\tilde{p}_i \to \phi_{ij},r} \big) \nn\\[-.5mm]
&\hspace{12mm} \times  \ist \mathcal{N} \big( \tilde{\bd{p}}_{j,s}; \bm{\mu}^{\internode{(q-1)}}_{\tilde{p}_j \to \phi_{ji},s},  \bm{\Sigma}^{\internode{(q-1)}}_{\tilde{p}_j \to \phi_{ji},s} \big) 
  \ist \mathrm{d}\tilde{\bd{p}}_{i,r} \ist \mathrm{d}\tilde{\bd{p}}_{j,s} \ist.
\label{eq:app_zeta_df_mixture_0}
\end{align}
Here, $\Psi_{ij,rs}(d_{ij})$ describes the $(r,s)$th Gaussian mixture \nolinebreak 
component \nolinebreak 
 and, e.g., $\tilde{\bd{p}}_{i,r} \sim \mathcal{N}\big( \tilde{\bd{p}}_{i,r}; \bm{\mu}^{\internode{(q-1)}}_{\tilde{p}_i \to \phi_{ij},r},\bm{\Sigma}^{\internode{(q-1)}}_{\tilde{p}_i \to \phi_{ij},r} \big)$ 
\nolinebreak 
corresponds \nolinebreak 
to the $r$th Gaussian component.
We can write $d_{ij} \rmv= \| \tilde{\bd{p}}_{i,r} \!-\rmv \tilde{\bd{p}}_{j,s} \|$ as a function $d_{ij} \rmv=\rmv \chi(\tilde{\bd{p}}_{ij,rs})$
of the stacked vector\linebreak 
$\tilde{\bd{p}}_{ij,rs} \rmv\triangleq\rmv \big[\tilde{\bd{p}}_{i,r}^{\text{T}} \,\ist\ist \tilde{\bd{p}}_{j,s}^{\text{T}} \big]^{\text{T}}\!$.
We have $\tilde{\bd{p}}_{ij,rs} \rmv\sim\rmv f(\tilde{\bd{p}}_{ij,rs}) \rmv=\rmv \mathcal{N}( \tilde{\bd{p}}_{ij,rs};$\linebreak 
$ \bm{\mu}_{ij,rs},\bm{\Sigma}_{ij,rs})$, where 
$\bm{\mu}_{ij,rs} \rmv\triangleq\rmv \big[\bm{\mu}^{\internode{(q-1)\text{T}}}_{\tilde{p}_i \to \phi_{ij},r}\;\, \bm{\mu}^{\internode{(q-1)\text{T}}}_{\tilde{p}_{j} \to \phi_{ji},s}\big]^{\text{T}}\!$ 
and\linebreak 
$\bm{\Sigma}_{ij,rs}$ has been specified in Section \ref{sssec:Msg_itMsgPass_zeta-phi-d}; 
furthermore,\linebreak 
$\mathcal{N}\big( \tilde{\bd{p}}_{i,r};\bm{\mu}^{\internode{(q-1)}}_{\tilde{p}_i \to \phi_{ij},r}, \bm{\Sigma}^{\internode{(q-1)}}_{\tilde{p}_i \to \phi_{ij},r} \big) \ist 
\mathcal{N} \big( \tilde{\bd{p}}_{j,s}; \bm{\mu}^{\internode{(q-1)}}_{\tilde{p}_j \to \phi_{ji},s},  \bm{\Sigma}^{\internode{(q-1)}}_{\tilde{p}_j \to \phi_{ji},s} \big)$\linebreak 
$=\rmv \mathcal{N}( \tilde{\bd{p}}_{ij,rs}; \bm{\mu}_{ij,rs},  \bm{\Sigma}_{ij,rs})$.
Therefore, we can rewrite
\eqref{eq:app_zeta_df_mixture_0} 
as
\begin{align}
&\hspace{-2mm}\Psi_{ij,rs}(d_{ij})\nn \\[0mm]
&\hspace{-1mm}\triangleq \! \int \rmv \delta\big( \chi(\tilde{\bd{p}}_{ij,rs})
\rmv-\rmv d_{ij} \big)
\, \mathcal{N}( \tilde{\bd{p}}_{ij,rs}; \bm{\mu}_{ij,rs},  \bm{\Sigma}_{ij,rs}) \, \mathrm{d}\tilde{\bd{p}}_{ij,rs}.\! \label{eq:app_zeta_df_mixture_1} 
\end{align}
For an approximate evaluation of this integral,
we linearize the function $\chi(\tilde{\bd{p}}_{ij,rs})$
around 
$\bm{\mu}_{ij,rs}$.
This yields
\begin{align}
&\hspace*{-2mm}\chi(\tilde{\bd{p}}_{ij,rs}) \nn \\[0mm]
&\hspace*{-2mm}\,\approx\ist \tilde{\chi}_{rs}(\tilde{\bd{p}}_{ij,rs}) 
  \ist\triangleq\ist \big\| \bm{\mu}_{d_{ij},rs}^{(q-1)} \big\| + \bar{\bm{\mu}}_{d_{ij},rs}^{(q-1)\text{T}} (\tilde{\bd{p}}_{ij,rs} \!- \rmv \bm{\mu}_{ij,rs} ) \ist , \!\! \label{eq:app_z_def}
\end{align}
with $\bm{\mu}_{d_{ij},rs}^{(q-1)}$ and $\bar{\bm{\mu}}_{d_{ij},rs}^{(q-1)}$ as defined in Section \ref{sssec:Msg_itMsgPass_zeta-phi-d}.
Inserting \eqref{eq:app_z_def} into \eqref{eq:app_zeta_df_mixture_1}, we 
obtain the approximation
\begin{align}
&\hspace*{-1mm}\Psi_{ij,rs}(d_{ij})\nn\\[.5mm] 
&\hspace*{1mm}\,\approx\ist \tilde{\Psi}_{ij,rs}(d_{ij}) \nn\\[.5mm]
&\hspace*{1mm}\,\triangleq\! \int \rmv \delta\big( \tilde{\chi}_{rs}(\tilde{\bd{p}}_{ij,rs}) \rmv-\rmv d_{ij} \big) \, 
  \mathcal{N}( \tilde{\bd{p}}_{ij,rs}; \bm{\mu}_{ij,rs},  \bm{\Sigma}_{ij,rs}) \,  \mathrm{d}\tilde{\bd{p}}_{ij,rs} \ist .\nn \\[-2mm]
\label{eq:app_psi_0} \\[-7.1mm]
\nn
\end{align}
Within our approximation $d_{ij} \!\rmv\approx\rmv \tilde{\chi}_{rs}(\tilde{\bd{p}}_{ij,rs})$, $\delta\big( \tilde{\chi}_{rs}(\tilde{\bd{p}}_{ij,rs}) -$\linebreak 
$d_{ij} \big)$ can be interpreted as 
$f(d_{ij} | \tilde{\bd{p}}_{ij,rs})$. 
Hence, \eqref{eq:app_psi_0} becomes
\be
\hspace{-.1mm}\tilde{\Psi}_{ij,rs}(d_{ij}) =\! \int \!\rmv f(d_{ij} | \tilde{\bd{p}}_{ij,rs}) \ist f(\tilde{\bd{p}}_{ij,rs}) \ist \mathrm{d}\tilde{\bd{p}}_{ij,rs} 
 \rmv=\rmv f_{rs}(d_{ij}) \ist , \!\!\!\!\!
  \label{eq:app_psi_1}
\ee
where $f_{rs}(d_{ij})$ denotes the pdf of $d_{ij}$ under our approximation $d_{ij} \!\approx\rmv \tilde{\chi}_{rs}(\tilde{\bd{p}}_{ij,rs})$.
Because $\tilde{\bd{p}}_{ij,rs} \sim \mathcal{N}( \tilde{\bd{p}}_{ij,rs}; \bm{\mu}_{ij,rs},  \bm{\Sigma}_{ij,rs})$ and 
$\tilde{\chi}_{rs}(\cdot)$ is an affine function (see \eqref{eq:app_z_def}),
$f_{rs}(d_{ij})$ is 
again Gaussian, i.e., $f_{rs}(d_{ij}) = \mathcal{N}\big( d_{ij}; \mu_{d,rs},  \sigma^2_{d,rs} \big)$, with
\begin{align*}
\mu_{d,rs}\rmv &=\ist \mathbb{E}[\tilde{\chi}_{rs}(\tilde{\bd{p}}_{ij,rs})] \\
&= \big\| \bm{\mu}_{d_{ij},rs}^{(q-1)} \big\| + \bar{\bm{\mu}}_{d_{ij},rs}^{(q-1)\text{T}} \big(\mathbb{E}[\tilde{\bd{p}}_{ij,rs}] \!- \rmv \bm{\mu}_{ij,rs} \big) \nn\\[.5mm]
&= \big\| \bm{\mu}_{d_{ij},rs}^{(q-1)} \big\| \nn\\[-7mm]
\end{align*}
and
\vspace{-2mm}
\begin{align*}
\sigma^2_{d,rs}\rmv &=\ist \mathrm{var} [ \tilde{\chi}_{rs}(\tilde{\bd{p}}_{ij,rs}) ] \\[.7mm] 
&=\ist \bar{\bm{\mu}}_{d_{ij},rs}^{(q-1)\text{T}} \ist \mathrm{cov} [ \tilde{\bd{p}}_{ij,rs} \!- \rmv \bm{\mu}_{ij,rs} ] \, \bar{\bm{\mu}}_{d_{ij},rs}^{(q-1)} \\[.7mm]
&=\ist \bar{\bm{\mu}}_{d_{ij},rs}^{(q-1)\text{T}} \ist \bm{\Sigma}_{ij,rs} \, \bar{\bm{\mu}}_{d_{ij},rs}^{(q-1)} \ist. 
\end{align*}
Thus, because of \eqref{eq:app_psi_1}, we also have
$\tilde{\Psi}_{ij,rs}(d_{ij}) \rmv=\rmv \mathcal{N}\big( d_{ij};$\linebreak 
$\mu_{d,rs}, \sigma^2_{d,rs} \big)$. 
Substituting this for $\Psi_{ij,rs}(d_{ij})$ in \eqref{eq:app_zeta_df_mixture_sum} yields
\[ 
\zeta^{\internode{(q)}}_{\phi_{ij}}(d_{ij})  \ist\approx\rmv  \sum_{r=1}^{S_{i\to j}^{(q-1)}} \rmv \sum_{s=1}^{S_{j\to i}^{(q-1)}} \!\!w_{i\to j,r}^{(q-1)} \ist w_{j\to i,s}^{(q-1)} 
  \,\ist \mathcal{N}\big( d_{ij};\mu_{d,rs}, \sigma^2_{d,rs} \big) \ist.
\]
This is a mixture of up to four Gaussian components. Finally, we use moment matching \cite{orguner07_momentmatching} to
approximate this Gaussian mixture by a single Gaussian. The resulting mean and variance
are given in \eqref{eq:IU_d3} and \eqref{eq:IU_d4}, respectively.

\bibliographystyle{ieeetr_noParentheses}
\bibliography{references}

\end{document}